\RequirePackage{fix-cm}
\documentclass[smallextended]{svjour3}       
\smartqed  
\usepackage{graphicx}
\usepackage{cite}
\usepackage{amsmath,amsfonts,amssymb}
\usepackage{subfigure}
\usepackage{float}
\usepackage{morefloats}
\usepackage{color}
\usepackage{multirow}
\usepackage{algorithm}
\usepackage{algpseudocode}

\extrafloats{100}

\begin{document}

\title{Neural network closures for nonlinear model order reduction  
}

\titlerunning{Machine learning for reduced order models}        

\author{Omer San         \and
        Romit Maulik 
}


\institute{O. San \at
              Department of Mechanical and Aerospace Engineering \\
              Oklahoma State University \\
              Stillwater, OK 74078, USA \\
              Tel.: +1 (405) 744-2457\\
              Fax: +1 (405) 744-7873\\
              \email{osan@okstate.edu}
           \and
           R. Maulik \at
              Department of Mechanical and Aerospace Engineering \\
              Oklahoma State University \\
              Stillwater, OK 74078, USA
}

\date{Received: \today }

\maketitle

\begin{abstract}
Many reduced order models are neither robust with respect to the parameter changes nor cost-effective enough for handling the nonlinear dependence of complex dynamical systems. In this study, we put forth a robust machine learning framework for projection based reduced order modeling of such nonlinear and nonstationary systems. As a demonstration, we focus on a nonlinear advection-diffusion system given by the viscous Burgers equation, which is a prototype setting of more realistic fluid dynamics applications with the same quadratic nonlinearity. In our proposed methodology the effects of truncated modes are modeled using a singe layer feed-forward neural network architecture. The neural network architecture is trained by utilizing both the Bayesian regularization and extreme learning machine approaches, where the latter one is found to be computationally more efficient. A particular effort is devoted to the selection of basis functions considering the proper orthogonal decomposition and Fourier bases. It is shown that the proposed models yield significant improvements in the accuracy over the standard Galerkin projection models with a negligibly small computational overhead and provide reliable predictions with respect to the parameter changes.


\end{abstract}

\section{Introduction}
\label{sec:intro}

The ever-increasing need for accurate prediction of many complex nonlinear processes leads to very large-scale dynamical systems whose simulations and analyses make overwhelming and unmanageable demands on computational resources. Since the computational cost of traditional full-order numerical simulations is extremely prohibitive, many successful model order reduction approaches have been introduced \cite{roychowdhury1999reduced,fang2009pod,buffoni2006low,lucia2004reduced,fortuna2012model,silveira1996efficient,freund1999reduced,noack2011reduced,dyke1996modeling,kazantzis2010new,akhtar2015using}. The purpose of such approaches is to reduce this computational burden and serve as surrogate models for efficient computational analysis of such systems, especially in settings where the traditional methods require repeated model evaluations over a large range of parameter values. Simplifying computational complexity of the underlying mathematical model these reduced order models offer promises in many prediction, identification, design, optimization, and control applications. However, they are neither robust with respect to the parameter changes nor low-cost to handle nonlinear dependence for complex nonlinear dynamical systems encountered in a wide range of physical phenomena. Therefore, reduced order modeling (ROM) remains an open challenge and the development of efficient and reliable model order reduction techniques is of paramount importance for both fundamental and applied science. In this study, we focus on the development of an accurate and robust machine learning framework for projection based nonlinear model order reduction of such transient nonlinear systems.

The basic philosophy of projection based approaches is the alleviation of the intense computational burden of the partial differential equations obtained from governing laws \cite{benner2015survey}. These approaches endeavor to reduce the high degrees of freedom of a governing law through an expansion in a transformed space, traditionally with orthogonal bases. The choice of the transformation aids the user in the identification of sparseness and thus allows them to obtain a dense (but low dimensional) representation of the same governing law. Among the large variety of reduced order modeling strategies, \emph{proper orthogonal decomposition} (POD) has emerged as a popular technique for the study of dynamical systems \cite{amsallem2008interpolation,aubry1988dynamics,banyay2014proper}. With respect to terminology, POD is also referred to as the Karhunen-Lo\'{e}ve expansion \cite{loeve1955probability}, principal component analysis \cite{hotelling1933analysis} or the empirical orthogonal function \cite{lorenz1956empirical}. A successful construction of quality POD bases requires a collection of high fidelity numerical simulations of the governing equations of the dynamic system being studied if an exact solution is not available. This information is used to devise optimal bases for the transformed space. Indeed, the construction of these bases also serve as a post-processing tool for instance as a method of extraction of the large scale coherent structures in statistical pattern recognition \cite{lumley1967structures}. This is because the global POD modes may be spanned through a considerably truncated number of bases to resolves attractors and transients as well. This study, however, is oriented more towards the feedback flow control community which attempts to used POD as a degree of freedom reduction technique. We would like to emphasize here that there exist several noteworthy approaches to ROM implementation such as Dynamic Mode Decomposition (or DMD) which is a technique for determining a low-dimensional subspace from observed data similar to POD but with a description of the dynamics on that subspace \cite{schmid2010dynamic}. Another popular technique for determining a reduced subspace is the extended DMD (or EDMD). The reader is direct to \cite{rowley2017model,taira2017modal} for an excellent discussion of these closely related approaches.

In our investigation, we implement the POD for our ROM analysis using a Galerkin projection (GP) approach. POD-GP has extensively been used to provide fast and accurate simulations of large nonlinear systems \cite{borggaard2007interval,kunisch2001galerkin,weller2010numerical}. This approach is devised so as to generate a reduced model where the truncated modes are evolved through time as a set of ordinary differential equations. This may be considered to be a \emph{projection} type approach as it is obtained by projecting the truncated modes (obtained from our POD) onto the governing equations. We must remark here that the an \emph{orthogonal} projection has been used to move our problem to the reduced subspace. A non-orthogonal projection may also be employed which is generally known as the Petrov-Galerkin projection \cite{antoulas2005approximation}. Another approach one may use for nonlinear systems is given by Koopman operator theory \cite{gaspard2005chaos} which identifies a subspace where the nonlinear dynamics are linearized and uncoupled. However, in this investigation we limit ourselves to the GP approach alone. The process of truncation (which is necessary for degree of freedom reduction) introduces a limitation to our chosen approach. The truncation procedure causes a significant loss of information particularly for highly non-stationary, convective and nonlinear problems \cite{cordier2010calibration,el2016new}. Therefore, there have been considerable efforts to mitigate the drawbacks of this loss of information such as through the process of closure modeling \cite{wang2011two,wang2012reduced,san2013proper,borggaard2016goal} where arguments are made in favor of modeling the effect of the discarded modes on the preserved ones. Another promising approach is to develop strategies for constructing more representative bases \cite{san2015stabilized,bui2007goal,carlberg2011low,iollo2000stability,kunisch2010optimal,buffoni2006low}. We note that bases obtained from orthogonal Fourier modes have also been utilized in this investigation for added comparison.

By analogy with the large eddy simulations of turbulent flows, several \emph{eddy viscosity} type of closure models have been introduced for ROMs \cite{aubry1988dynamics,noack2002low,ullmann2010pod,sirisup2004spectral,bergmann2009improvement,borggaard2011artificial,wang2011two,wang2012proper,akhtar2012new,san2013proper,san2015stabilized}.  Recently, a new closure modeling framework for stabilization of ROMs has been proposed by using a robust Lyapunov control theory. It has been demonstrated that the Lyapunov-based closure model is robust to parametric uncertainties and the free parameters are optimized using a data-driven multi-parametric extremum seeking algorithm \cite{benosman2016learning,benosman2017learning}. Other stabilization models for ROMs have been also suggested in literature \cite{noack2011reduced,balajewicz2012stabilization,amsallem2012stabilization,cordier2010calibration,lassila2013model,rowley2005model,imtiaz2016closure,benosman2016learning,xie2017approximate,wells2017evolve}. In addition to these models, Cazemier's closure modeling approach \cite{cazemier1997proper,cazemier1998proper,san2013proper} uses the concept of energy conservation to account for the truncated modes.

The focus of the present investigation will be to utilize a data-driven supervised \emph{machine learning} framework for the closure (and stabilization) of a highly truncated POD implemented using the GP approach. A single layer feed-forward \emph{artificial neural network} is used to estimate the effect of the discarded modes on the retained ones during the time integration of the ordinary differential equations obtained using POD-GP. In brief, an artificial neural network (ANN) can be used to setup a nonlinear relationship between a desired set of inputs and targets provided a large set of benchmark data for the underlying statistical relationship is available. It is therefore no surprise that this subset of the machine learning field has seen wide application in function approximation, data classification, pattern recognition and dynamic systems control applications \cite{widrow1994neural,demuth2014neural}. Before an ANN is deployed, it must be trained to accurately capture the nonlinear relationship between its inputs and outputs. For this purpose, a high fidelity solution of the Burgers equation is used to generate the `true' modes. An optimization problem is solved to minimize a desired performance function (which generally represents a measure of the suitability of a network for prediction). While we shall provide information about the machine learning framework implemented here in far greater detail within, we note that we have used the Bayesian regularization \cite{foresee1997gauss} and extreme learning machine \cite{huang2006extreme} approaches for the training of the artificial neural network. Both methods have been developed in the machine learning community for the accurate capture of statistical trends in noisy data. While the former is famed for its accuracy, the latter is attractive due to its very low computational cost. Indeed, the extreme learning machine (or ELM) proves to be very exciting as a possible means to address the need for fast learning in dynamically adaptive systems (although the topic of which is left to a separate investigation).

While there have been investigations into the suitability of machine learning techniques for the purpose of ROM implementations \cite{narayanan1999low,khibnik2000analysis,sahan1997artificial,moosavi2015efficient,moosavi2017multivariate} and these approaches are quite well studied for use in feedback flow control where they are used to generate a direct mapping of flow measurements to actuator control systems \cite{gillies1995low,gillies1998low,gillies2001multiple,faller1997unsteady,hocevar2004experimental,efe2004modeling,efe2005control,lee1997application}, it is our belief that the work presented in this document represents the first utilization of an ANN as closure model in the POD-GP framework. As mentioned previously, information from the time evolution of our PDE (from its analytical solution) is leveraged to provide a supervised learning framework for the ANN to approximate the error in the modal evolution through GP. A regularized training ensures that no localized behavior is observed from the trained ANN with respect to the governing equations and the data they have produced. Training data sets are generated through different realizations of the viscosity of the viscous Burgers equation with a high degree of freedom. The trained ANN architecture is then tested for its ability to stabilize modal evolution for values of viscosity that are not in the training data set. Our investigations can be summarized by the following bullet points
\begin{itemize}
  \item A machine learning based closure model is implemented in the ROM framework (using a Galerkin projection) through the use of a single layer feedforward artificial neural network. Both optimal POD and Fourier modes are used as the bases of the transformation space.
  \item The closure is `trained' using high fidelity data obtained from the true solution of the time evolution of our partial differential equation. The Bayesian regularization and extreme learning machine approaches are used for a regularized training to capture underlying statistical data.
  \item The proposed framework is tested through a ROM framework for the viscous Burgers equation for physical parameters which were not included in the training data set.
\end{itemize}

The rest of this paper is devoted to a mathematical development of the concepts we have introduced as well as a documentation of our results.

\section{Mathematical model}
\label{sec:mat}

The viscous Burgers equation is our governing law of choice for this study. We employ the conservative form of the equation given by
\begin{equation}
\label{e:1}
    \frac{\partial u}{\partial t} +  \frac{\partial}{\partial x} \left( \frac{1}{2} u^2 \right) = \frac{1}{Re} \frac{\partial^2 u}{\partial x^2}\,,
\end{equation}
where $Re$ is the Reynolds number. This equation is generally considered a lower-dimensional equivalent for the full Navier-Stokes equations due to its advective and diffusive behavior. It is therefore, quite common, to use the Burgers equation for the preliminary evaluation of ROM frameworks prior to deployment for fluid flow problems \cite{kunisch1999control,san2013proper,imtiaz2016closure}. Indeed, it may be considered a challenging convective system for ROM assessments as it characterizes localized flow structures such as shock waves.

\section{Basis selection}
\label{sec:basis}

\begin{figure}[!ht]
\centering
\mbox{
\subfigure[Fourier bases]{\includegraphics[width=0.5\textwidth]{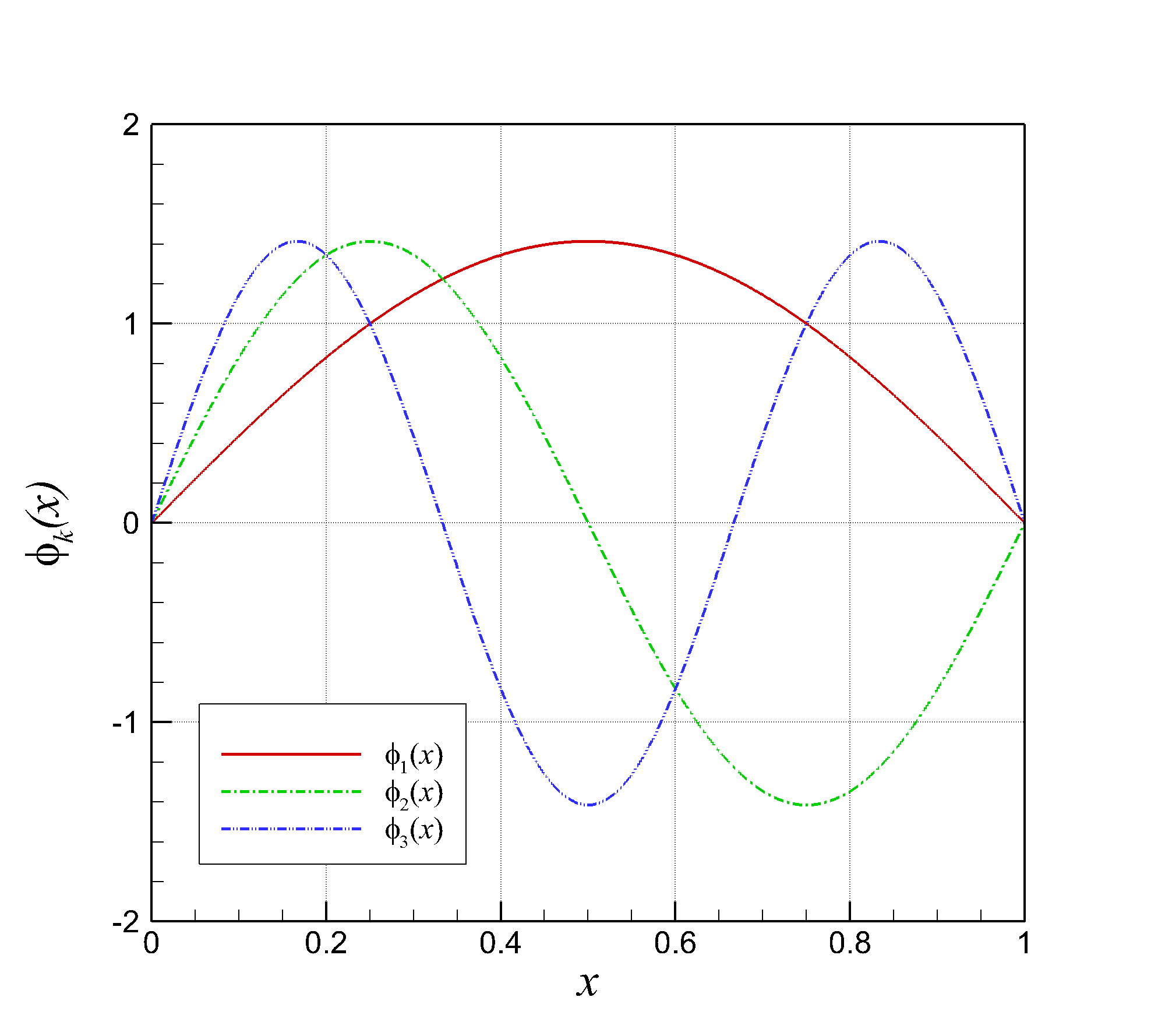}}
\subfigure[POD bases]{\includegraphics[width=0.5\textwidth]{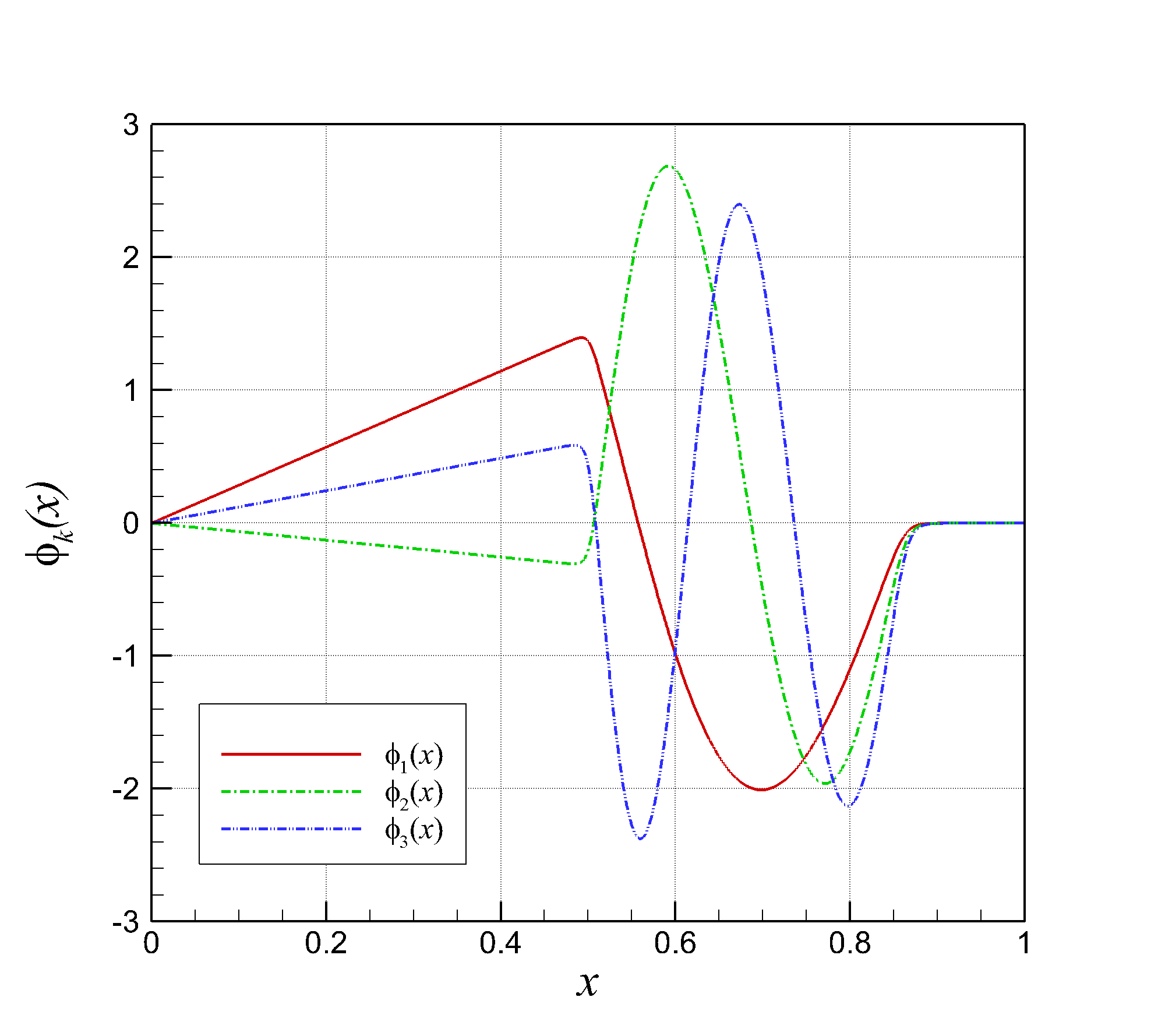}}
}
\caption{Representative basis functions.}
\label{fig:3}
\end{figure}

In this section, we detail the two types of bases chosen for the transformation space of our PDE. Some of the bases obtained using our two approaches are shown in \ref{fig:3} and we note that they are orthogonal to each other.

\subsection{Fourier basis functions}

As a first choice for our orthogonal bases, we utilize the Fourier series coefficients. In order to ensure that the boundary conditions of the PDE are respected we discard the Cosine components of the series and are left with
\begin{align}\label{FourierBases}
  \phi_k = \sqrt{\frac{2}{L}} \sin(k \pi x/L).
\end{align}
where $x \in (0,L)$ is our extent of the physical domain. We remark here that these bases may \emph{not} be considered optimal as they are devised without any information from our exact solutions but through intuition. The condition for orthonormality between bases can be expressed as:
\begin{align}
\label{ortho}
    \left( \phi_k,\phi_l \right)=\left\{
                      \begin{array}{ll}
                        1, & k=l \\
                        0, & k\neq l \, .
                      \end{array}
                    \right.
\end{align}

\subsection{POD basis functions}
In this subsection, we detail the POD approach for ROM to capture unsteady, convective and non-periodic dynamics of governing PDEs. The reader is directed to \cite{san2013proper} for a far more detailed discussion of this topic.

A POD can be constructed from the scalar field $f$ at different times (also known as snapshots). These snapshots are either obtained by solving the governing equations through a DNS or (as in this case) through the exact solution. In the following, we will utilize the index $i$ to indicate a particular snapshot in time. For the POD approach we utilize a total of $N$ snapshots for the field variable, i.e., $f^i(\mathbf{x})$ for $i=1,2,\hdots,N$. The flow field data for our governing laws may thus be represented as
\begin{align}\label{POD_1}
  f(\mathbf{x},t) = \bar{f}(\mathbf{x}) + \hat{f}(\mathbf{x},t), \quad \bar{f}(\mathbf{x}) = \frac{1}{N} \sum_{i=1}^{N} f(\mathbf{x},t_i),
\end{align}
where $\bar{f}$ implies a temporal averaging of a particular point value in the field and $\hat{f}$ contains the fluctuating quantity. We must remark here that $\bar{f}$ is a function of space alone whereas $\hat{f}$ is a function of both space and time. A correlation matrix may be constructed using the fluctuating components of the snapshots to give
\begin{align}\label{POD_2}
  C_{ij} = \int_{\Omega} \hat{f}^{i}(\textbf{x}) \hat{f}^{j}(\textbf{x}) d\textbf{x},
\end{align}
where $\Omega$ is the entire spatial domain and $i$ and $j$ refer to the $i^{th}$ and $j^{th}$ snapshots. The correlation matrix $C$ is a non-negative symmetric square matrix of side $N$. If we define the inner product of any two fields $f_1$ and $f_2$ as
\begin{align}\label{POD_3}
  (f_1,f_2) = \int_{\Omega} f_1(\textbf{x}) f_2(\textbf{x}) d\textbf{x}
\end{align}
we may express the correlation matrix as $C_{ij} = (\hat{f}^i,\hat{f}^j)$. In this study, we use the well-known Simpson's 3/8 integration rule for a numerical computation of the inner products. The optimal POD basis functions are obtained on performing an eigendecomposition for the $C$ matrix. This has been shown in detail in the POD literature (see, e.g., \cite{sirovich1987turbulence,holmes1998turbulence,ravindran2000reduced}). The eigenvalue problem can be written in the following form:
\begin{align}
\label{POD_4}
    CW=W\Lambda \, ,
\end{align}
where $\Lambda=\mbox{diag}[\lambda_1, \lambda_2, ..., \lambda_N ]$ is a diagonal matrix containing the eigenvalues of this decomposition and
$W$ =[ $\boldsymbol w^{1}$ , $\boldsymbol w^{2}$ , ...,$\boldsymbol w^{N}$], $\lambda_j$ is an orthogonal basis consisting of the eigenvectors of this decomposition.
The eigenvalues are stored in descending order, $\lambda_1 \geq \lambda_2 \geq ... \geq \lambda_N$. The POD basis functions can be written as
\begin{align}
\label{POD_5}
    \phi_1(\textbf{x}) = \sum_{i=1}^{N}w^{1}_{i}\hat{f}(\textbf{x},t_i), \quad \phi_2(\textbf{x}) = \sum_{i=1}^{N}w^{2}_{i}\hat{f}(\textbf{x},t_i), \quad ..., \quad \phi_N(\textbf{x}) = \sum_{i=1}^{N}w^{N}_{i}\hat{f}^{i}(\textbf{x}) \, ,
\end{align}
where $w^{j}_{i}$ is the $i$th component of eigenvector $\boldsymbol w^{j}$. Therefore we emphasize that the POD modes are ranked according to the magnitude of their eigenvalue. The eigenvectors must also be normalized in order to satisfy the condition of orthonormality between bases given by Eq. (\ref{ortho}). It can be shown that, for Eq.~(\ref{ortho}) to be true for the POD bases, the eigenvector $\boldsymbol w^{j}$ must satisfy the following equation:
\begin{align}
\label{POD_7}
\sum_{i=1}^{N}w^{j}_{i}w^{j}_{i}=\frac{1}{\lambda_j}.
\end{align}
In practice, most of the subroutines for solving the eigensystem given in Eq.~(\ref{POD_4}) return the eigenvector matrix $W$ having all the eigenvectors normalized to unity. In that case, the orthogonal POD bases are given by
\begin{align}
\label{POD_8}
    \phi_j(\textbf{x}) = \frac{1}{\sqrt{\lambda_j}}\sum_{i=1}^{N}w^{j}_{i}\hat{f}(\textbf{x},t_i)
\end{align}
where $\phi_j(\textbf{x})$ is the $j$th POD basis function. The main motivation behind the construction of a ROM using the optimal POD bases is due to the fact that modes with high magnitudes of eigenvalues retain a greater proportion of the energy of the system. Hence, it is possible to construct a lower degree of freedom approximation of our sparse PDE in the space transformed by the aforementioned eigenvectors.

\section{Projection based model order reduction}
\label{sec:GP}
For our test case given by the Burgers equation, the Galerkin projection can be carried out in the following manner.
\begin{align}
\label{e:14}
    \hat{u}(x,t)=\sum_{k=1}^{M}a_k(t)\phi_{k}(x) \, ,
\end{align}
where $a_k$ are the time dependent coefficients, and $\phi_{k}$ are the space dependent modes.
To derive the Galerkin projection, we first rewrite the Burgers equation (i.e., Eq.~(\ref{e:1})) in the following form
\begin{align}
\label{e:15}
    \frac{\partial u}{\partial t} = L[u] + N[u;u] \, ,
\end{align}
where
\begin{align}\label{linearOP}
L[f]=\frac{1}{Re}\frac{\partial^2 f}{\partial x^2}
\end{align}
is the linear operator, and
\begin{align}\label{nlop}
N[f;g]= -  \frac{\partial}{\partial x}\left(\frac{1}{2}f g\right)
\end{align}
is the nonlinear operator. By applying this projection to our nonlinear system (i.e., multiplying Eq.~(\ref{e:15}) with the basis functions and integrating over the domain), we obtain our ROM, denoted POD-GP-ROM when using the optimal POD modes and Fourier-GP-ROM when using Fourier modes:
\begin{align}
\label{e:16}
    \left(\frac{\partial u}{\partial t},\phi_k\right) = (L[u],\phi_k) + (N[u;u],\phi_k), \quad \mbox{for} \quad k=1,2, ..., M \, ,
\end{align}
where we use Eq.~(\ref{POD_1}) and Eq.~(\ref{e:14}) to yield
\begin{align}
\label{e:166}
    u(x,t)=\bar{u}(x) + \sum_{k=1}^{M}a_k(t)\phi_{k}(x) \, ,
\end{align}
and substituting Eq.~(\ref{e:166}) into Eq.~(\ref{e:14}), and simplifying the resulting equation by using the condition of orthogonality given in Eq.~(\ref{ortho}), the ROM implementation can be written as follows:
\begin{align}
\label{e:17}
    \frac{d a_{k}}{dt} = B_{k} + \sum_{i=1}^{M} \mathfrak{L}_{ik} a_i + \sum_{i=1}^{M}\sum_{j=1}^{M} \mathcal{N}_{ijk}a_{i}a_{j}, \quad \mbox{for} \quad k=1,2, ..., M \, ,
\end{align}
where
\begin{eqnarray}
B_{k}  &=&(  L[\bar{u}],\phi_{k}) \label{e:18} + ( N[\bar{u};\bar{u}],\phi_{k}) \\
\mathfrak{L}_{ik} &=&( L[\phi_{i}],\phi_{k}) + ( N[\bar{u};\phi_{i}] + N[\phi_{i} ; \bar{u}],\phi_{k}) \label{e:20}  \\
\mathcal{N}_{ijk}    &=&( N[\phi_{i};\phi_{j}],\phi_{k}) \label{e:22} .
\end{eqnarray}

The GP-ROM given by Eq.~(\ref{e:17}) consists of $M$ coupled ODEs and can be solved by a standard numerical method (such as the third-order Runge-Kutta scheme that was used in this study). The number of degrees of freedom of the system is now significantly lower. The vectors, matrices and tensors in Eqs.~(\ref{e:18})-(\ref{e:22}) are also precomputed quantities, which results in a dynamical system that can be solved very efficiently. To complete the dynamical system given by Eq.~(\ref{e:17}), the initial condition is given by using the following projection:
\begin{equation}
\label{e:23}
a_{k}(t=0)= \left( u(x,t=0)-\bar{u}(x),\phi_{k} \right) \, ,
\end{equation}
where $u(x,t=0)$ is the physical initial condition of the problem given in Eq.~(\ref{e:1}).

\section{Artificial neural networks}
\label{sec:ANN}

\subsection{Network architecture}
\label{sec:arch_ann}

The basic structure of the simple feed-forward artificial neural network consists of $\mathbb{L}$ layers with each layer possessing a predefined number of unit cells called neurons. Each of these layers has an associated transfer function and each unit cell has an associated bias. Any input to the neuron has a bias added to it followed by activation through the transfer function. To describe this process using equations, we have for a single neuron in the $l^{th}$ layer receiving a vector of inputs $\mathbf{S}^l$ from the $(l-1)^{th}$ layer given by \cite{demuth2014neural}
\begin{align}\label{ANN_1}
  \mathbf{S}^l = \mathbf{W}^l \mathbf{X}^{l-1},
\end{align}
where $\mathbf{W}^l$ stands for a matrix of weights linking the $l-1$ and $l$ layers with $\mathbf{X}^{l-1}$ being the output of the $(l-1)^{th}$ layer. The output of the $l^{th}$ layer is now given by
\begin{align}\label{ANN_2}
  \mathbf{X}^l = G(\mathbf{S}^l + \mathbf{B}^l),
\end{align}
where $\mathbf{B}^l$ is the vector of biasing parameters for the $l^{th}$ layer. Every node (or unit cell) has an associated transfer function which acts on its input and bias to produce an output which is `fed forward' the network. The nodes which take the raw input value of our training data set (i.e., the nodes of the first layer in the network) perform no computation (i.e., they do not have any biasing or activation through a transfer function). The next layers are a series of unit cells which have an associated bias and activation function which perform computation on their inputs. These are called the hidden layers with the individual unit cells known as neurons. Note that it is common in literature to consider these the only layers in the network. The final layer in the network is that of the outputs. The output layers generally have a linear activation function with a bias which implies a simple summation of inputs incident to a unit cell with its associated bias. In this investigation, we have used one hidden layer of neurons between the set of inputs and targets with a Tan-Sigmoid activation function. The Tan-Sigmoid function can be expressed as
\begin{align}\label{ANN_3}
  G(a) = \frac{2}{1+\exp(-2a)}-1
\end{align}
The transfer function $G$ calculates the neuron's output given its net input. In theory, any differentiable function can qualify as an activation function \cite{zhang1998forecasting}, however, only a small number of functions which are bounded, monotonically increasing and differentiable are used for this purpose. The choice of using ANN is motivated by its excellent performance as a forecasting tool \cite{dawson1998artificial,kim2003nonlinear} and its general suitability in the machine learning and function estimation domain (e.g., see \cite{haykin2009neural} and references therein).

\subsection{ANN closure modeling for ROM}

Eq.~(\ref{e:17}) can be rewritten as
\begin{align}\label{closure00}
  \frac{d a_{k}}{dt} = R_k
\end{align}
where
\begin{align}\label{closure01}
  R_k = B_{k} + \sum_{i=1}^{M} \mathfrak{L}_{ik} a_i + \sum_{i=1}^{M}\sum_{j=1}^{M} \mathcal{N}_{ijk}a_{i}a_{j},
\end{align}
where we have truncated our modal quantities to a reduced number ($M$). At this point, a closure term must be introduced to account for the effect of truncation. This gives us
\begin{align}\label{closure02}
  \frac{d a_{k}}{dt} = R_k + \tilde{R}_k
\end{align}
where $\tilde{R}_k$ accounts for the residual effects of the discarded modes. An ANN architecture is introduced to model this term. For the purpose of training, we need to assess what our modal quantities would evolve like if there was \emph{no} truncation (i.e., a pure evolution of the PDE in a transformed space). We may obtain this by using our underlying PDE
\begin{align}\label{closure03}
  \frac{\partial u}{\partial t} = L[u] + N[u;u]
\end{align}
and applying our familiar projection to get
\begin{align}\label{closure04}
  \left( \phi_k, \frac{\partial u}{\partial t} \right) = \left( \phi_k, L[u] + N[u;u]\right)
\end{align}
which gives us
\begin{align}\label{closure05}
  \frac{d a_k}{d t} = \left( \phi_k, L[u] + N[u;u]\right)
\end{align}
where our true projected right hand side becomes
\begin{align}\label{closure06}
  \bar{R}_k = \left( \phi_k, L[u] + N[u;u]\right).
\end{align}
One may now compare Eqs.~(\ref{closure02}) and (\ref{closure06}) to obtain
\begin{align}\label{closure07}
  R_k + \tilde{R}_k = \bar{R}_k.
\end{align}
Thus to summarize: $R_k$ is our standard truncated GP obtained using either POD or Fourier bases, $\tilde{R}_k$ is the closure we would like to model, $\bar{R}_k$ is the true projection obtained by transforming the exact solution to the new space spanned by the POD or Fourier bases.

\begin{figure}[!ht]
\centering
\includegraphics[width=0.5\textwidth]{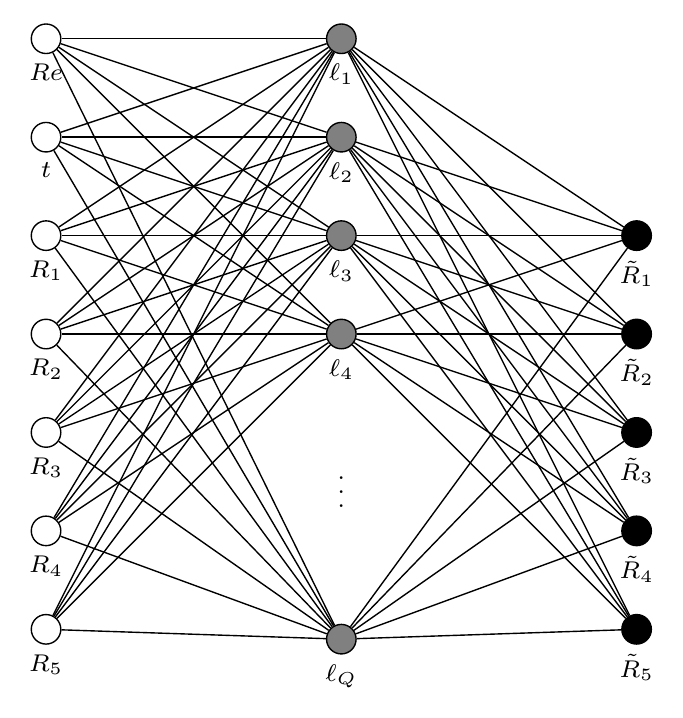}
\caption{The proposed ANN architecture for closure modeling of ROMs.}
\label{fig:4}
\end{figure}

An ANN architecture is devised which aids us in developing a nonlinear relationship between a set of inputs given by the Reynolds number ($Re$), the present time ($t$) and the truncated GP projections (i.e., $R_1,R_2,\hdots,R_M$). The outputs are given by the closures (i.e., $\tilde{R}_1,\tilde{R}_2,\hdots,\tilde{R}_M$). The architecture of our ANN is shown in Fig.~(\ref{fig:4}) and $Q$ is our number of hidden neurons.

\subsection{Bayesian regularization}
\label{sec:eml}

The training of a desired ANN is carried out by minimizing the error between the target and the inputs to determine a set of best fit parameters. These best fit parameters are the biases and linear weights that have captured the underlying relationship between the targets and inputs and may now be used to predict target data for inputs a posteriori. The advantage of using the ANN approach over traditional statistical regression models is that comparatively smaller data sets for training are suitable. The objective function of a neural network training is given by:
\begin{align}\label{OF}
  F \doteq E_D = \sum_{i=1}^{n_s} \left| \left|\mathbf{t}_i - \mathbf{x}^{0}_{i}\right|\right|_2,
\end{align}
which is a classic mean squared error of the outputs and the targets. Note that our input data may be defined as:
\begin{align}
    \mathbf{X}^{0} =
    \begin{bmatrix}
    \mathbf{x}_{1}^{0} & \mathbf{x}_{2}^{0} & \hdots & \mathbf{x}^{0}_{n_s}
    \end{bmatrix}
\end{align}
where $n_s$ is the number of sample data. Our output data can also be represented similarly as:
\begin{align}
    \mathbf{T} =
    \begin{bmatrix}
    \mathbf{t}_{1} & \mathbf{t}_{2} & \hdots & \mathbf{t}_{n_s}
    \end{bmatrix}.
\end{align}
There is an important caveat to the `mean-squared-error' based training of neural networks: regularization techniques are imperative for the prevention of \emph{overfitting} noise in our data. In our investigation, we shall use the Bayesian regularization (BR) and extreme learning machine (ELM) approaches which have become popular for their regularization ability.

The Bayesian regularization training procedure augments the performance function of the ANN training by penalizing the sum of the squared weights of the network \cite{mackay1992bayesian,foresee1997gauss}. This is in contrast to the traditional methods which are oriented to minimizing the sum of squared errors alone (which is dangerous for noisy data as it promotes excessive localization). This modified objective function is then implemented in a Bayesian framework where the parameters in the network (i.e., the weights and the biases) are considered as random variables. Mathematically we have,
\begin{align}\label{OF2}
  F = \beta E_D + \alpha E_W,
\end{align}
where $E_W$ is the sum of squared weights and $\alpha,\beta$ are the weight coefficients which are dynamically adjusted during training. The scalar $E_W$ may be mathematically represented as the sum of the Frobenius norm of the layer 1 and layer 2 weight matrices:
\begin{align}\label{E_W}
  E_W = \left| \left| \mathbf{W}^{1} \right| \right|_F + \left| \left| \mathbf{W}^{2} \right| \right|_F.
\end{align}
If $\alpha << \beta$, the training aims to reduce weight sizes at the expense of network errors thereby producing a smoother network response. The main challenge is the setting of the weight coefficients after which a regular gradient based optimization technique may be used to adjust the weights of the ANN for one iteration. This is done through the computation of a quantity known as the effective number of parameters ($\gamma$) \cite{foresee1997gauss}. Once one step of the standard Levenberg-Marquardt algorithm is completed from the initial condition where $\alpha = 0$, $\beta=1$ and weights are initialized according to the Nguyen-Widrow method \cite{nguyen1990improving}, the effective number of parameters is computed from the knowledge of the Hessian matrix available from the Levenberg-Marquardt algorithm\cite{levenberg1944method,marquardt1963algorithm}:
\begin{align}\label{BR1}
  \gamma = P^{tot} - 2\alpha \textnormal{tr}(\mathbf{L})^{-1}
\end{align}
where $P^{tot}$ is the total number of parameters (i.e., weights and biases of the entire network) and $\mathbf{L}$ is given by:
\begin{align}\label{BR2}
  \mathbf{L} = \nabla^2 F(\mathbf{z}) \approx 2 \beta \mathbf{J}^\intercal \mathbf{J} + 2 \alpha \mathbf{I}
\end{align}
where $\mathbf{J}$ is the Jacobian matrix of the training set errors \cite{hagan1994training} and $\mathbf{z}$ is the vector of all network parameters (refer equation 12.36 in \cite{demuth2014neural}). The superscript $\intercal$ represents the matrix transpose operation. The Levenberg-Marquardt algorithm, may then be utilized to obtain a new estimate of the network parameters:
\begin{align}\label{BR3}
  \mathbf{z}^{new} = \mathbf{z} - [\mathbf{J}^\intercal \mathbf{J} + \mu I]^{-1} \mathbf{J}^\intercal \textbf{e}.
\end{align}
Note that this algorithm is a \emph{hybrid} between the Newton's method and steepest descent approach to function minimization with the gradient of the objective function given by:
\begin{align}\label{BR4}
  \nabla \mathbf{z} = \mathbf{J}^\intercal \mathbf{e},
\end{align}
with $\mathbf{e}$ being the vector of network errors. When the scalar $\mu$ is small, the algorithm behaves more like the Newton's method of root finding using the Hessian Matrix $\mathbf{J^\intercal J}$. When $\mu$ is large, the algorithm becomes steepest descent with a small step size. In brief, the parameter $\mu$ is decreased after each successful step that reduced the objective function and increased only if a tentative step increases the objective function. Following this step, new estimates for the weighting coefficients are obtained from
\begin{align}\label{BR5}
  \alpha = \frac{\gamma}{2 E_W(\mathbf{z}^{new})} \quad \textnormal{and } \beta = \frac{P^{tot} - \gamma}{2 E_D (\mathbf{z}^{new})}.
\end{align}

Now that the weighting coefficients are updated one may iterate through the aforementioned steps till a desired convergence is reached. An updated value of $\alpha$ and $\beta$, gives us a newly weighted performance function which may be minimized by a variety of gradient based optimization algorithms (here we have used the Levenberg-Marquardt algorithm) to obtain a new set of optimal weights and biases at the end of each iteration. Algorithm \ref{Algo1} summarizes the BR training methodology. The BR algorithm is equipped with a convergence criteria given by a maximum value of $\mu$ (implying that the minima of the objective function has been reached) or a value of the gradient is obtained which is lower than the minimum gradient threshold specified prior to the start of the minimization process.

\begin{algorithm}
  \caption{Bayesian Regularization}\label{Algo1}
  \begin{algorithmic}[1]
      \State Given $\mathbf{X}^{0} \textnormal{ and } \mathbf{T}$                                                                  \Comment{Given inputs and targets}
      \State $\alpha = 0 \textnormal{ and } \beta = 1$                                                            \Comment{Our regularization initial condition}
      \State $\mu = 0.001 \textnormal{ and } (\nabla\mathbf{z})_{min} = 10^{-7}$                                  \Comment{Our training initial condition}
      \State Initialize $\mathbf{W}^1$, $\mathbf{W}^2$, $\mathbf{B}^1$, $\mathbf{B}^2$                                                                \Comment{Initialize non-zero random parameters}
        \While{Not converged}                                                                                     \Comment{Convergence criteria}
        \State Calculate new $\alpha$ and $\beta$ from Eqs.~(\ref{BR1}-\ref{BR5})                                 \Comment{To ensure generalization}
        \State Minimize objective function $F$                                                             \Comment{One iteration of Levenberg-Marquardt}
        \State Update $\mathbf{W}^1 = \mathbf{W}^1 + \Delta \mathbf{W}^1$                                         \Comment{Levenberg-Marquardt calculates $\Delta\mathbf{W}^1$}
        \State Update $\mathbf{W}^2 = \mathbf{W}^2 + \Delta \mathbf{W}^2$                                         \Comment{Levenberg-Marquardt calculates $\Delta\mathbf{W}^2$}
        \State Update $\mathbf{B}^1 = \mathbf{B}^1 + \Delta \mathbf{B}^1$                                         \Comment{Levenberg-Marquardt calculates $\Delta\mathbf{B}^1$}
        \State Update $\mathbf{B}^2 = \mathbf{B}^2 + \Delta \mathbf{B}^2$                                         \Comment{Levenberg-Marquardt calculates $\Delta\mathbf{B}^2$}
      \EndWhile
  \end{algorithmic}
\end{algorithm}

\subsection{Extreme learning machine}

In this section we detail the extreme learning machine approach to generalized single layer feedforward ANN training. This methodology was proposed in \cite{huang2006extreme} for extremely fast training of a single layer feedforward ANN based on the principles of the least squares approximation. For the ease of description, let us define a few matrices for the single layer feedforward network. We remark that is a generalization of the architecture introduced in Section \ref{sec:arch_ann}. Our input matrix is
\begin{align}
    \mathbf{X}^{0} =
    \begin{bmatrix}
    \mathbf{x}^{0}_{1} & \mathbf{x}^{0}_{2} & \hdots & \mathbf{x}^{0}_{n_s}
    \end{bmatrix}
\end{align}
where $\mathbf{p}_{i}$ is the $i^{th}$ sample (out of a total of ${n_s}$ samples) of a multidimensional input vector. Our weights connecting the inputs to the middle (hidden) layer are given by
\begin{align}
    \mathbf{W}^{1} =
    \begin{bmatrix}
    \mathbf{w}^{1}_{1} \\
    \mathbf{w}^{1}_{2} \\
    \vdots      \\
    \mathbf{w}^{1}_{Q}
    \end{bmatrix}
\end{align}
where $\mathbf{w}_{i}$ is the $i^{th}$ neuron (out of a total of $Q$ neurons) in the hidden layer. These are initialized to be small non-zero random numbers to enforce generalization. The extreme learning machine methodology prescribes biases \emph{only} for hidden layer neurons and these may be given by can be given by
\begin{align}
    \mathbf{B}^{1} =
    \begin{bmatrix}
    \mathbf{b}^{1}_{1} \\
    \mathbf{b}^{1}_{2} \\
    \vdots      \\
    \mathbf{b}^{1}_{Q}
    \end{bmatrix}.
\end{align}
The output of the hidden layer neurons becomes
\begin{align}
    \label{eqH}
    \mathbf{H}^{\intercal} = G(\mathbf{W}^{1} \mathbf{X}^{0} + \mathbf{B}^{1})
\end{align}
where $G(\mathbf{X})$ implies an tan-sigmoid activation procedure on each element of a matrix $\mathbf{X}$. The weight matrix of the second layer may be given as
\begin{align}
    \mathbf{W}^{2} =
    \begin{bmatrix}
    \mathbf{w}^{2}_{1} & \mathbf{w}^{2}_{2} & \hdots & \mathbf{w}^{2}_{Q}
    \end{bmatrix}.
\end{align}
Our outputs of the ELM may thus be represented as
\begin{align}
    \mathbf{S}^{1} = \mathbf{W}^{2} \mathbf{H}^{\intercal},
\end{align}
which must be trained against a set of targets corresponding to each input vector given by
\begin{align}
    \mathbf{T} =
    \begin{bmatrix}
    \mathbf{t}_1 & \mathbf{t}_2 & \hdots & \mathbf{t}_{n_s}
    \end{bmatrix}.
\end{align}
The ELM training mechanism is given as follows. In order to calculate the matrix $\mathbf{W}^{2}$, we must recognize that its optimal solution should satisfy
\begin{align}
    \mathbf{W}^{2}_{op} \mathbf{H}^{\intercal} = \mathbf{T}
\end{align}
or by taking a transpose of both sides
\begin{align}
    \mathbf{H} \mathbf{W}^{2^\intercal}_{op} = \mathbf{T}^{\intercal}
\end{align}
which leads us to the following expression for the optimal weights
\begin{align}
    \label{Trained}
    \mathbf{W}^{2^\intercal}_{op} = \mathbf{H}^{\dagger} \mathbf{T}^{\intercal}.
\end{align}
The weights $\mathbf{W}^{1}$ and biases $\mathbf{B}^{1}$ are generated initially using random numbers. The matrix given by $\mathbf{H}^{\dagger}$ is calculated using a generalized Moore-Penrose pseudoinverse \cite{serre2002matrices}. Once the optimal weights of the second layer are obtained, our network is trained for deployment. Due to the random number values chosen for the weights in the first layer, our network is well suited to highly effective abstraction of the training process due to a smaller degree of freedom of the overall ANN. The obvious advantage of this approach compared to the BR method is the substitution of an iterative minimization to a direct least squares approximation using the pseudoinverse. The procedure for the ELM method is given in Algorithm \ref{Algo2}.

\begin{algorithm}
  \caption{Extreme Learning Machine}\label{Algo2}
  \begin{algorithmic}[1]
        \State Given $\mathbf{X}^{0} \textnormal{ and } \mathbf{T}$                                           \Comment{Given inputs and targets}
        \State Initialize $\mathbf{W}^{1} \textnormal{ and } \mathbf{B}^{1}$                                          \Comment{Initialize non-zero random parameters}
        \State Calculate $\mathbf{H}^\intercal$                                                                   \Comment{From Eq.~(\ref{eqH})}
        \State Calculate pseudoinverse $\mathbf{H}^\dagger$                                                       \Comment{Moore-Penrose pseudoinverse}
        \State Calculate layer 2 weights $\mathbf{W}^2 = \mathbf{H}^{\dagger} \mathbf{T}^{\intercal}$
        \Comment{Least squares solution for optimal weights}
  \end{algorithmic}
\end{algorithm}

\subsection{Performance comparison between BR and ELM}
\label{BR_ELM_Cost}

\begin{figure}[!ht]
\centering
\mbox{
\subfigure[ELM]{\includegraphics[width=0.5\textwidth]{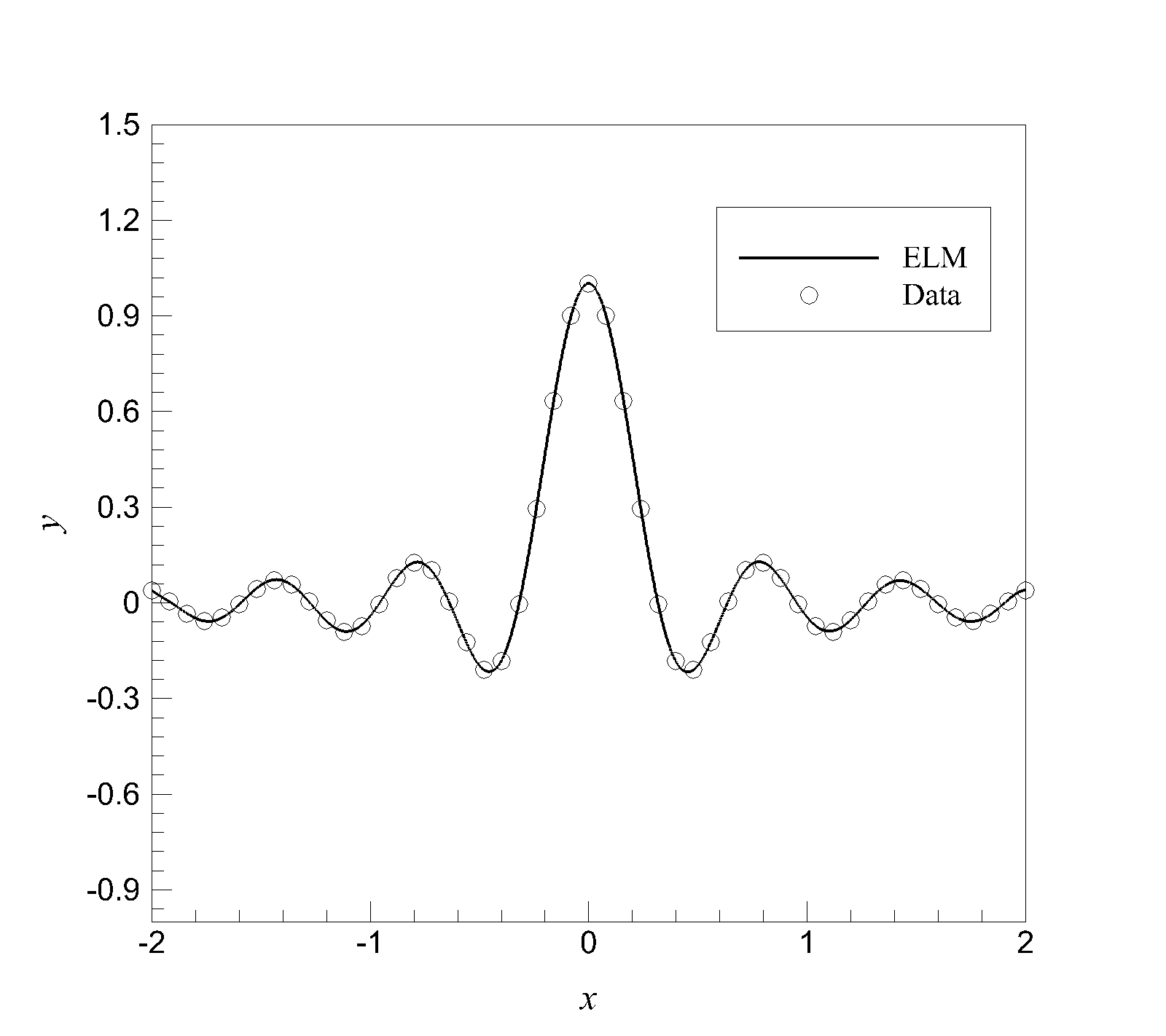}}
\subfigure[BR]{\includegraphics[width=0.5\textwidth]{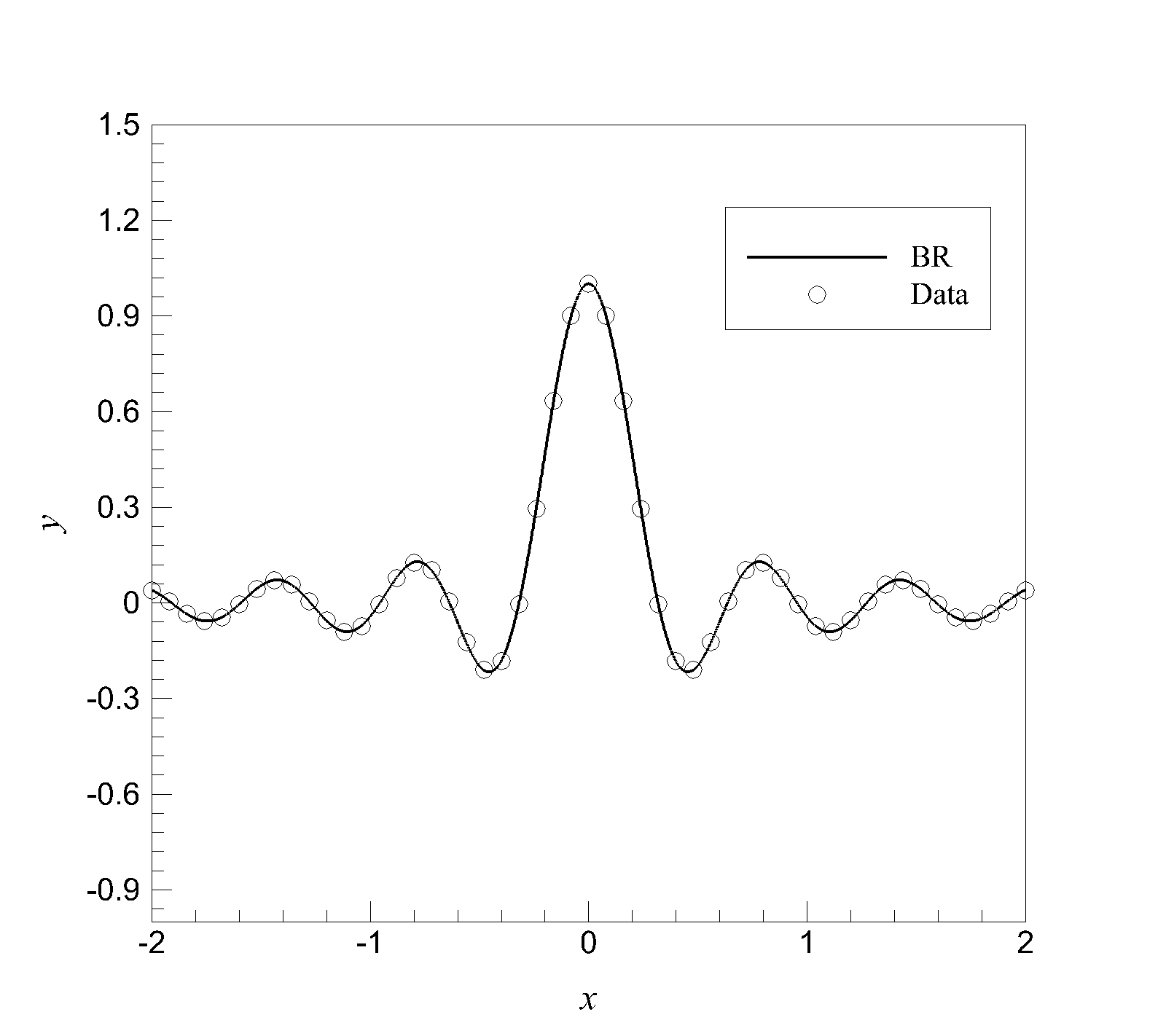}}
}\\
\mbox{
\subfigure[ELM]{\includegraphics[width=0.5\textwidth]{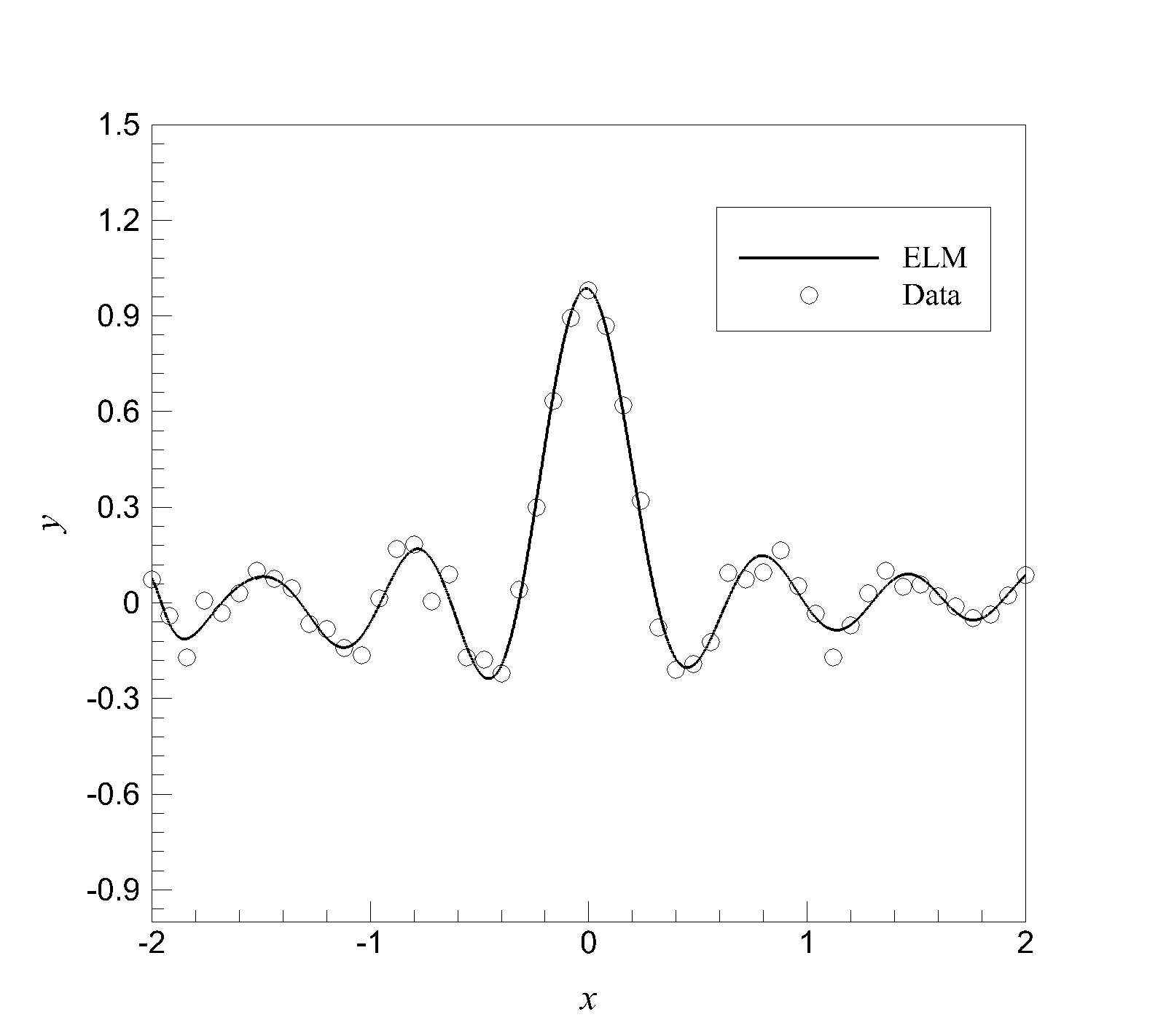}}
\subfigure[BR]{\includegraphics[width=0.5\textwidth]{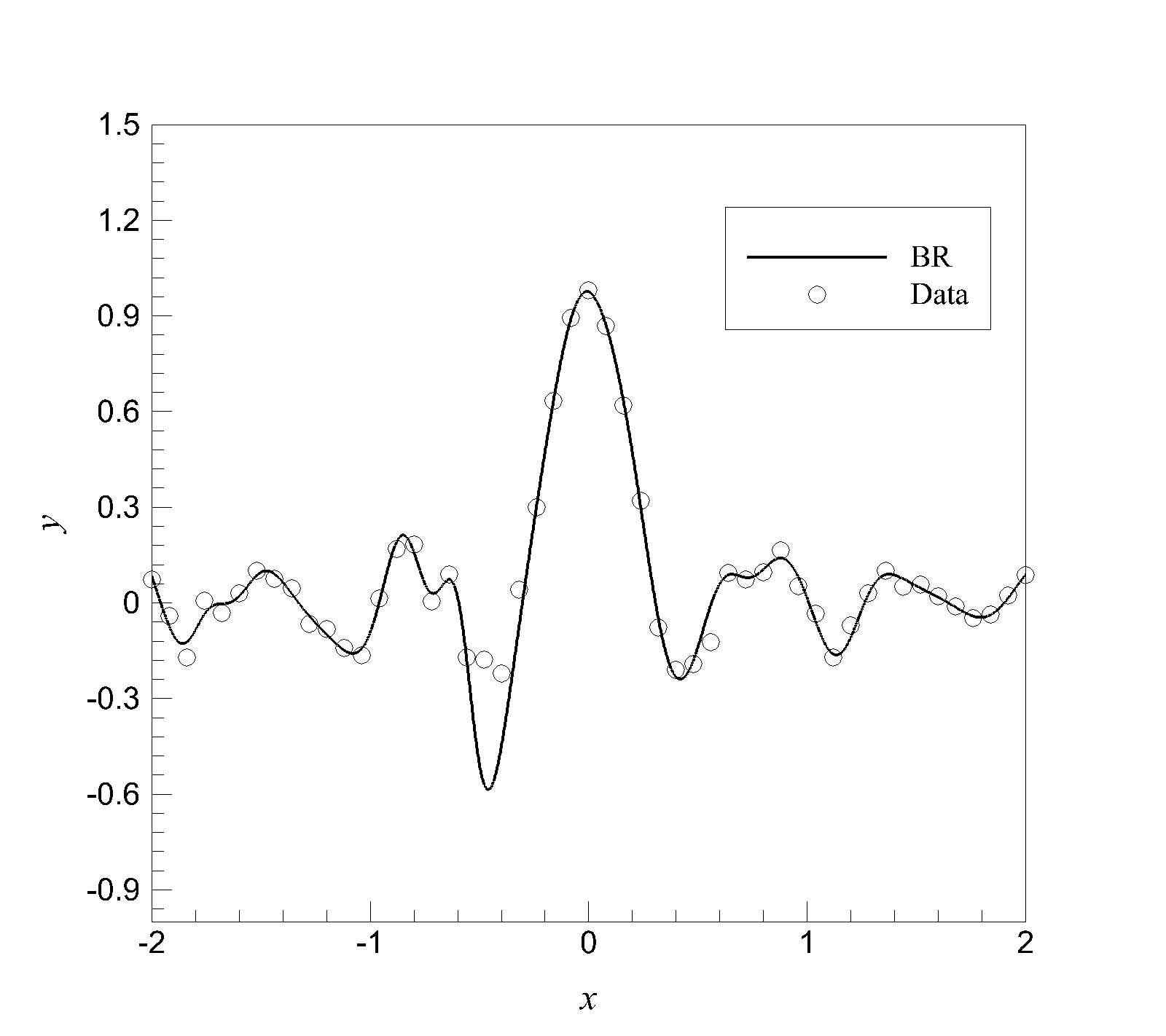}}
}\\
\mbox{
\subfigure[ELM]{\includegraphics[width=0.5\textwidth]{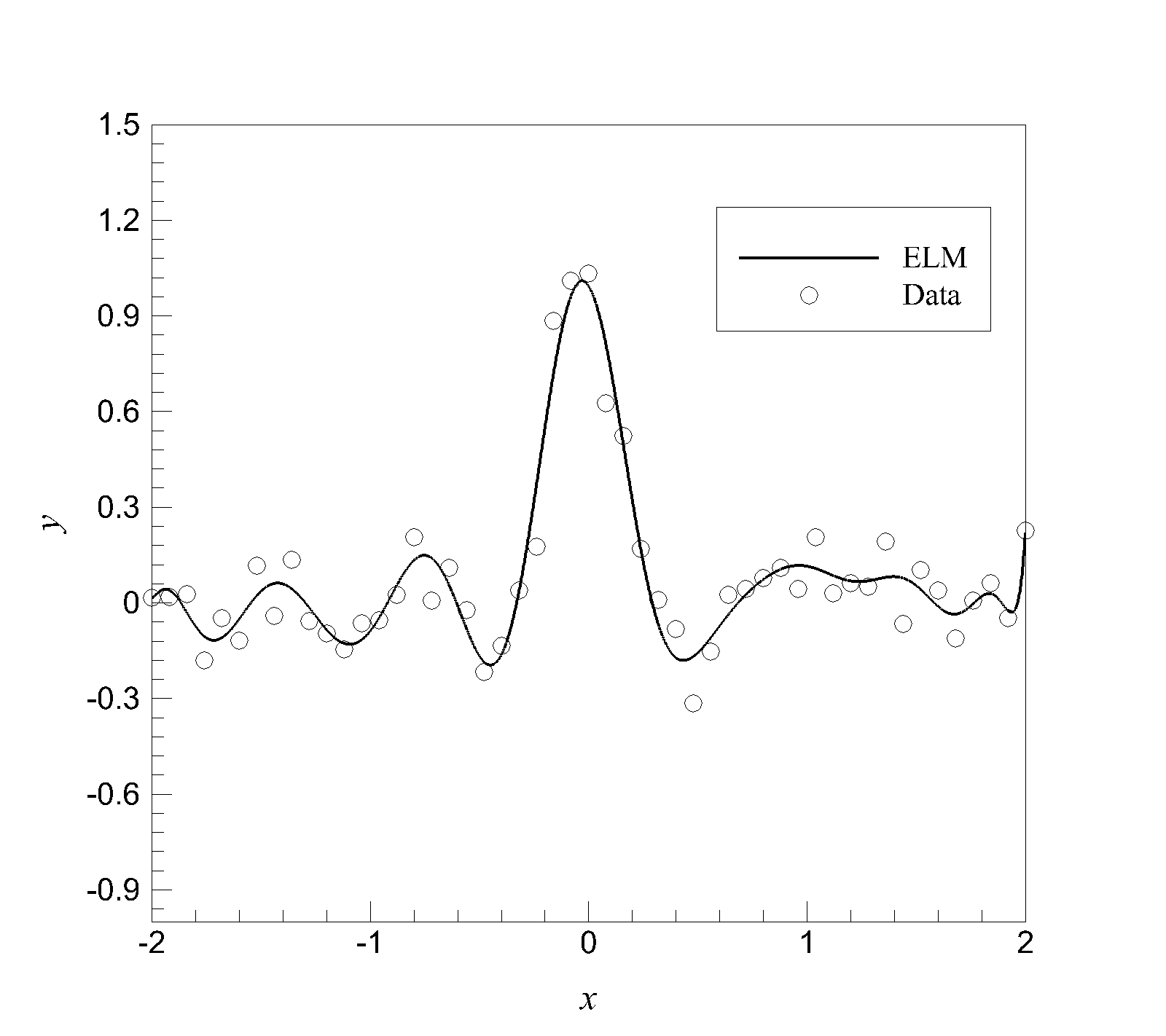}}
\subfigure[BR]{\includegraphics[width=0.5\textwidth]{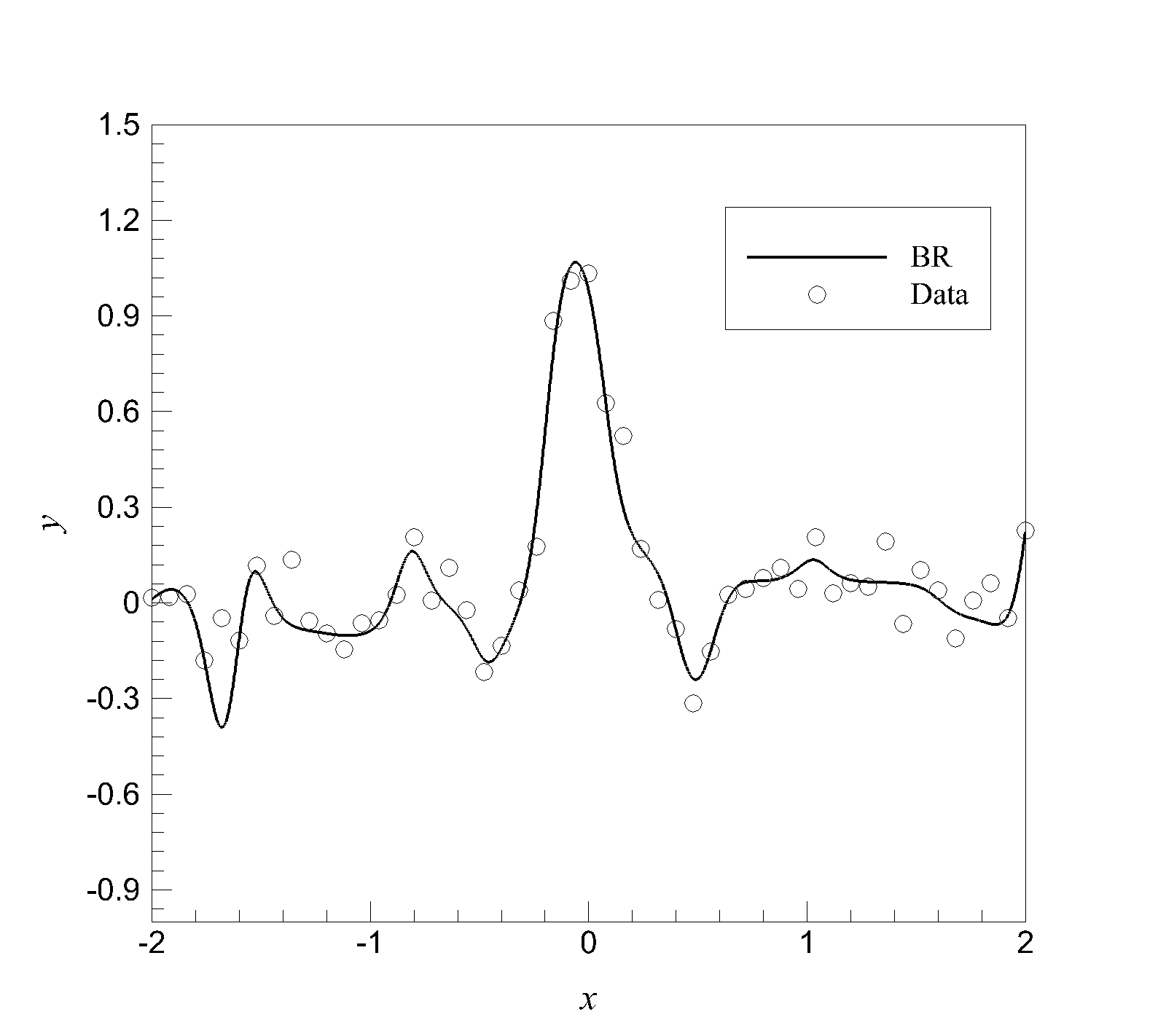}}
}
\caption{Comparison of the regularization ability of ELM and BR algorithms with no noise (top), a noise of amplitude 0.05 (middle) and a noise of amplitude 0.1 (bottom). Note how the ELM generally tends to capture the true function behavior better than BR.}
\label{fig:0}
\end{figure}

Before proceeding to the results of our investigation, it would be useful to compare the computational cost of the BR and ELM training algorithms. In addition, since both algorithms are designed for regularization ability, it would also be helpful to see their training performance for noisy data. For this purpose, we choose a simple one dimensional problem to test both approaches given by:
\begin{align}\label{test_case}
    f(x) =         \left\{
                      \begin{array}{ll}
                        \frac{\sin(\pi x)}{x}, & x \neq 0 \\
                        1, & x = 1 \, .
                      \end{array}
                    \right.
\end{align}
A training data set is devised by obtaining 51 samples of the above function at equal intervals within $x \in [-2,2]$. In addition two other training data sets are obtained with varying amounts of noise. We use a pseudorandom number generator and multiply the random number with an amplitude to mimic noise which is added to each training data point. To sum up, our test cases are those with zero noise, with an amplitude of 0.05 and 0.1. The ELM and BR algorithms were then tested on all training data sets. Five trials for each data set were run and training times were recorded as given in Table~(\ref{table1}). It is apparent that an increased noise in the data causes an increased duration for the convergence of the BR algorithm. On the other hand, the ELM method relies on the calculation of a pseudoinverse of a matrix, the dimensions of which depend solely on the number of hidden neurons and the size of the input data set. The regularization ability of the algorithms can be examined through the performance of the trained ANN output as shown in Fig.~(\ref{fig:0}). It can be observed that the ELM algorithm generally does a better job at extracting the underlying function behavior. However, we caution the reader that this behavior is not conclusive of either algorithm and a variety of test cases must be examined for stronger conclusions.

\begin{table}
\centering
\caption{Comparison of training times (in seconds) for the BR and ELM training algorithms for our model test case given by Eq.~(\ref{test_case}). Here amplitude refers to the approximate order of the random number that has been added to each data point. Note how increasing noise causes an increase in BR convergence times.}
\scalebox{0.85}{
\begin{tabular}{lllllll}
\hline\noalign{\smallskip}
& \multicolumn{2}{l}{\underline{Amplitude = 0 \hspace{0.5in}}} & \multicolumn{2}{l}{\underline{Amplitude = 0.05 \hspace{0.5in}}} & \multicolumn{2}{l}{\underline{Amplitude = 0.1\hspace{0.5in}}}\\
Trial Number & ELM & BR & ELM  & BR & ELM  & BR\\
\noalign{\smallskip}\hline\noalign{\smallskip}
Trial 1 & 0.002514 &  0.91155  & 0.002409 & 2.03174 & 0.023618  & 1.91132 \\
Trial 2 & 0.001187 &  0.51236  & 0.001286 & 2.06923 & 0.001208  & 5.29371\\
Trial 3 & 0.001373 &  0.70040  & 0.001565 & 2.86361 & 0.001638  & 1.64957 \\
Trial 4 & 0.001203 &  0.81693  & 0.001230 & 1.33273 & 0.001186  & 2.28243 \\
Trial 5 & 0.001335 &  1.24820  & 0.001099 & 2.26790 & 0.001328  & 2.06521 \\
\noalign{\smallskip}\hline
\end{tabular}}
\label{table1}
\end{table}

\section{Results}
\label{sec:results}

\begin{figure}[!ht]
\centering
\includegraphics[width=0.5\textwidth]{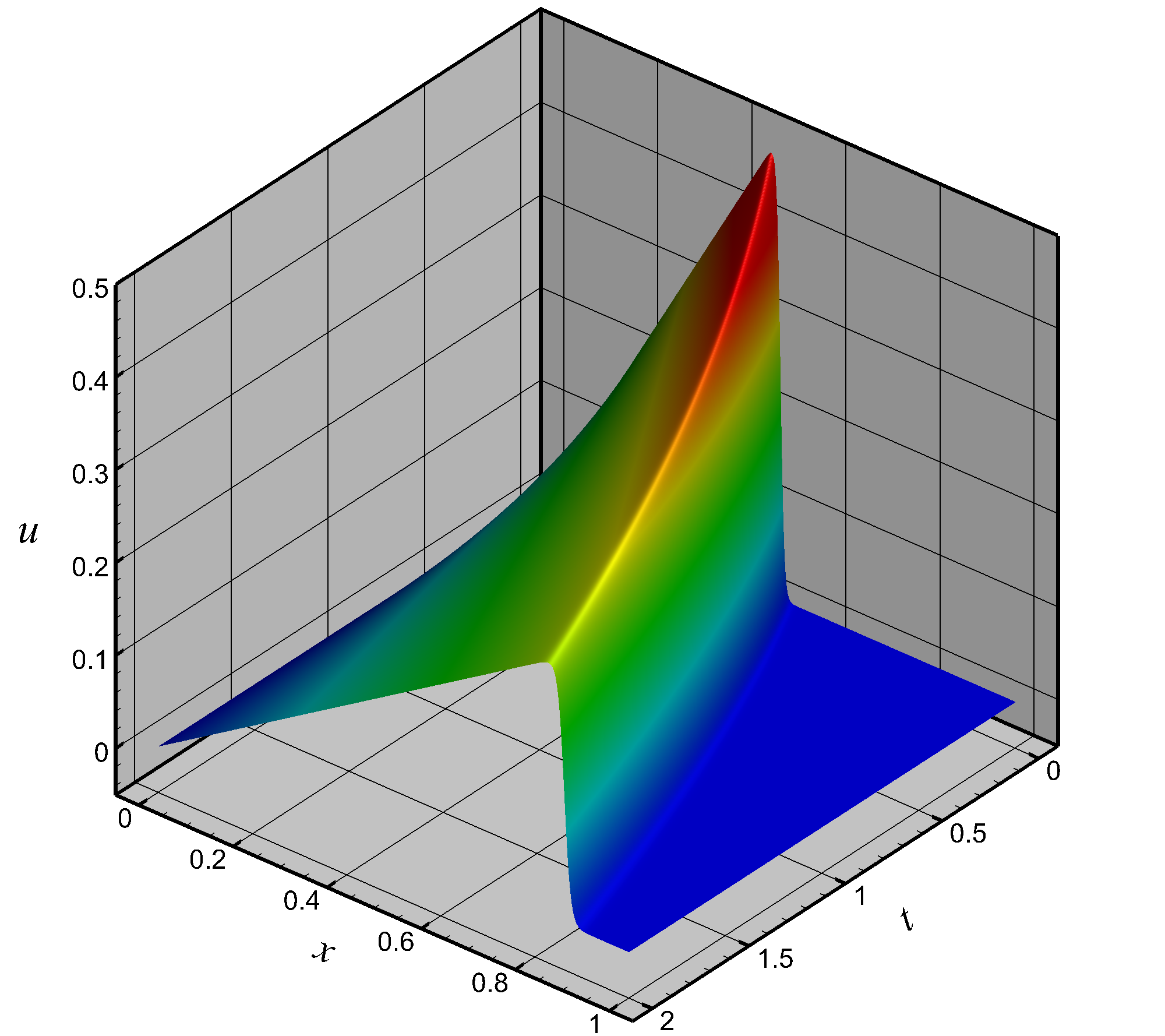}
\caption{Exact solution of Burgers equation at $Re=1000$.}
\label{fig:1}
\end{figure}

\begin{figure}[!ht]
\centering
\includegraphics[width=0.5\textwidth]{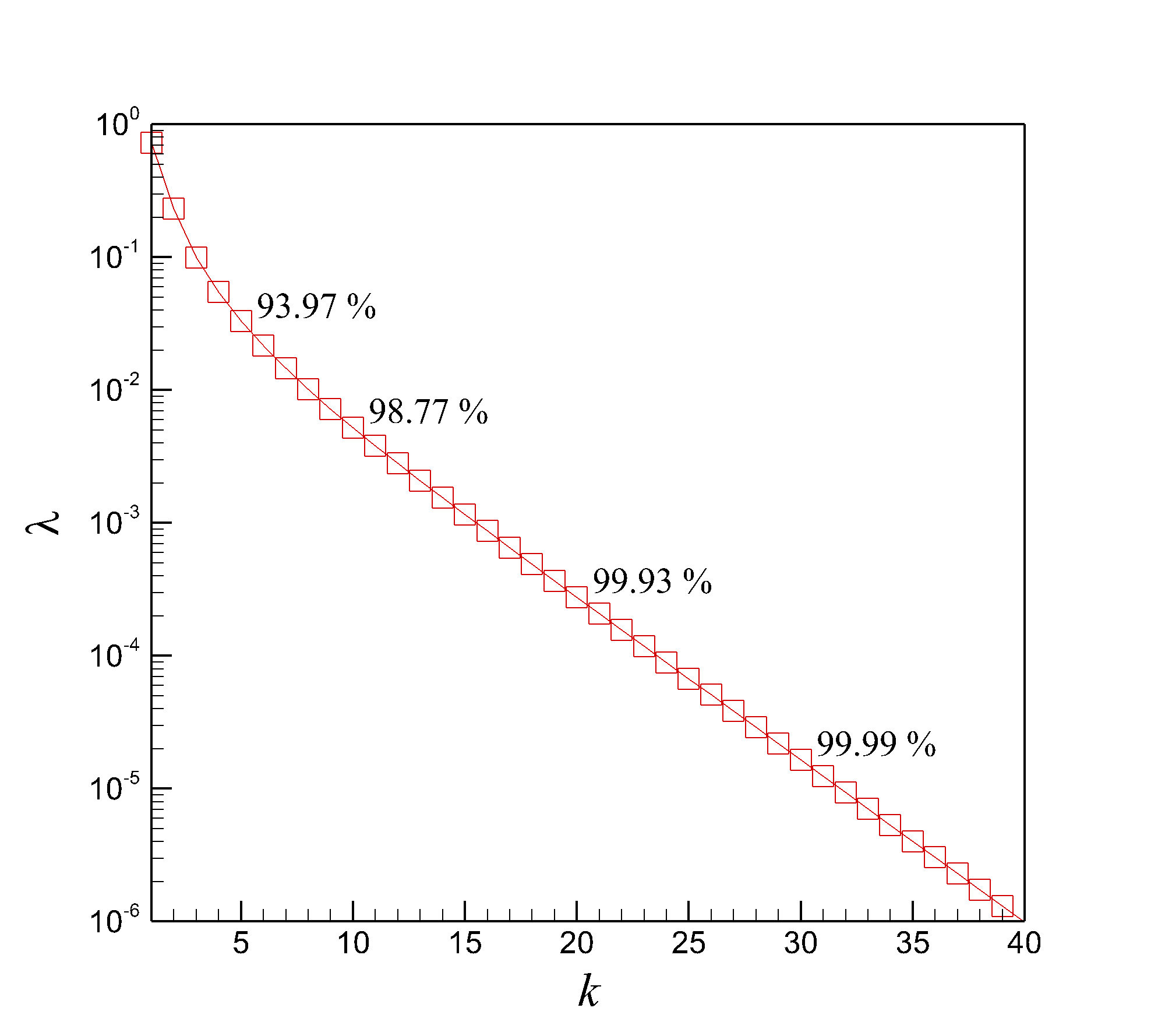}
\caption{Eigenvalues of the snapshot data matrix. Note that first five modes captures almost 94 \% of the energy.}
\label{fig:2}
\end{figure}

This section details the results of our investigation for the viscous Burgers equation problem. Our test case is given by the following initial and boundary conditions:
\begin{align}\label{results1}
  \begin{split}
  u(x,0) &= \frac{x}{1+\sqrt{\frac{1}{t_0}} \exp(Re \frac{x^2}{4})} \\
  u(0,t) &= 0 \\
  u(L,t) &= 0 \\
  \end{split}
\end{align}
where the length of our domain $L=1$ and maximum time $t_m=2$. The PDE for the Burgers equation with the aforementioned boundary and initial conditions may be solved exactly to obtain an analytical formulation for the time evolution for the field variable $u(x,t)$ given by \cite{maleewong2011line}
\begin{align}\label{VBE_Exact}
  u(x,t) = \frac{\frac{x}{t+1}}{1 + \sqrt{\frac{t+1}{t_0}} \exp(Re \frac{x^2}{4t+4})}
\end{align}
where $t_0 = \exp(Re/8)$. This exact expression is used to generate snapshot data for our ROM assessments. We compute inner products using the well-known Simpson's rule \cite{moin2010fundamentals} and a third-order Runge-Kutta scheme is utilized for solving the resulting ROM. The reader is directed to \cite{san2013proper} for a more detailed discussion of the numerical methods used in this investigation. A space-time evolution of our solution field is presented in Fig.~(\ref{fig:1}) for an $Re=1000$. Fig.~(\ref{fig:2}) shows the distribution of the eigenvalues as a function of the total energy captured. It is apparent that retaining the eigenvectors corresponding to the first 5 eigenvalues causes a 94\% capture of energy for the system and the first 20 modes contribute to 99.93\% of the total energy of the system at $Re=1000$. Note that a variation in $Re$ would manifest itself with a higher number of modes needed to capture an equivalent amount of energy (if $Re$ increases) and a lower number of modes (if $Re$ decreases).

Before proceeding with an elaboration of our results, it is worthwhile to elaborate on the definition of our models:
\begin{itemize}
  \item Fourier-GP-ROM: Reduced order model obtained by the standard Galerkin projection to Fourier modes.
  \item POD-GP-ROM: Reduced order model obtained by the standard Galerkin projection to POD modes.
  \item Fourier-ANN-ROM: Proposed reduced order model with ANN closure applied to Fourier modes.
  \item POD-ANN-ROM: Proposed reduced order model with ANN closure applied to POD modes.
  \item Fourier-True: Projection of exact solution of PDE into space spanned by Fourier modes.
  \item POD-True: Projection of exact solution of PDE into space spanned by POD modes.
\end{itemize}

\begin{figure}[!ht]
\centering
\mbox{
\subfigure[$M=5$]{\includegraphics[width=0.5\textwidth]{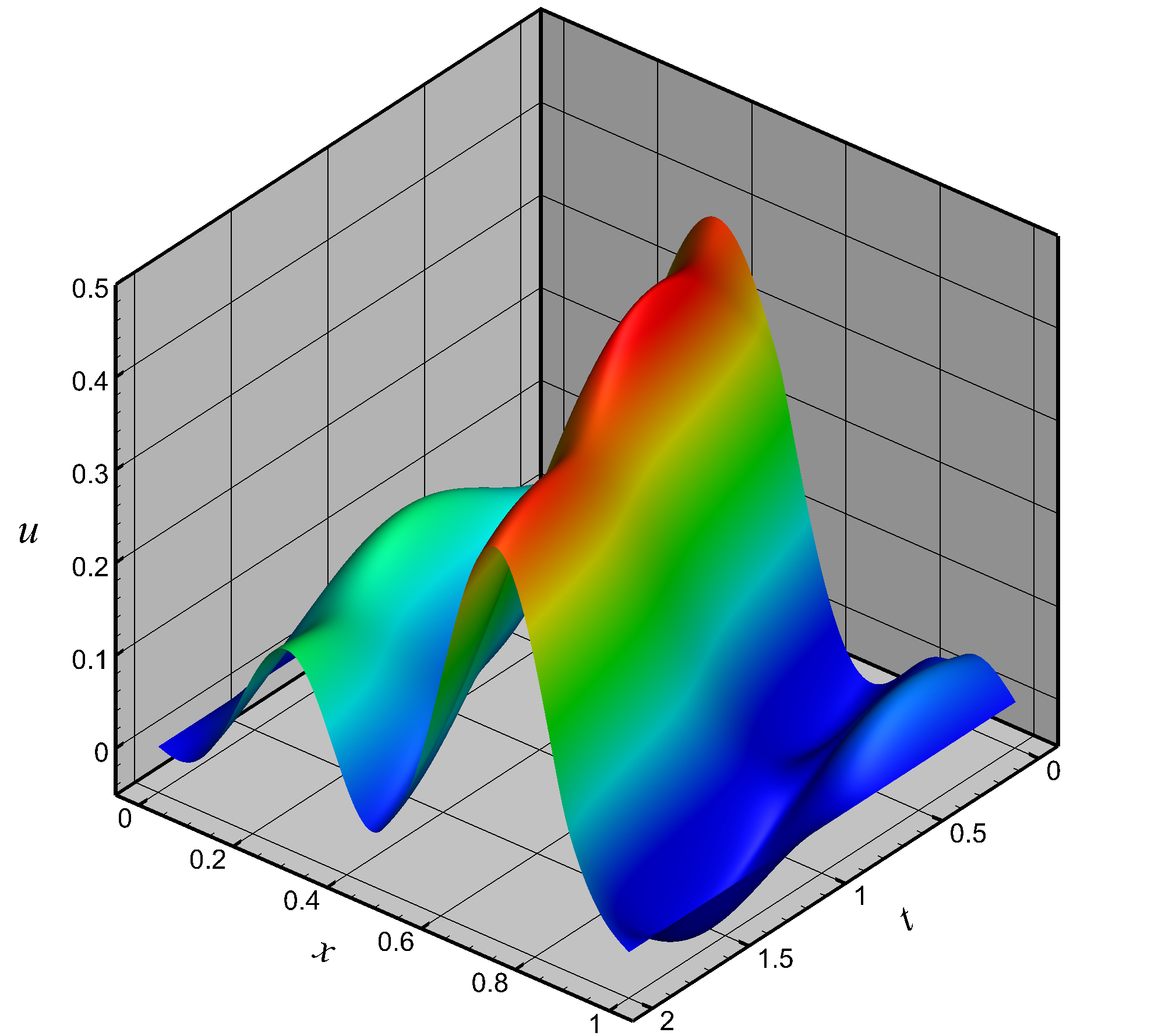}}
\subfigure[$M=10$]{\includegraphics[width=0.5\textwidth]{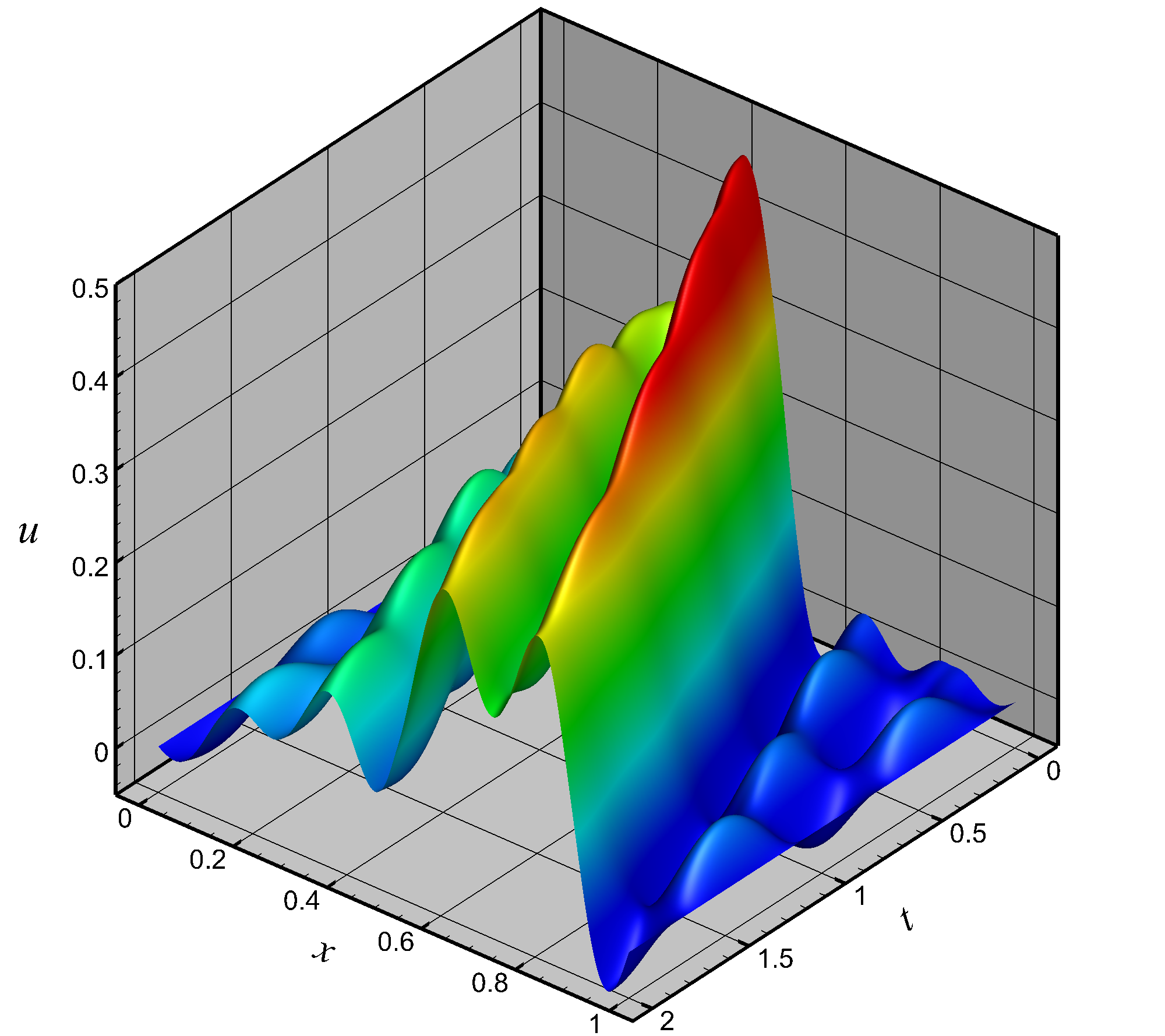}}
}\\
\mbox{
\subfigure[$M=20$]{\includegraphics[width=0.5\textwidth]{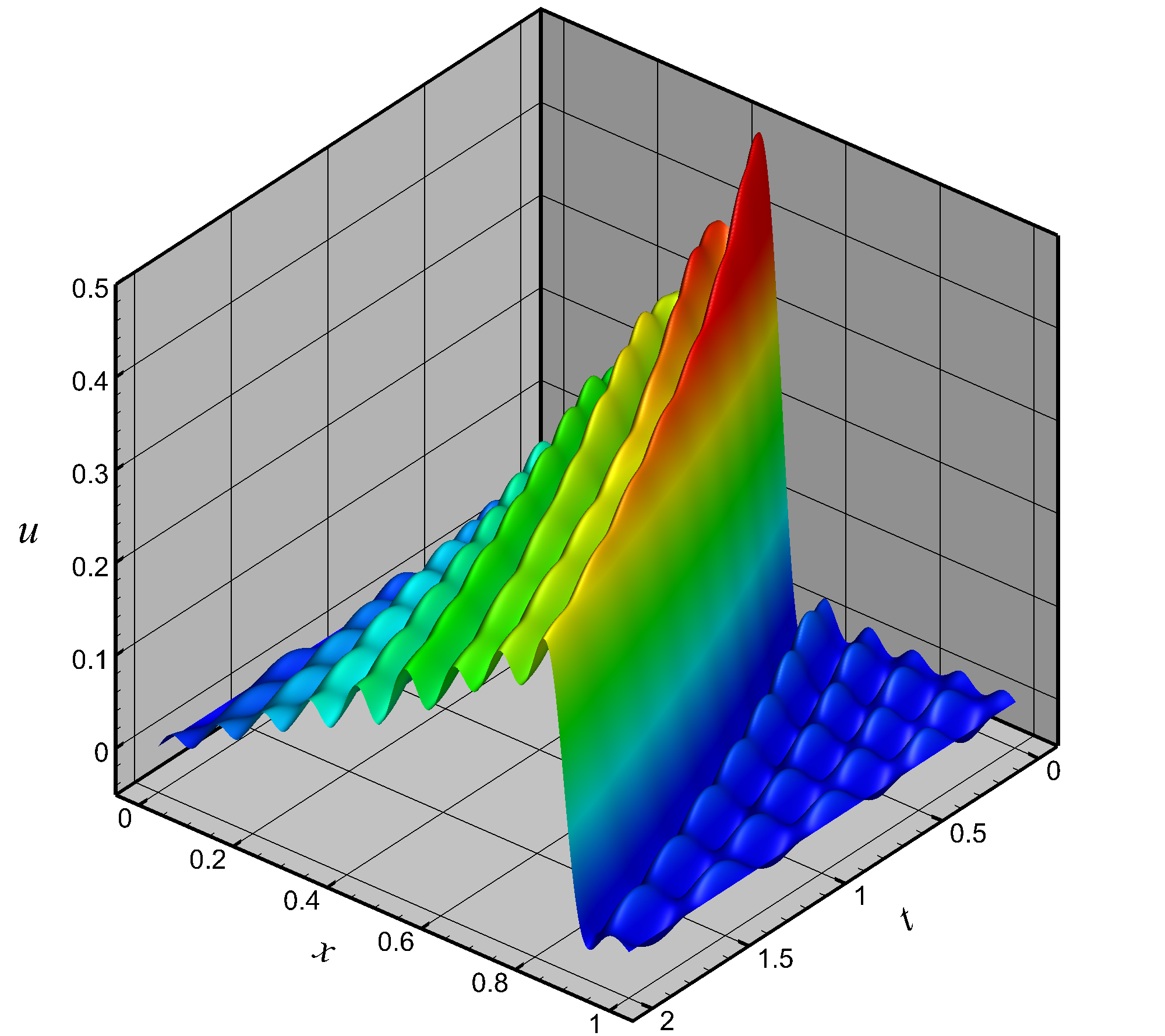}}
\subfigure[$M=30$]{\includegraphics[width=0.5\textwidth]{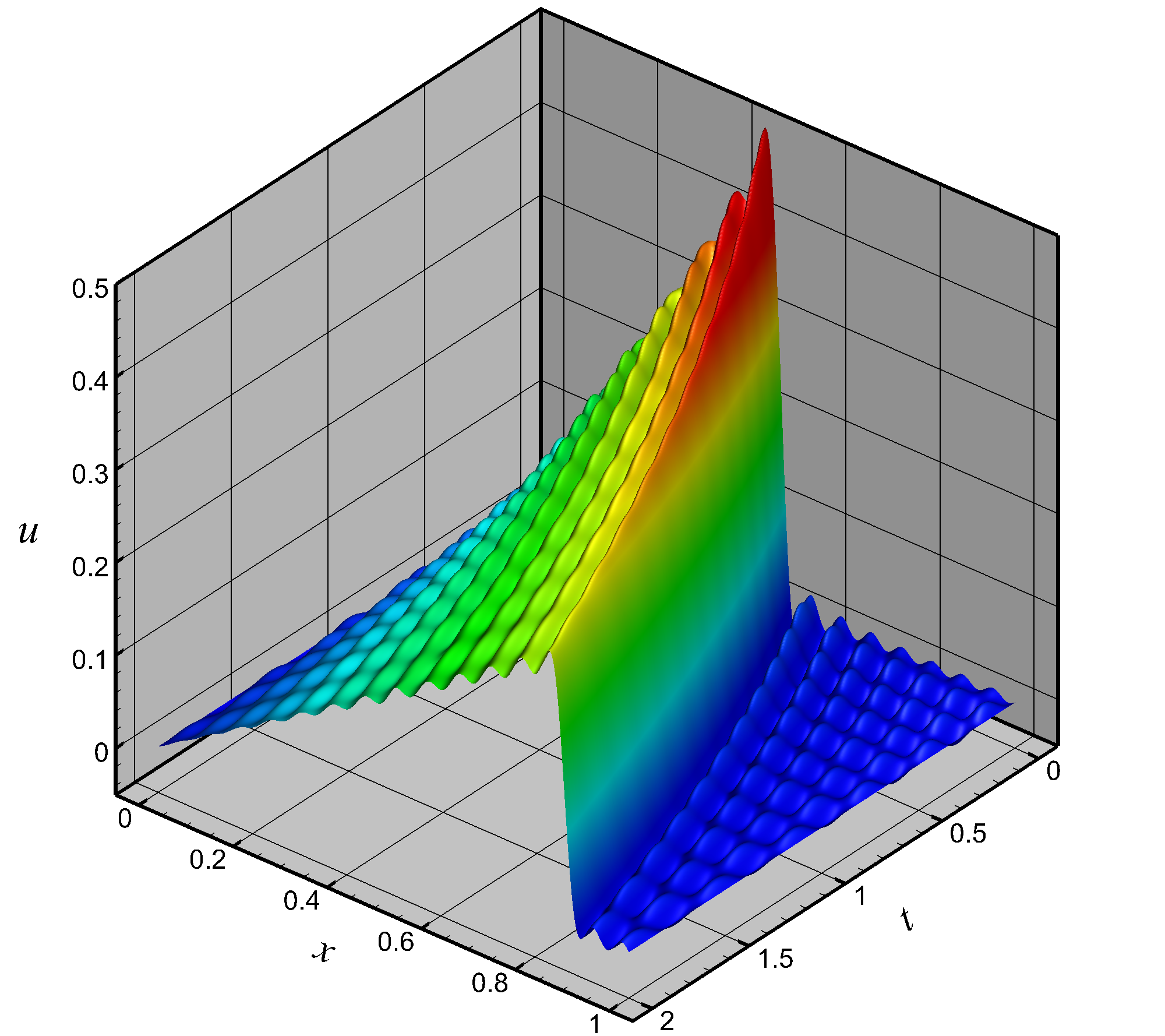}}
}
\caption{Fourier-GP-ROM with different number of modes at $Re=1000$.}
\label{fig:5a}
\end{figure}

\begin{figure}[!ht]
\centering
\mbox{
\subfigure[$M=5$]{\includegraphics[width=0.5\textwidth]{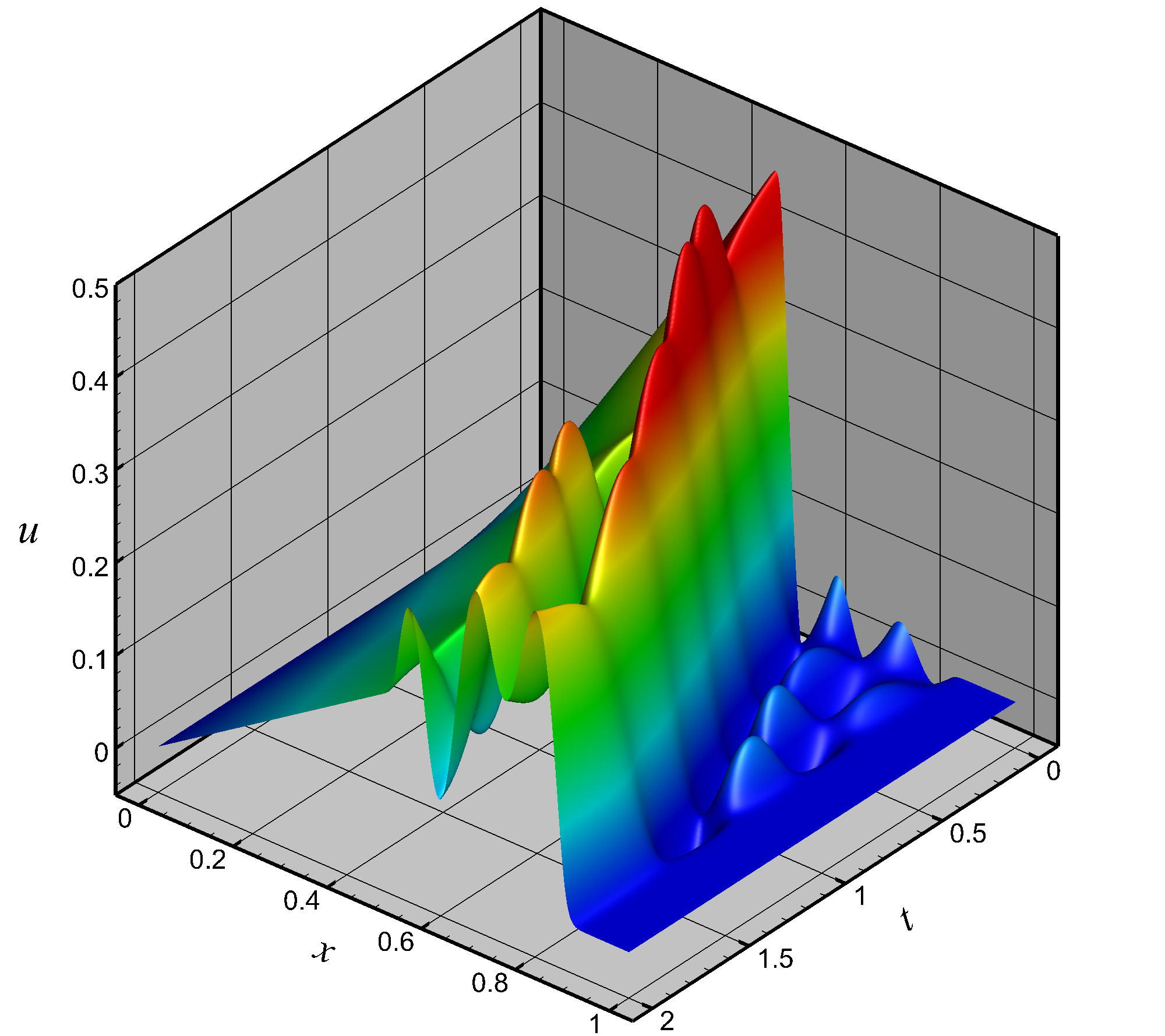}}
\subfigure[$M=10$]{\includegraphics[width=0.5\textwidth]{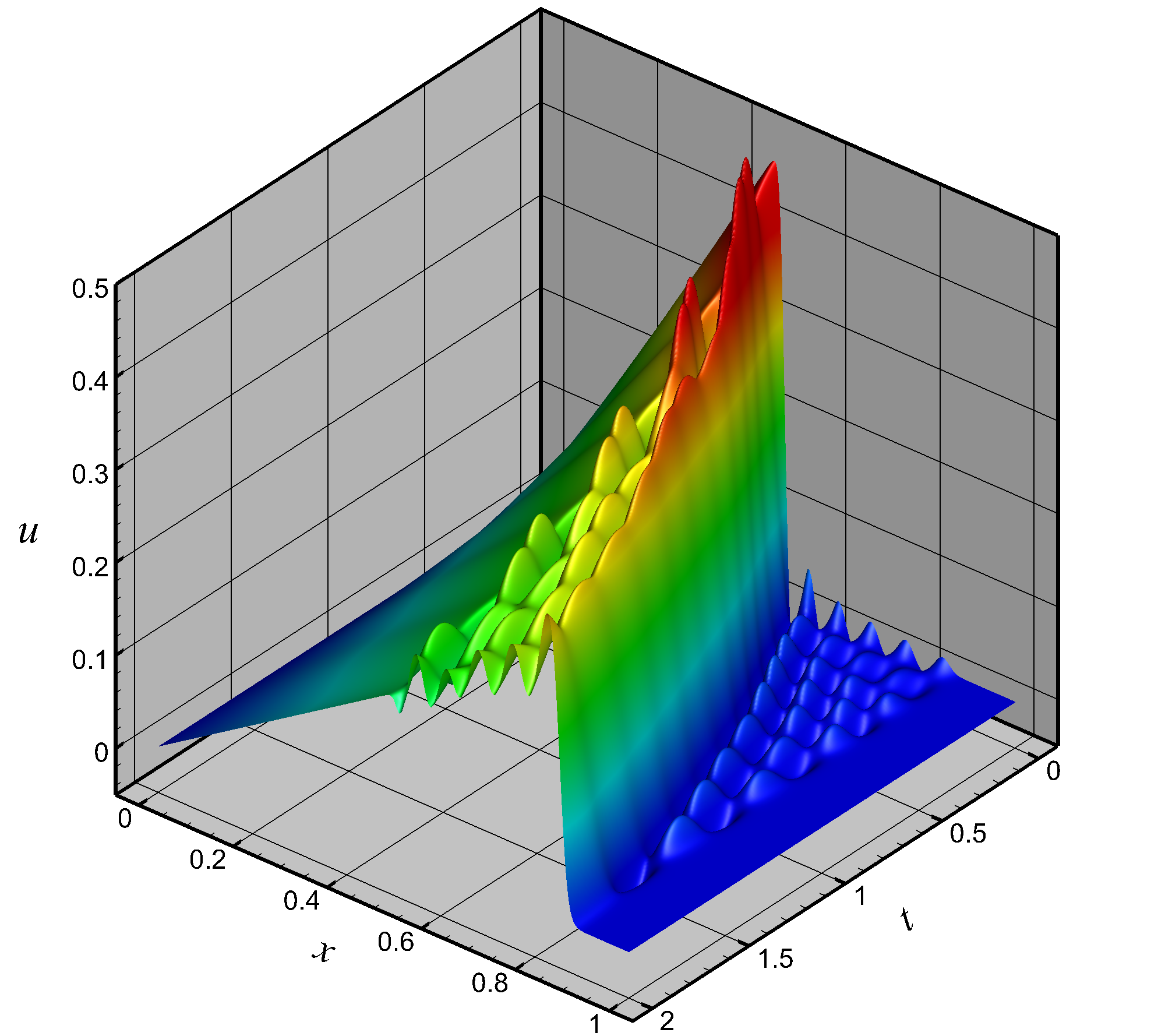}}
}\\
\mbox{
\subfigure[$M=20$]{\includegraphics[width=0.5\textwidth]{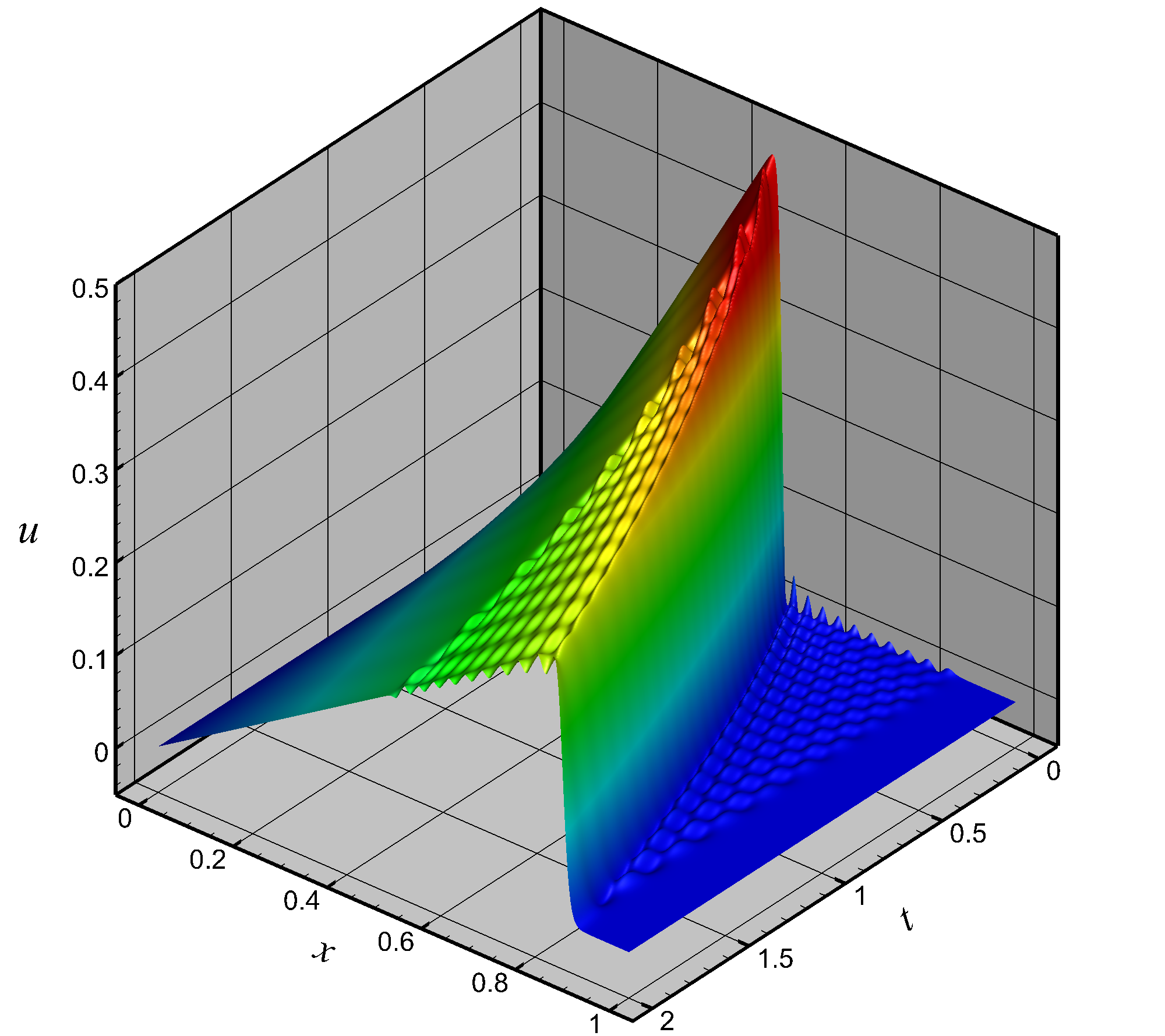}}
\subfigure[$M=30$]{\includegraphics[width=0.5\textwidth]{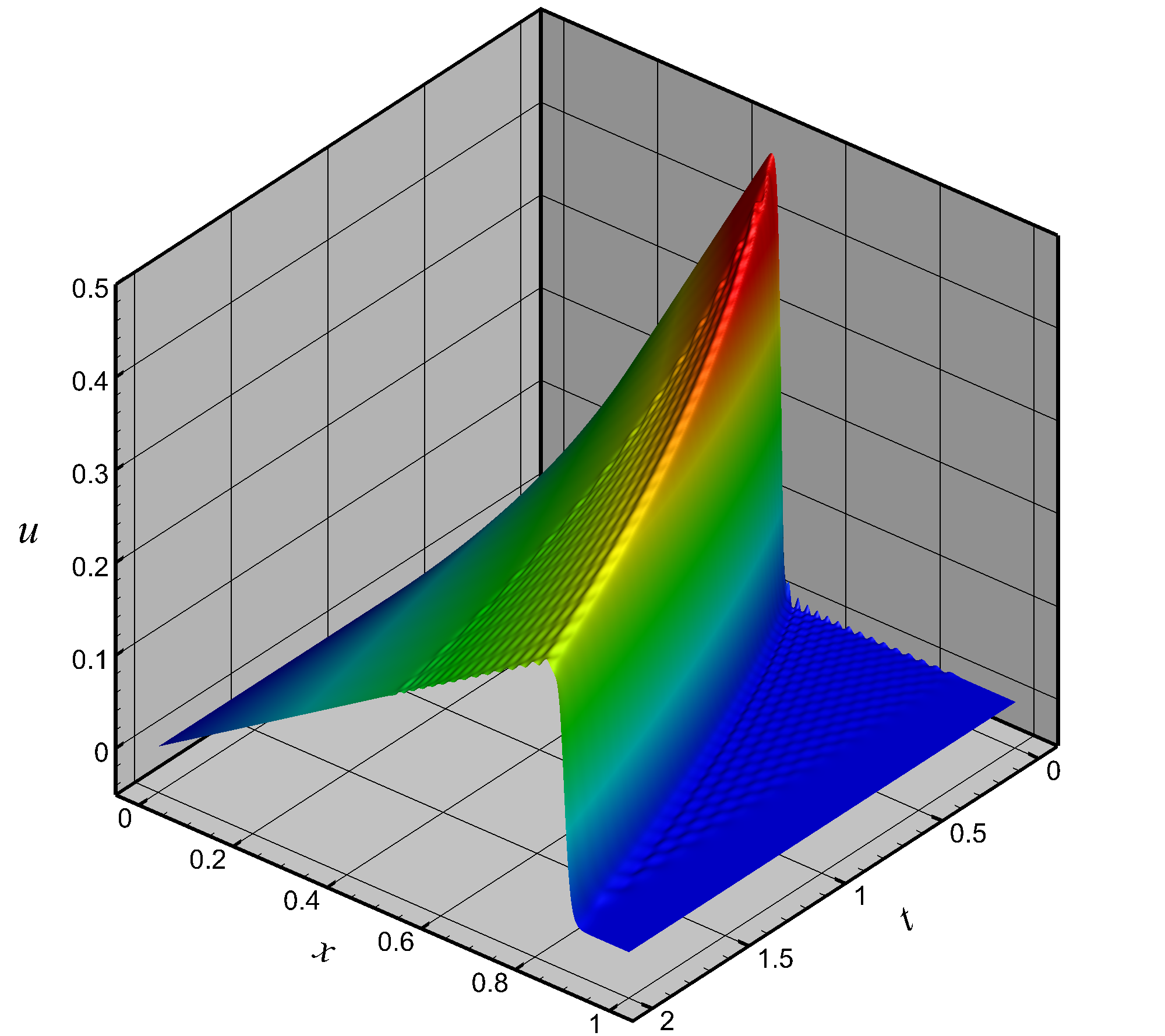}}
}
\caption{POD-GP-ROM with different number of modes at $Re=1000$.}
\label{fig:5b}
\end{figure}

Fig.~(\ref{fig:5a}) shows the time evolution of the solution field when using a ROM implementation of the Fourier modes. On increasing the number of retained modes $M$, one can see a convergence to the true behavior of the PDE. However, for $Re=1000$, some oscillations can still be observed for $M=30$ retained modes. When using the POD modes instead, as shown in Fig.~(\ref{fig:5b}), a quicker convergence is observed with increasing modes. This is expected since POD bases are generated from exact snapshot data.

\begin{figure}[!ht]
\centering
\mbox{
\subfigure[Fourier-GP-ROM]{\includegraphics[width=0.5\textwidth]{figures/b_GP_Fr_re1000.png}}
\subfigure[POD-GP-ROM]{\includegraphics[width=0.5\textwidth]{figures/b_GP_POD_re1000.png}}
}\\
\mbox{
\subfigure[Fourier-ANN-ROM]{\includegraphics[width=0.5\textwidth]{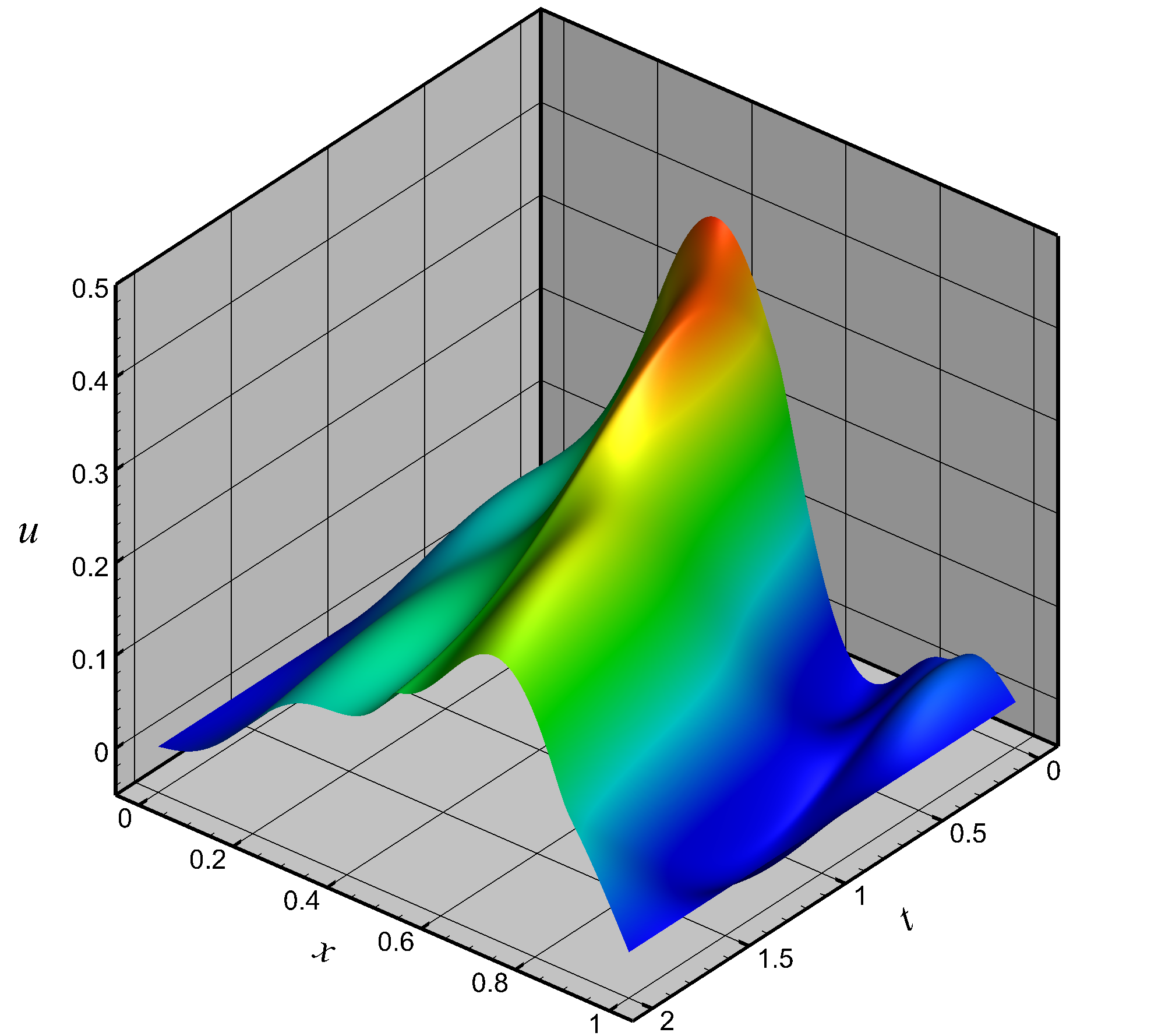}}
\subfigure[POD-ANN-ROM]{\includegraphics[width=0.5\textwidth]{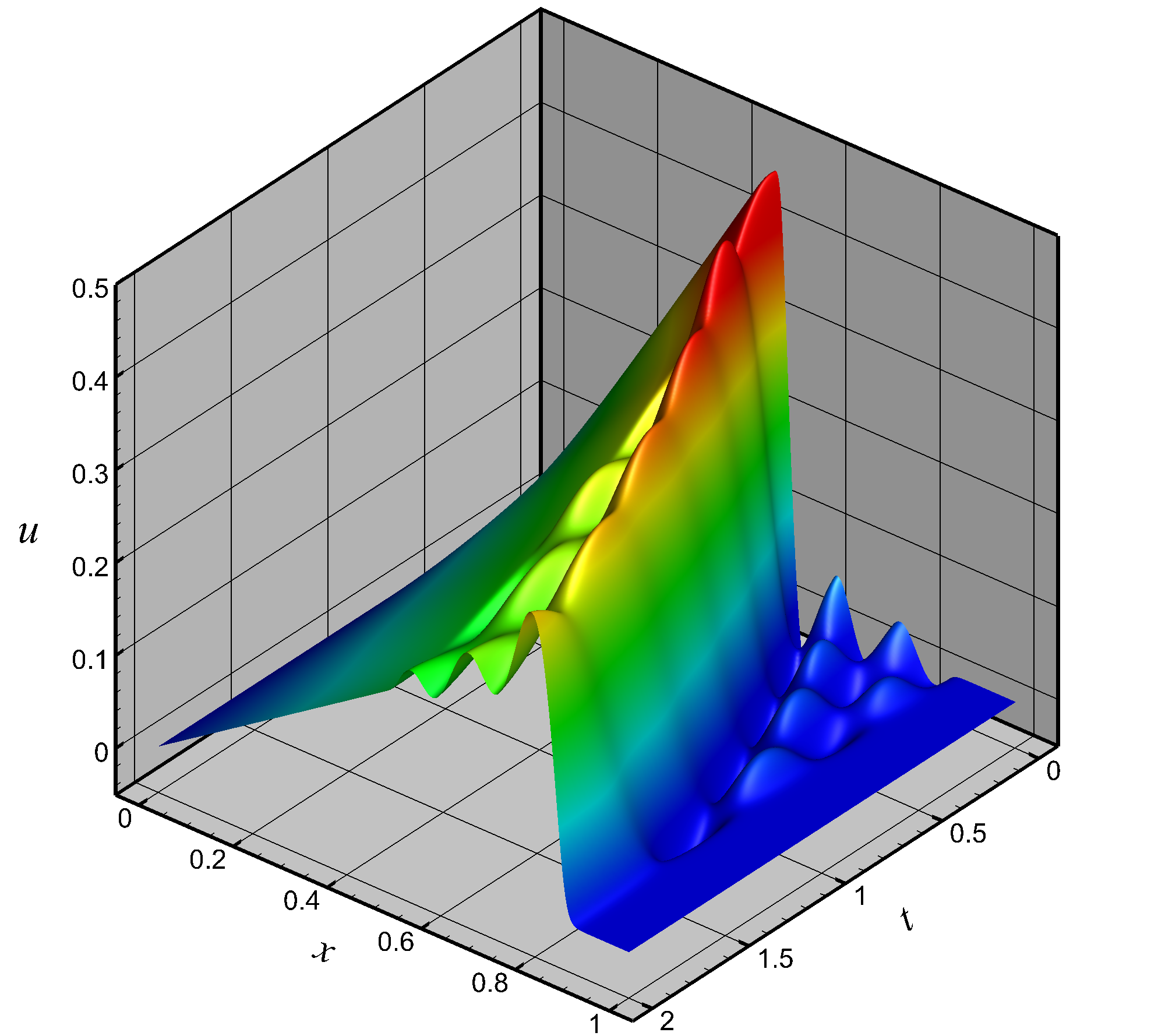}}
}\\
\mbox{
\subfigure[Fourier-True]{\includegraphics[width=0.5\textwidth]{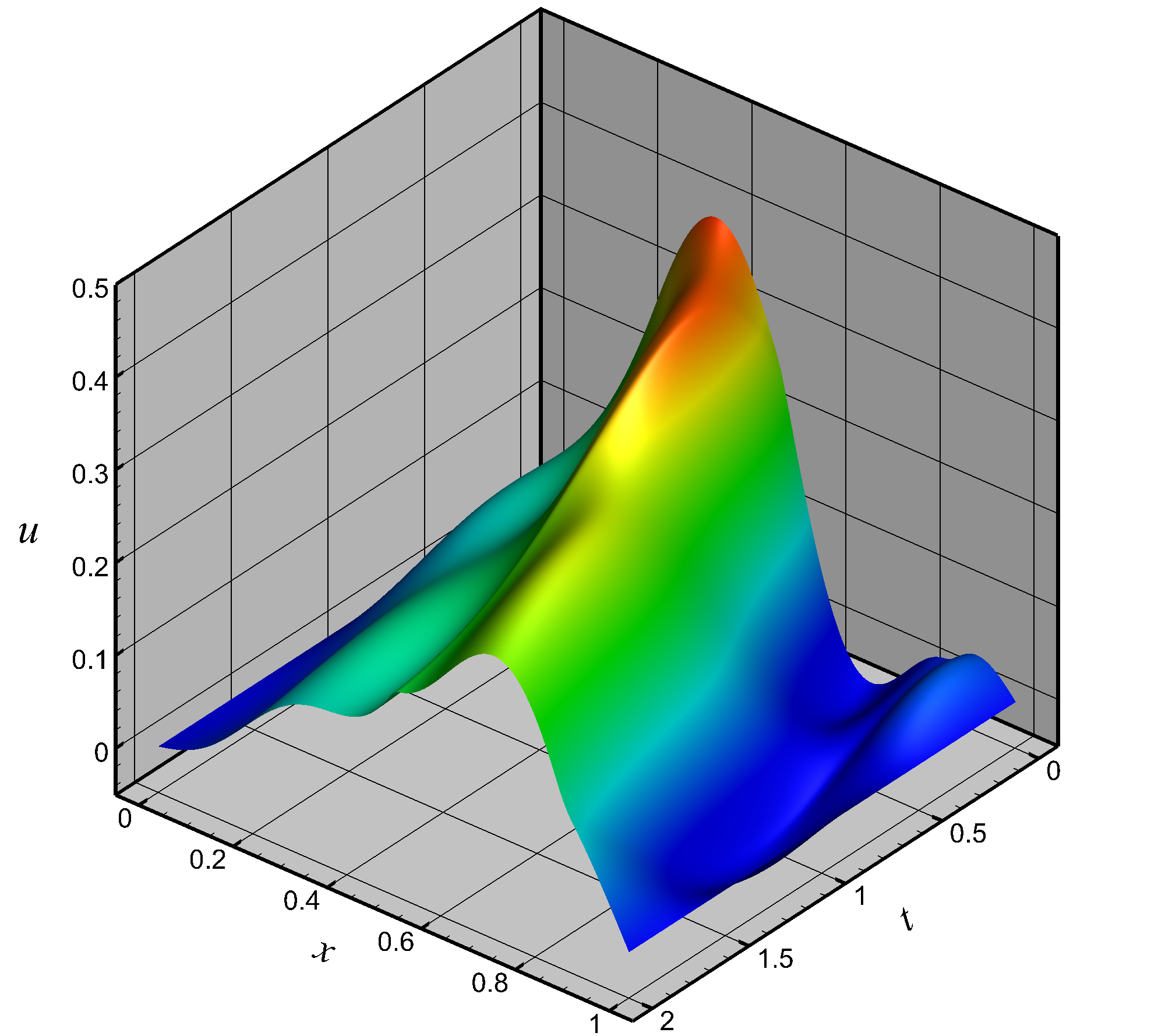}}
\subfigure[POD-True]{\includegraphics[width=0.5\textwidth]{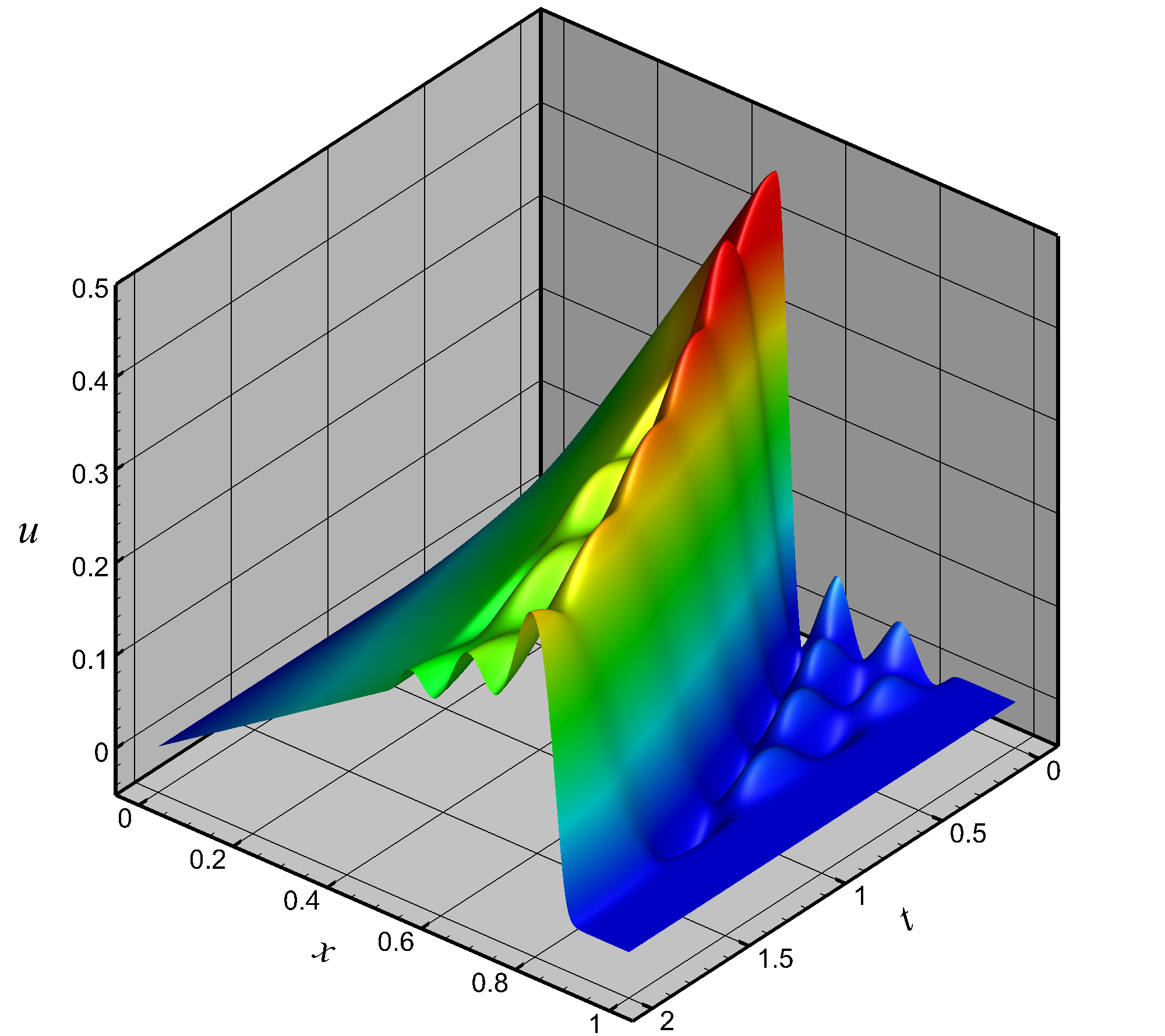}}
}
\caption{Predictive performance of the models using $M=5$ modes at $Re=1000$. ANN with Bayesian regularization with 10 neurons.}
\label{fig:6}
\end{figure}

Fig.~(\ref{fig:6}) shows the performance of the Fourier-GP-ROM and POD-GP-ROM methods against the Fourier-ANN-ROM and POD-ANN-ROM approaches for the first five modes. The true projections are also shown for the purpose of comparison. It is immediately apparent that the use of the ANN closures lends a `stabilization' effect to the modal evolution. We remark that the training data for the ANN architecture includes modal evolution data sets obtained by $Re \in [200,300,...,1200]$. At $Re=1000$, which is a value of Reynolds number within the data set, an excellent performance is observed with the ANN-ROM approaches virtually indistinguishable from the true modes. We note that the BR training technique has been used for these figures.

\begin{figure}[!ht]
\centering
\mbox{
\subfigure[Fourier-GP-ROM ($Re=250$)]{\includegraphics[width=0.5\textwidth]{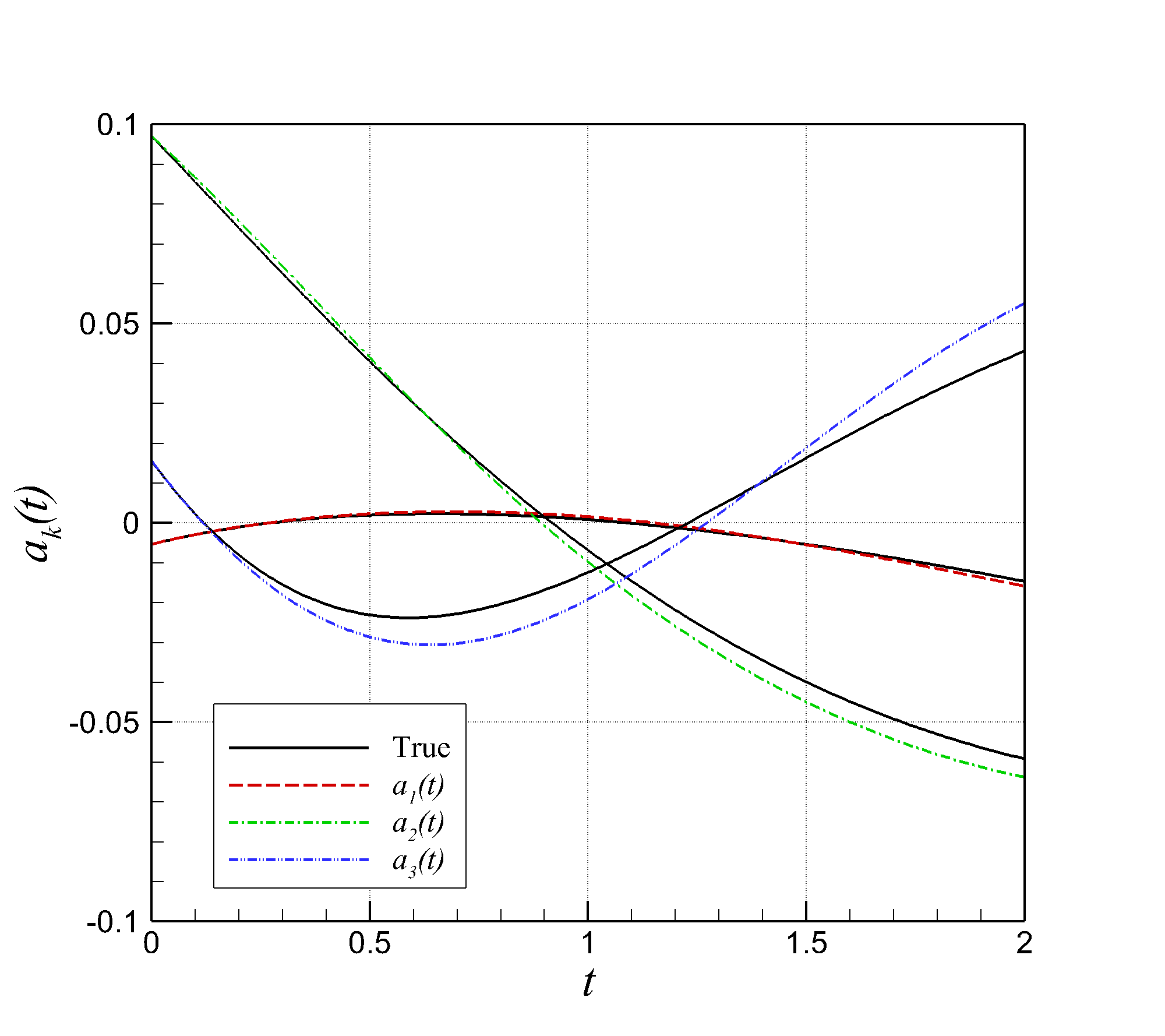}}
\subfigure[Fourier-ANN-ROM($Re=250$)]{\includegraphics[width=0.5\textwidth]{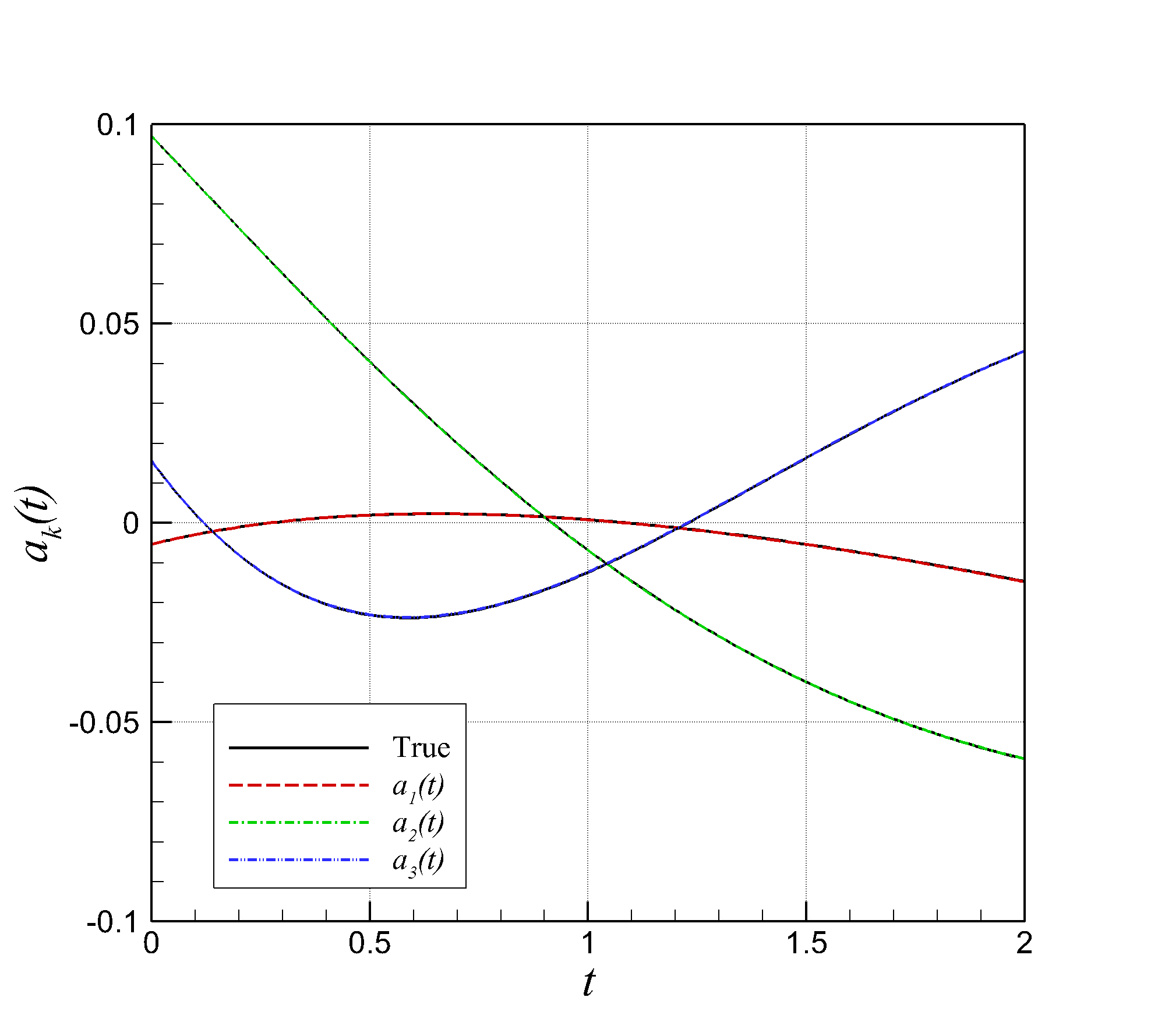}}
}\\
\mbox{
\subfigure[Fourier-GP-ROM ($Re=1000$)]{\includegraphics[width=0.5\textwidth]{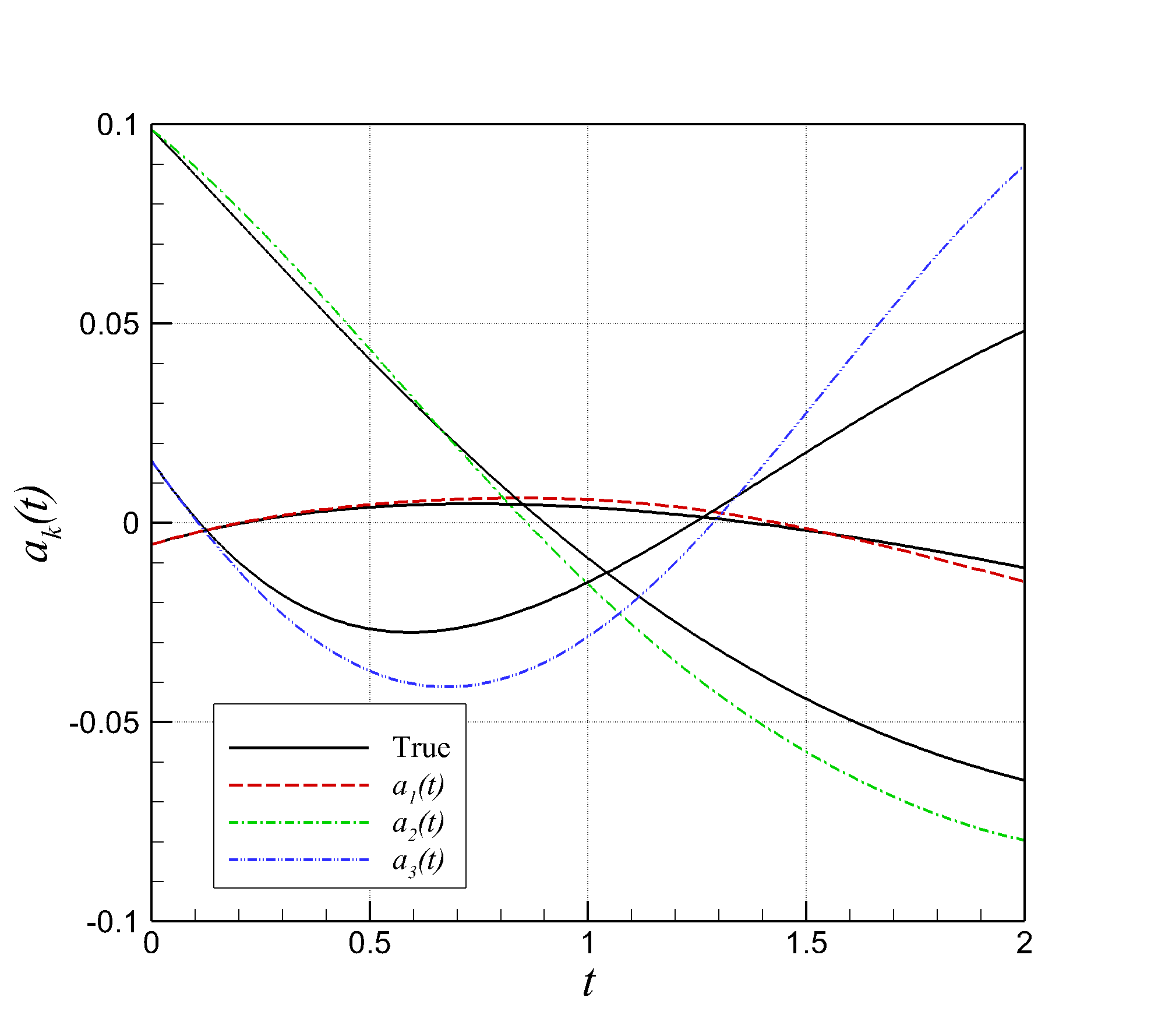}}
\subfigure[Fourier-ANN-ROM($Re=1000$)]{\includegraphics[width=0.5\textwidth]{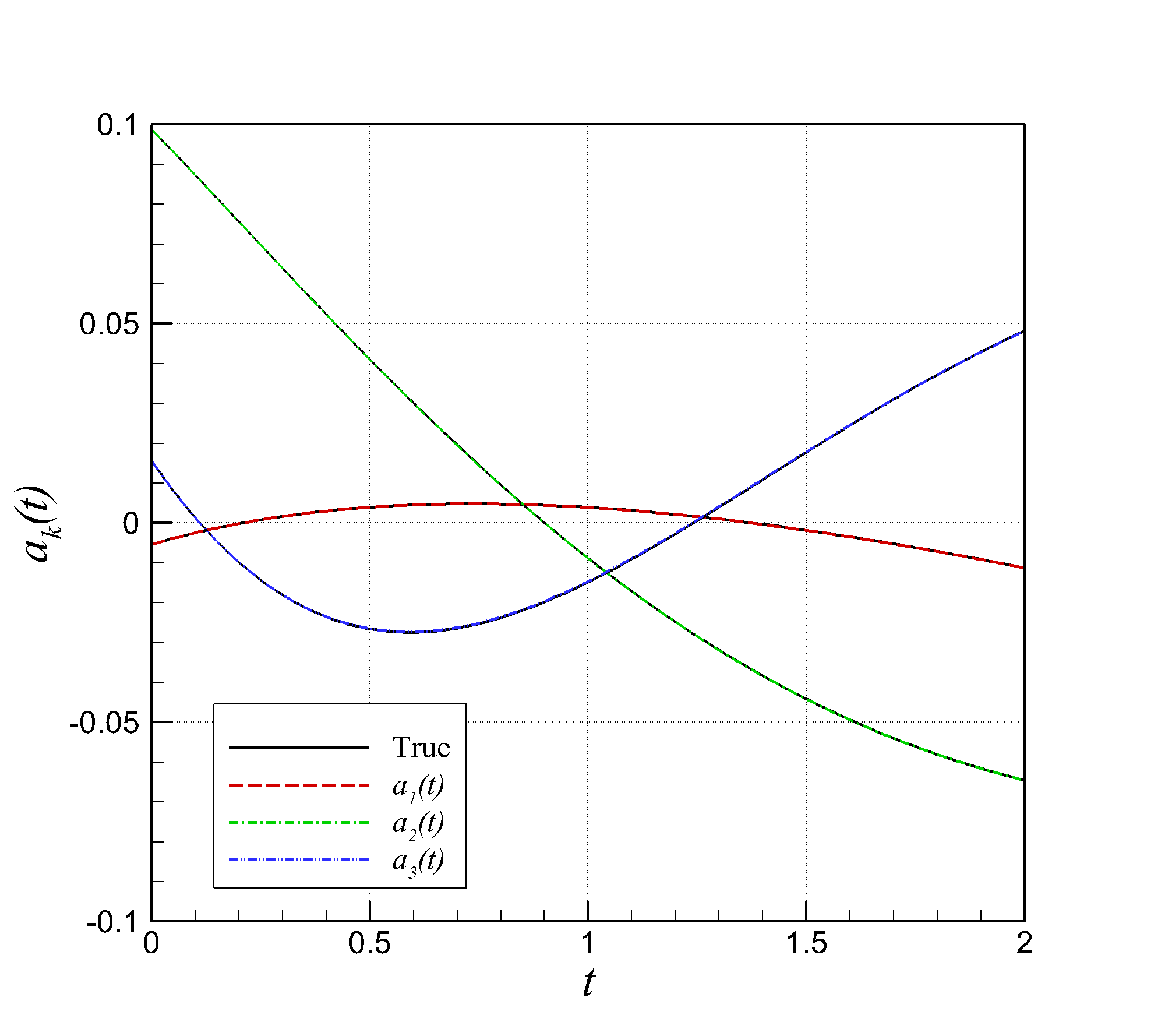}}
}\\
\mbox{
\subfigure[Fourier-GP-ROM ($Re=1500$)]{\includegraphics[width=0.5\textwidth]{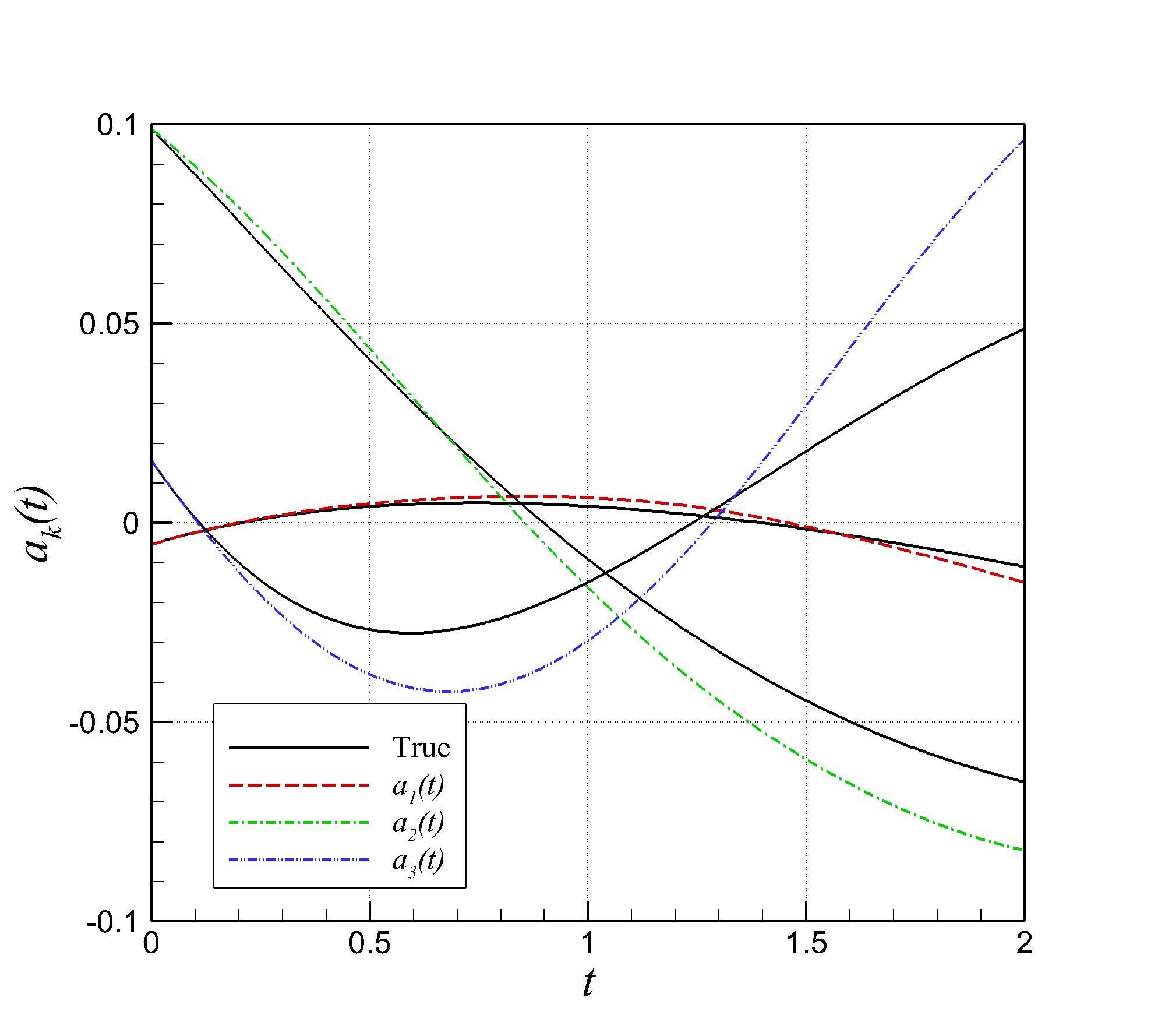}}
\subfigure[Fourier-ANN-ROM($Re=1500$)]{\includegraphics[width=0.5\textwidth]{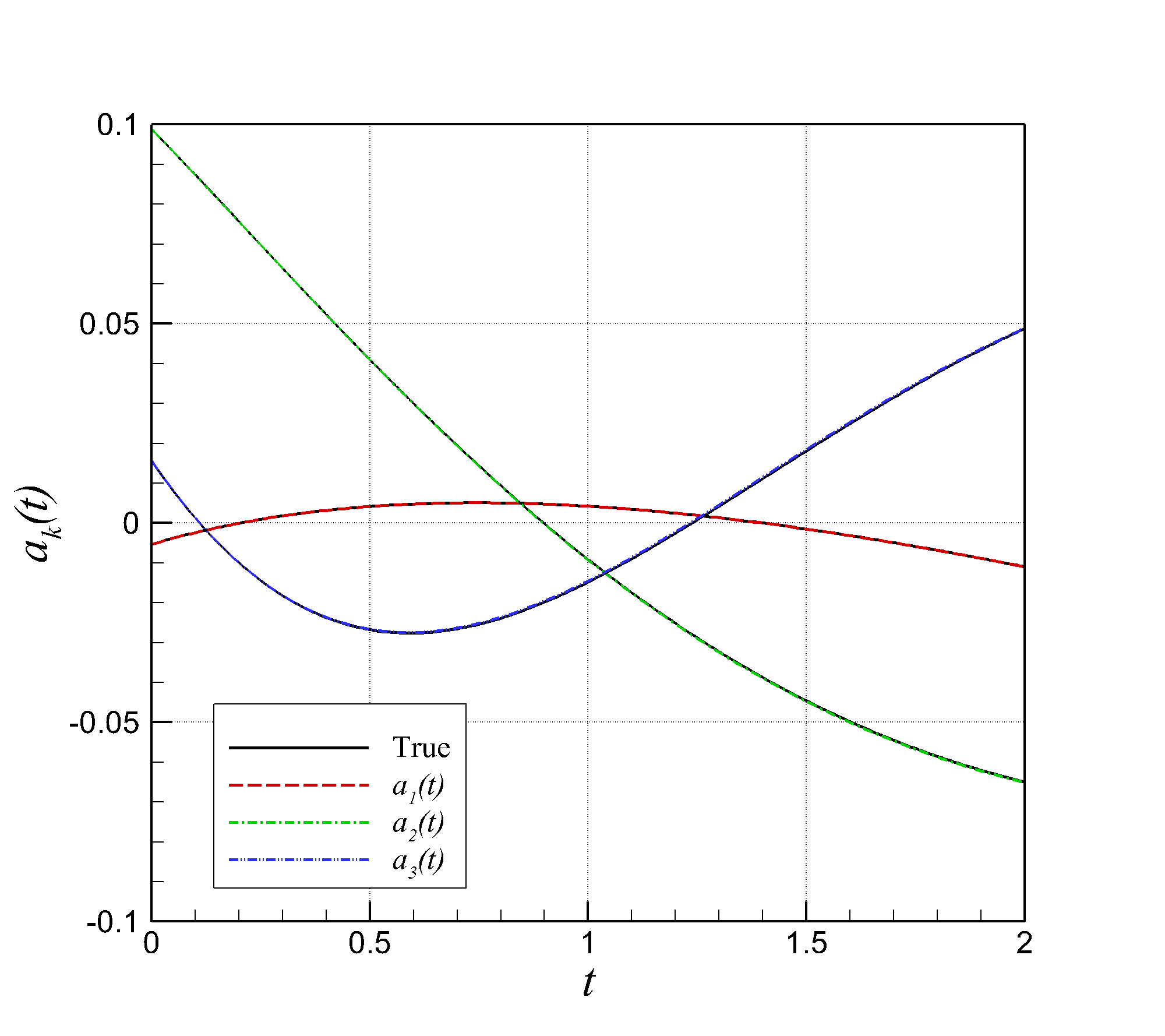}}
}
\caption{Time evolution of first few modes when using the Fourier basis. ANN with Bayesian regularization with 10 neurons. Note that $Re=250$ and $Re=1500$ cases are not included in our training set of $Re \in [200-1200]$. }
\label{fig:7}
\end{figure}

\begin{figure}[!ht]
\centering
\mbox{
\subfigure[POD-GP-ROM ($Re=250$)]{\includegraphics[width=0.5\textwidth]{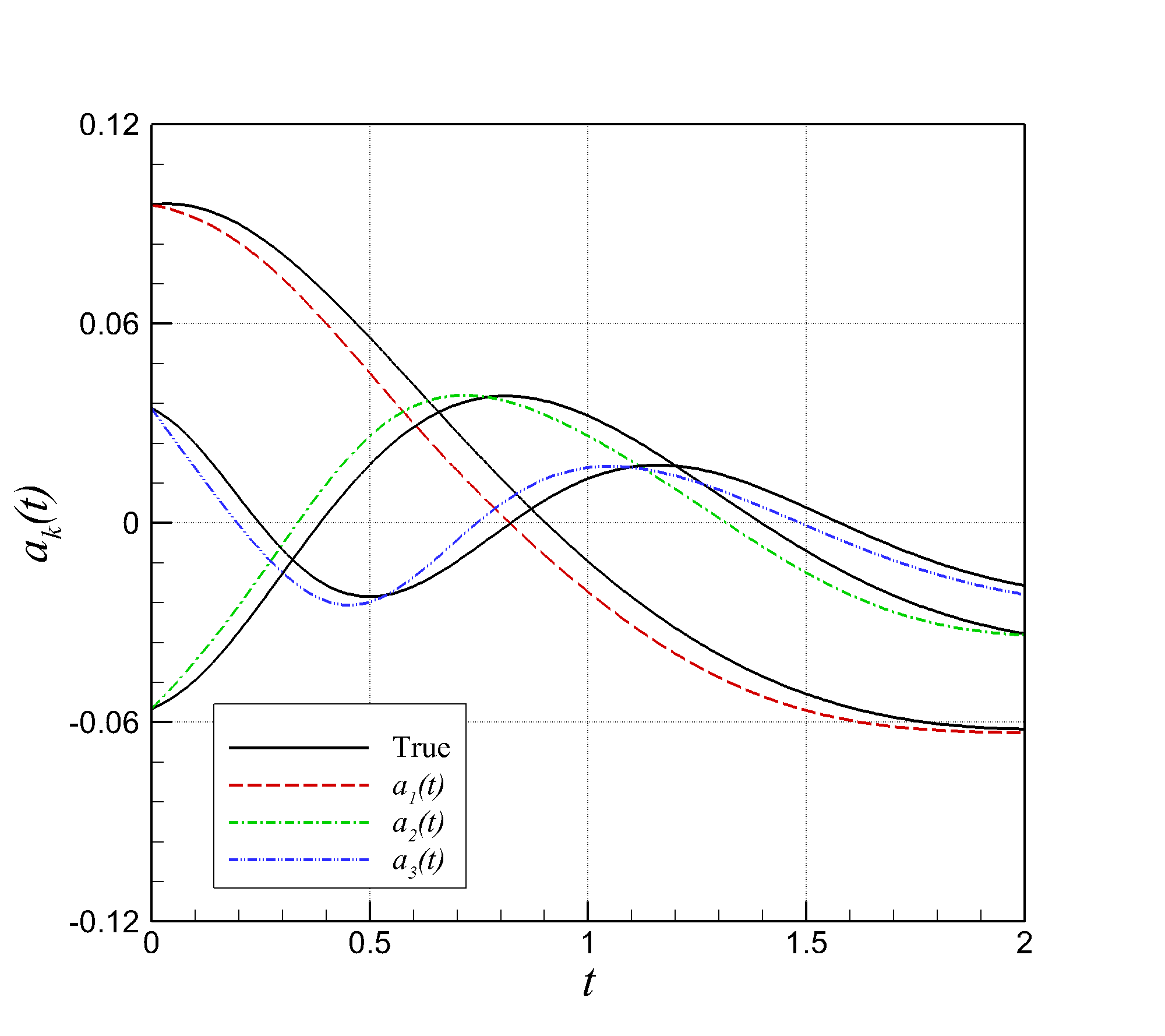}}
\subfigure[POD-ANN-ROM($Re=250$)]{\includegraphics[width=0.5\textwidth]{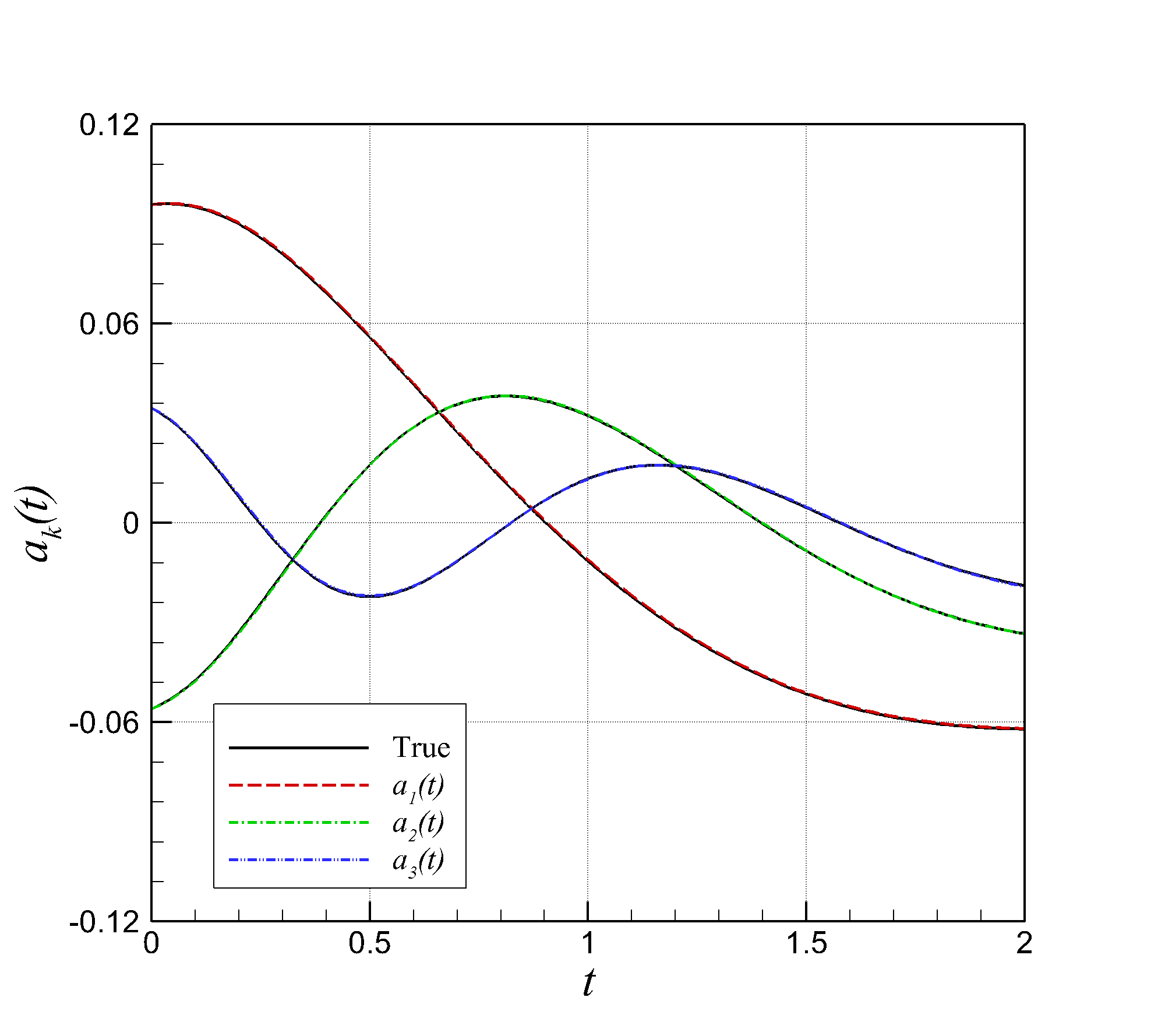}}
}\\
\mbox{
\subfigure[POD-GP-ROM ($Re=1000$)]{\includegraphics[width=0.5\textwidth]{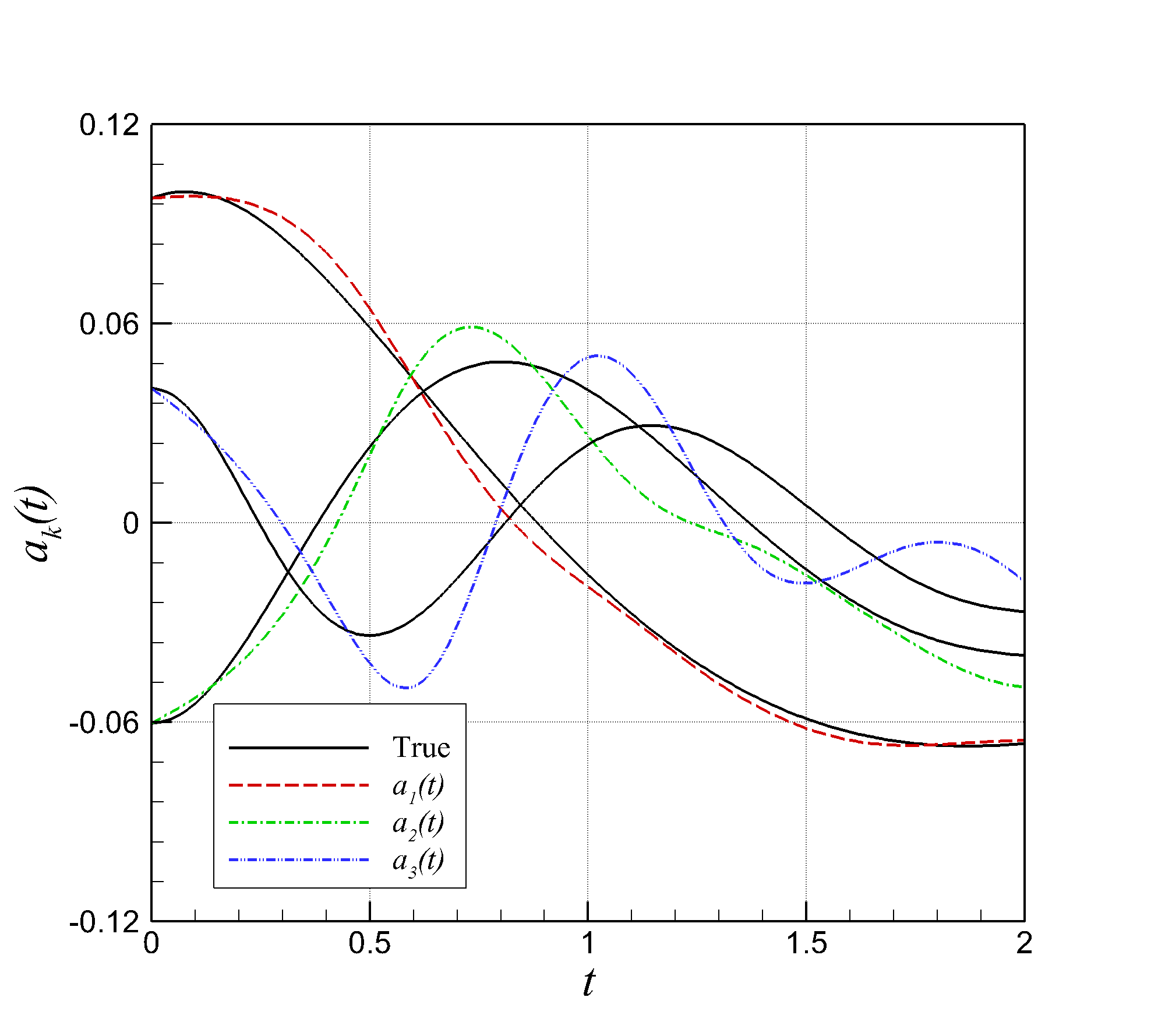}}
\subfigure[Fourier-ANN-ROM($Re=1000$)]{\includegraphics[width=0.5\textwidth]{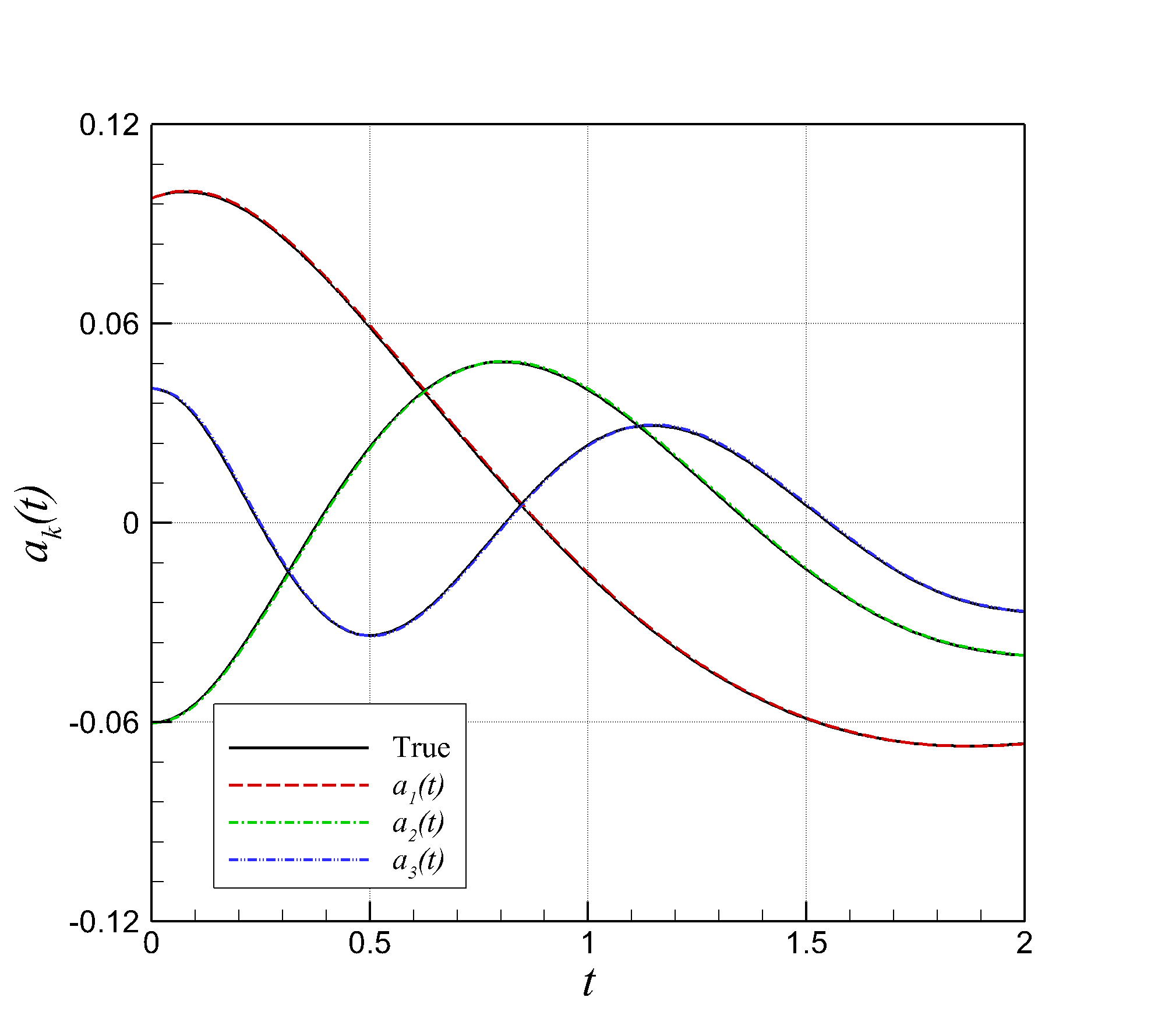}}
}\\
\mbox{
\subfigure[POD-GP-ROM ($Re=1500$)]{\includegraphics[width=0.5\textwidth]{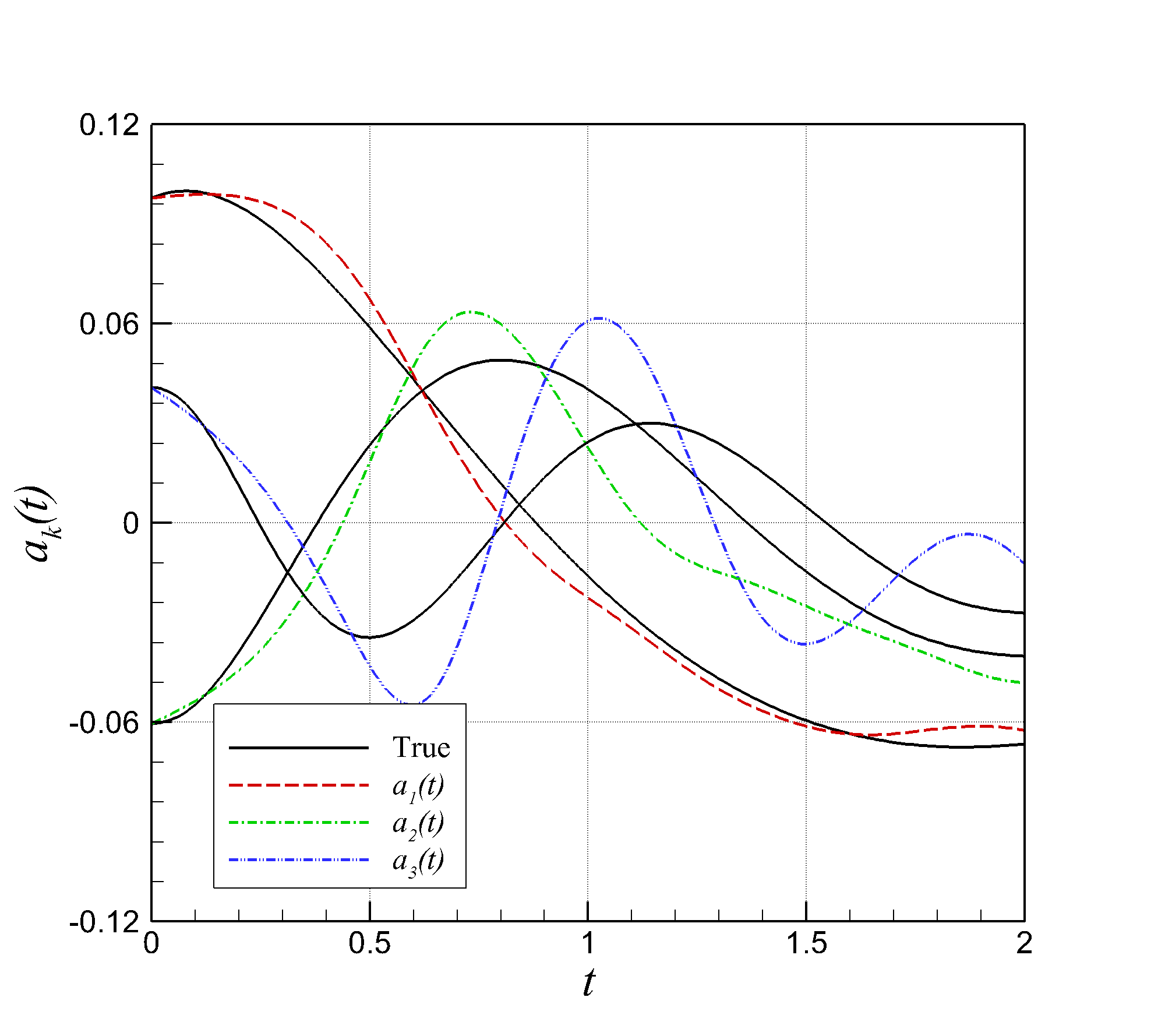}}
\subfigure[POD-ANN-ROM($Re=1500$)]{\includegraphics[width=0.5\textwidth]{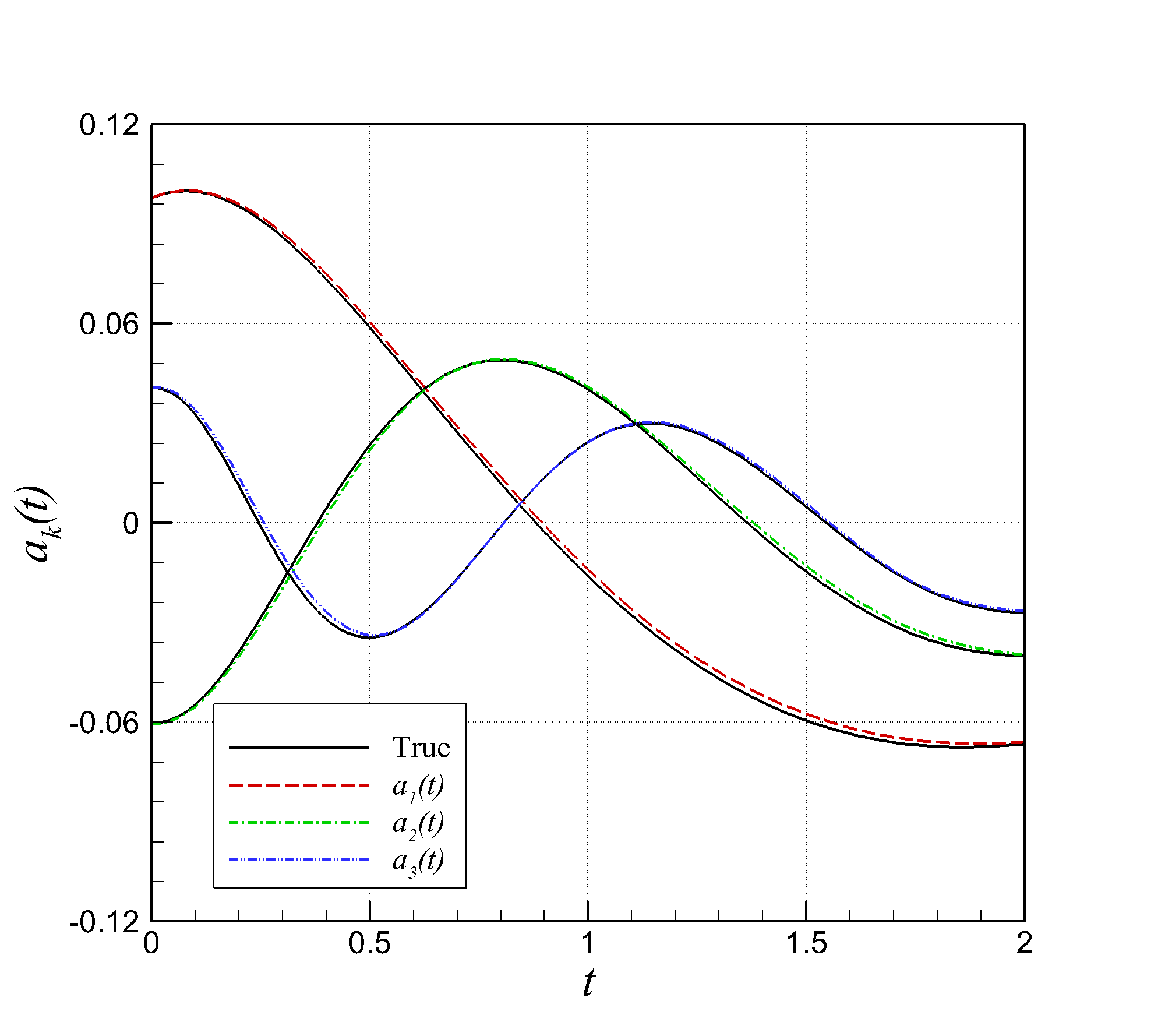}}
}
\caption{Time evolution of first few modes when using the POD basis. ANN with Bayesian regularization with 10 neurons. Note that $Re=250$ and $Re=1500$ cases are not included in our training set of $Re \in [200-1200]$.}
\label{fig:8}
\end{figure}

The real challenge of our proposed closure lies in its performance for values of physical parameters which do not lie in the training data set. The ability to predict modal evolution for values of the Reynolds number which are not a part of the training data set would be favorable for optimal control applications where small differences in physical parameters wouldn't require a complete high fidelity numerical simulation prior to ROM deployment. Fig.~(\ref{fig:7}) describe the time series evolution of the first three modes for the ROM implementations with and without the proposed closures. The bases used for the Galerkin projection here are the Fourier bases. The ANN closure clearly leads to a much closer alignment with the true evolution of the modes. What is notable here is that for $Re=250$ and $Re=1500$, the ANN closure performs excellently as well in comparison to the Fourier-GP-ROM which can be seen to deviate from the trends of the true evolution. A similar conclusion can also be drawn from Fig.~(\ref{fig:8}), where we have now used the POD bases for our transformation space. It can also be observed that the POD modes show a larger deviation from the true evolution without the use of the closure.

\begin{figure}[!ht]
\centering
\mbox{
\subfigure[BR ($Q=10$)]{\includegraphics[width=0.5\textwidth]{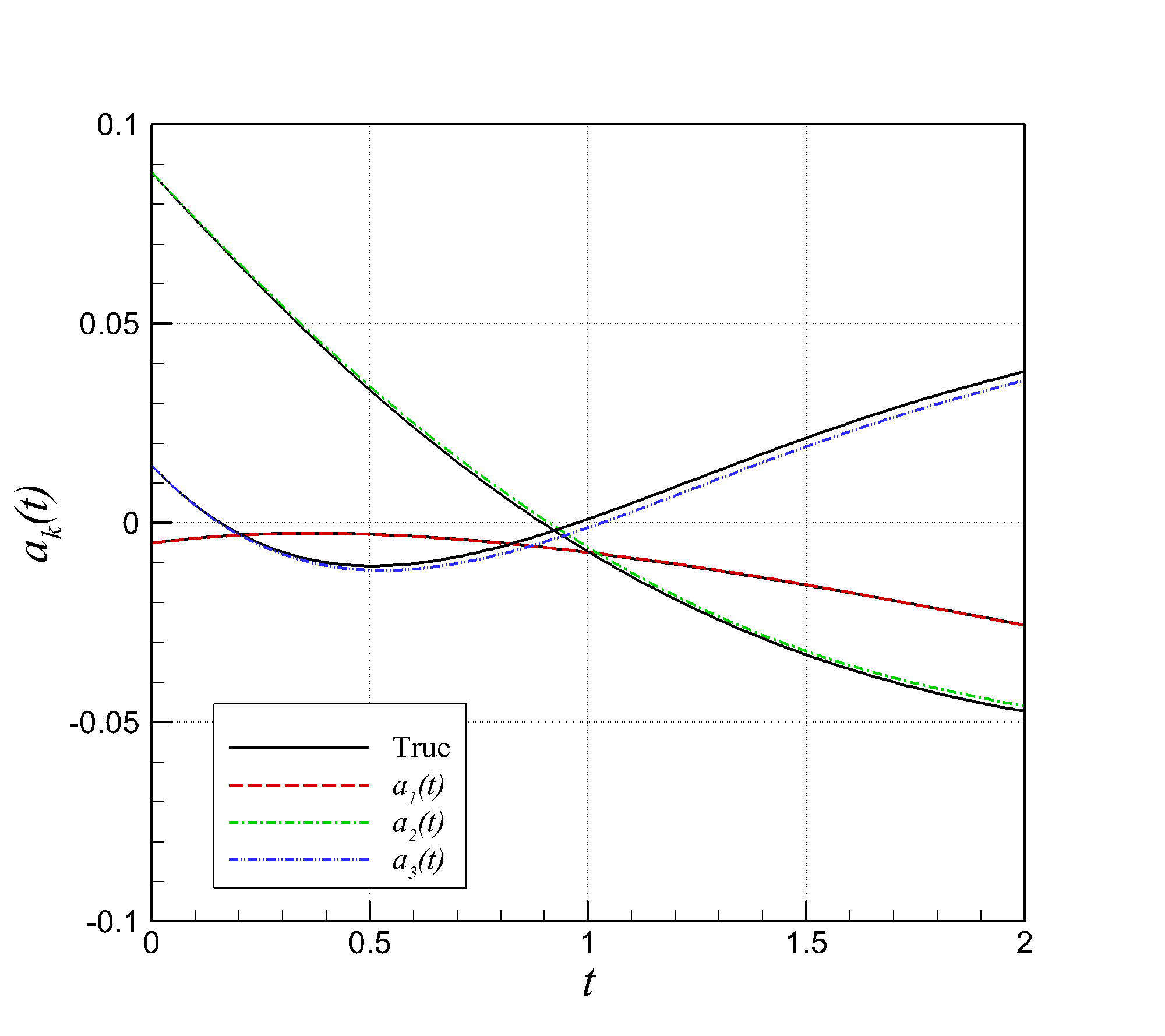}}
\subfigure[ELM ($Q=10$)]{\includegraphics[width=0.5\textwidth]{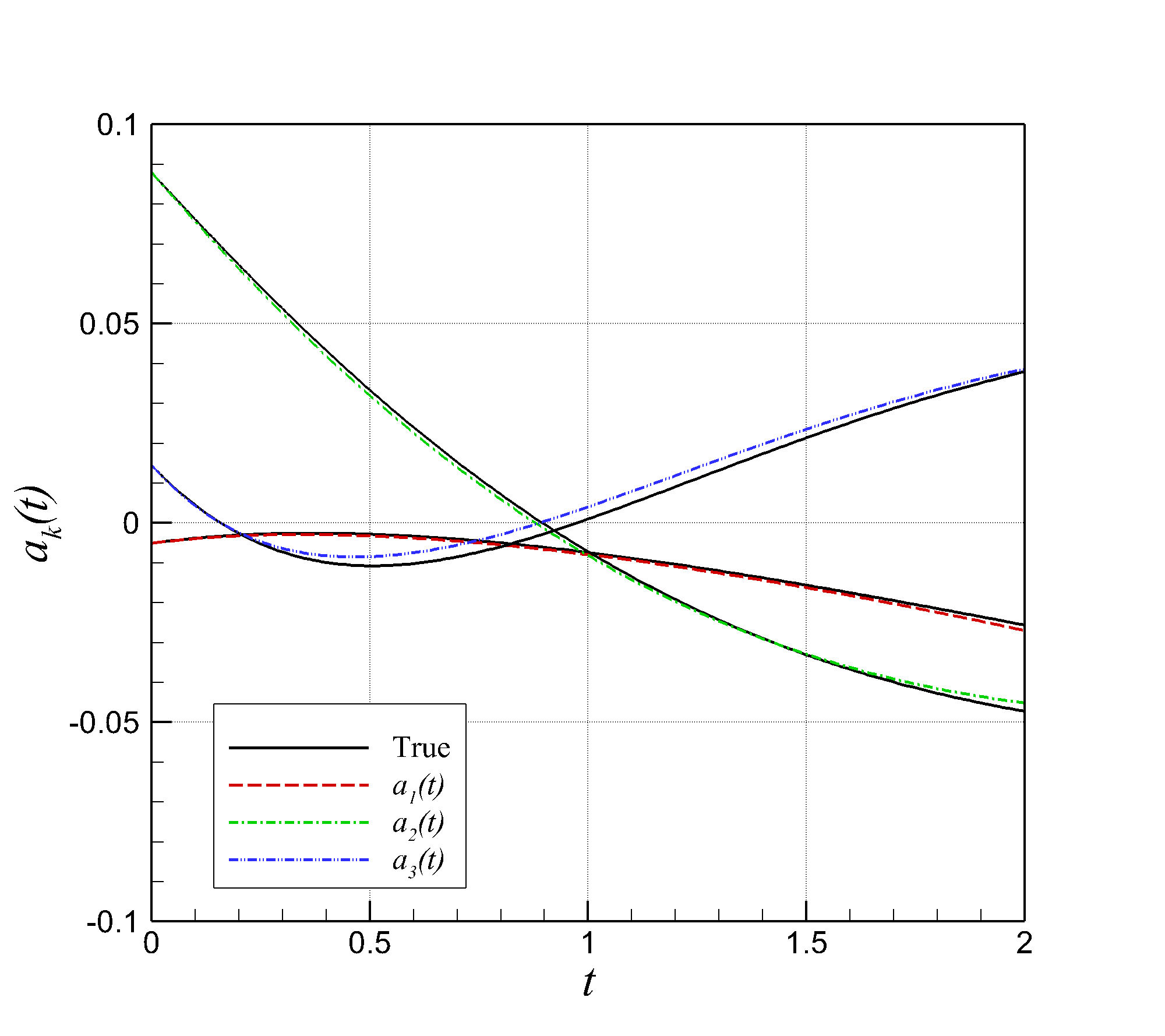}}
}\\
\mbox{
\subfigure[ELM ($Q=20$)]{\includegraphics[width=0.5\textwidth]{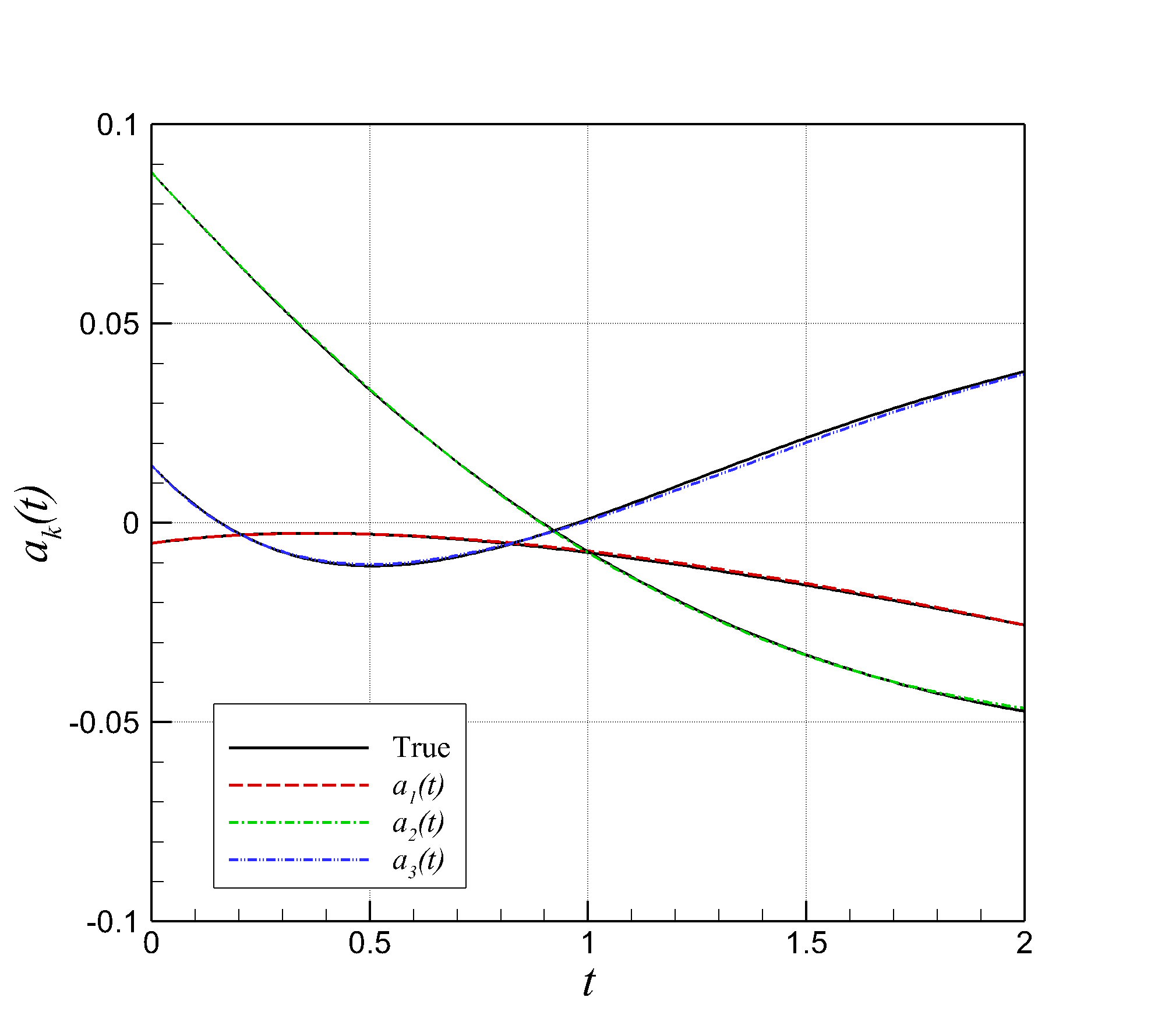}}
\subfigure[ELM ($Q=40$)]{\includegraphics[width=0.5\textwidth]{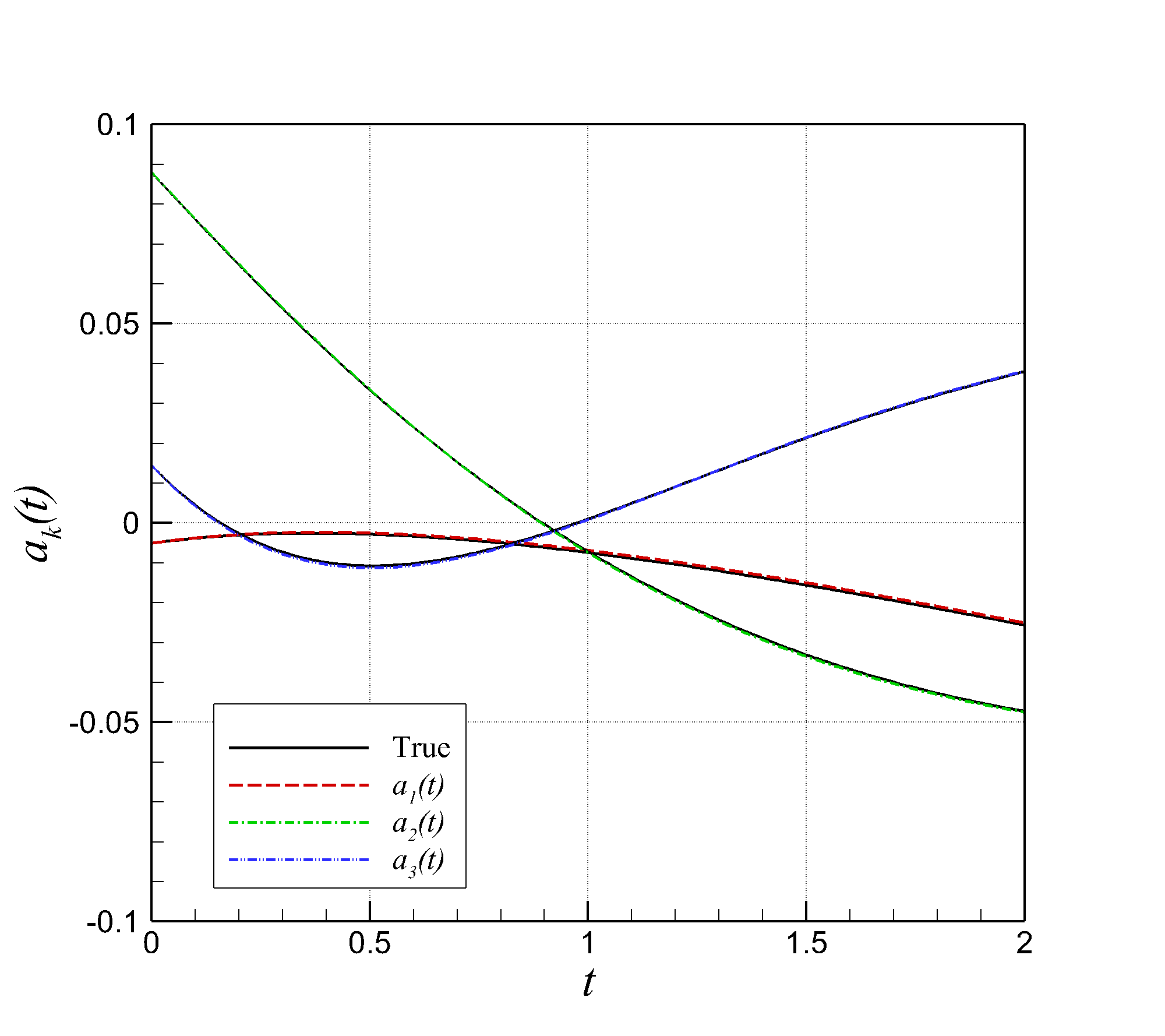}}
}
\caption{Time evolution of first few modes when using the Fourier basis at $Re=100$. Note that this Reynolds number is not included in our training set of $Re \in [200-1200]$. }
\label{fig:10}
\end{figure}

\begin{figure}[!ht]
\centering
\mbox{
\subfigure[BR ($Q=10$)]{\includegraphics[width=0.5\textwidth]{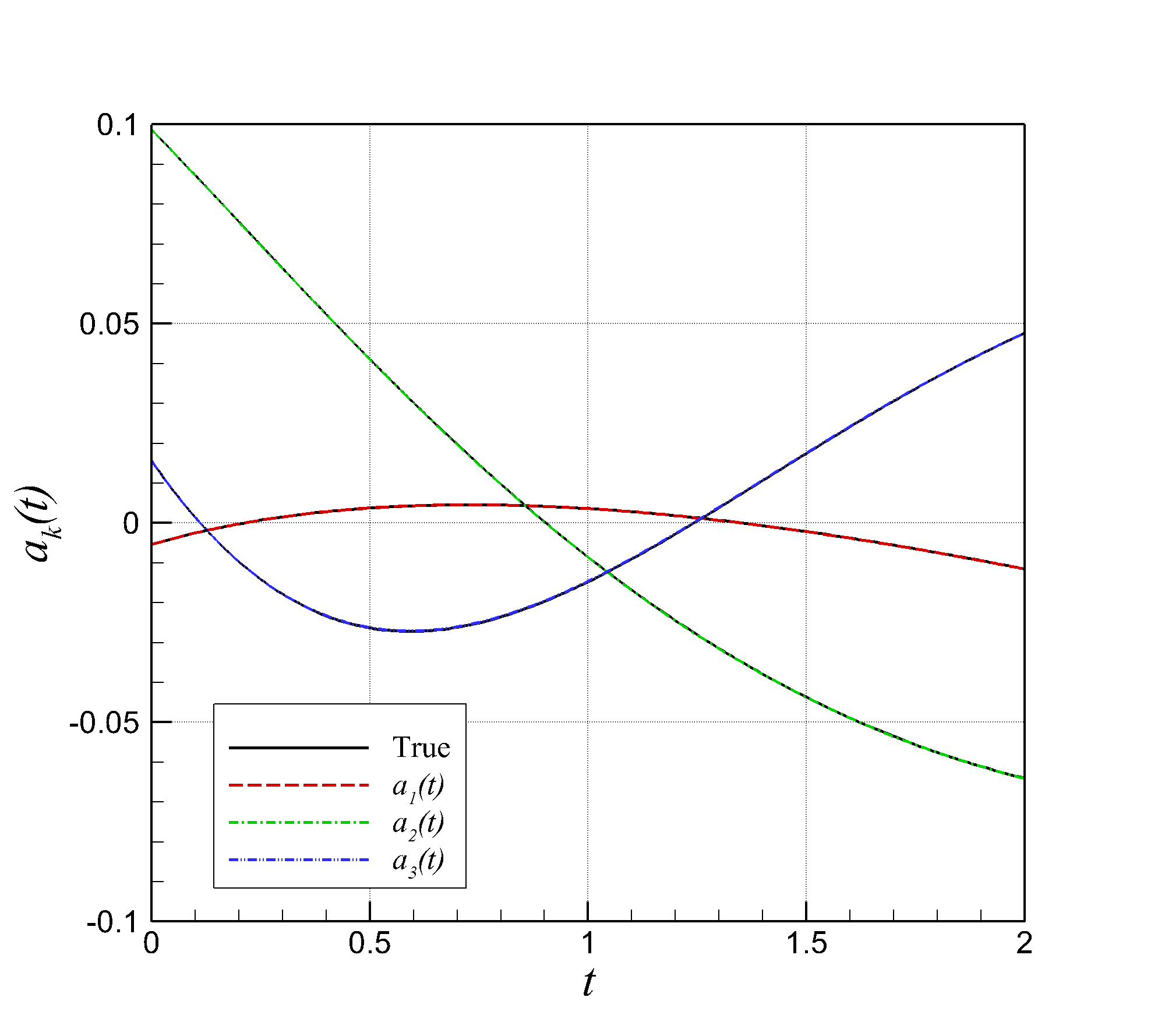}}
\subfigure[ELM ($Q=10$)]{\includegraphics[width=0.5\textwidth]{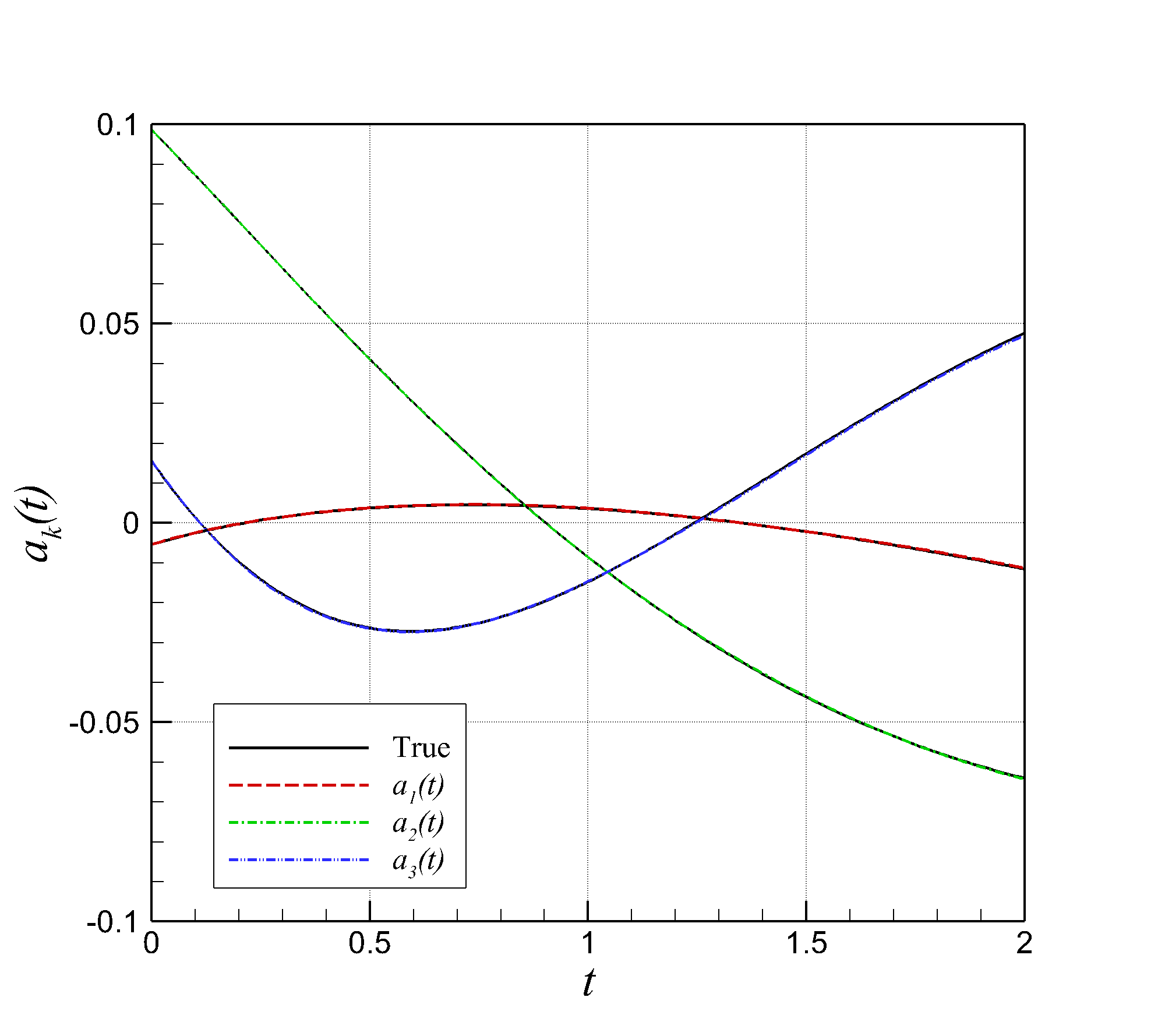}}
}\\
\mbox{
\subfigure[ELM ($Q=20$)]{\includegraphics[width=0.5\textwidth]{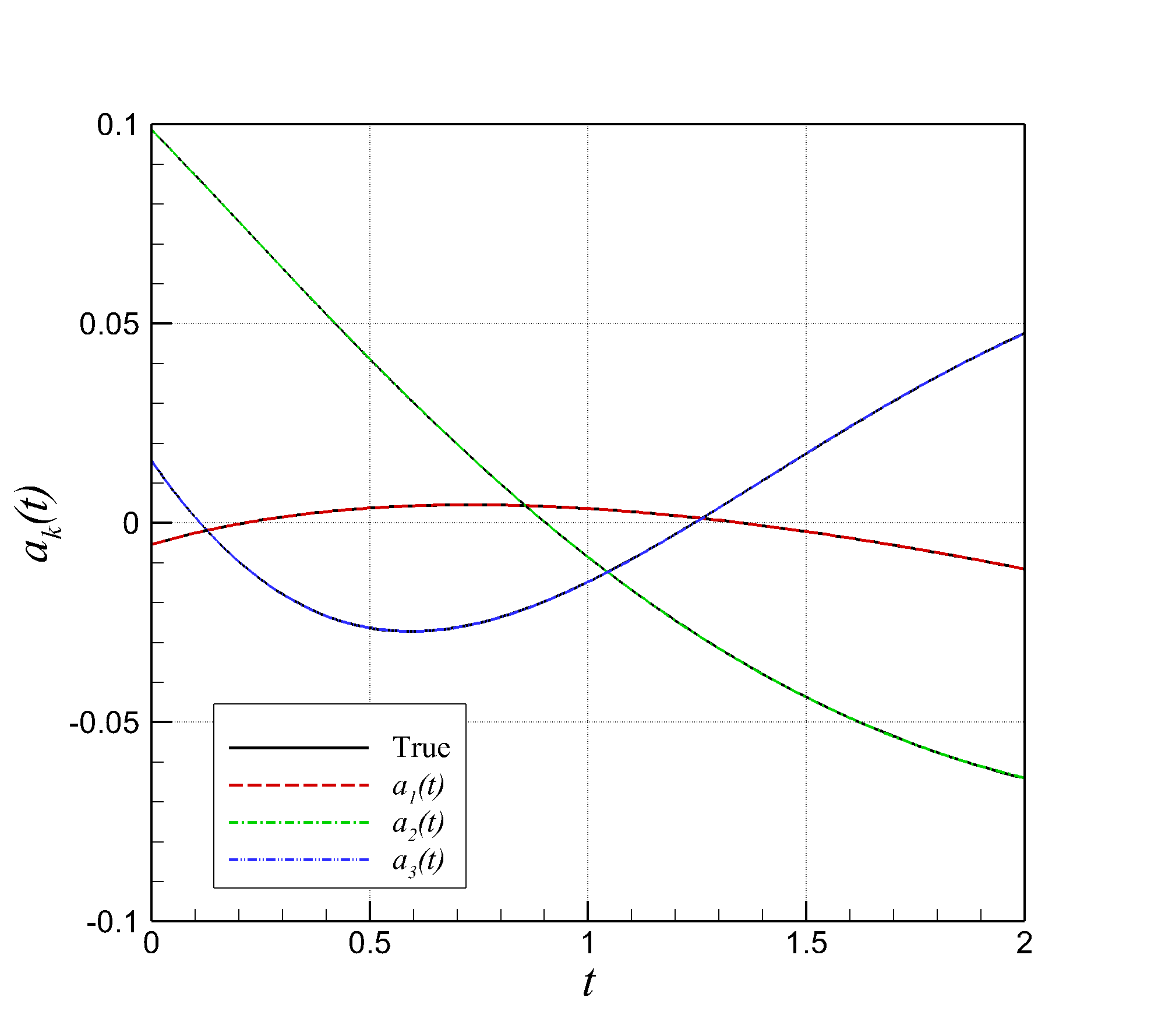}}
\subfigure[ELM ($Q=40$)]{\includegraphics[width=0.5\textwidth]{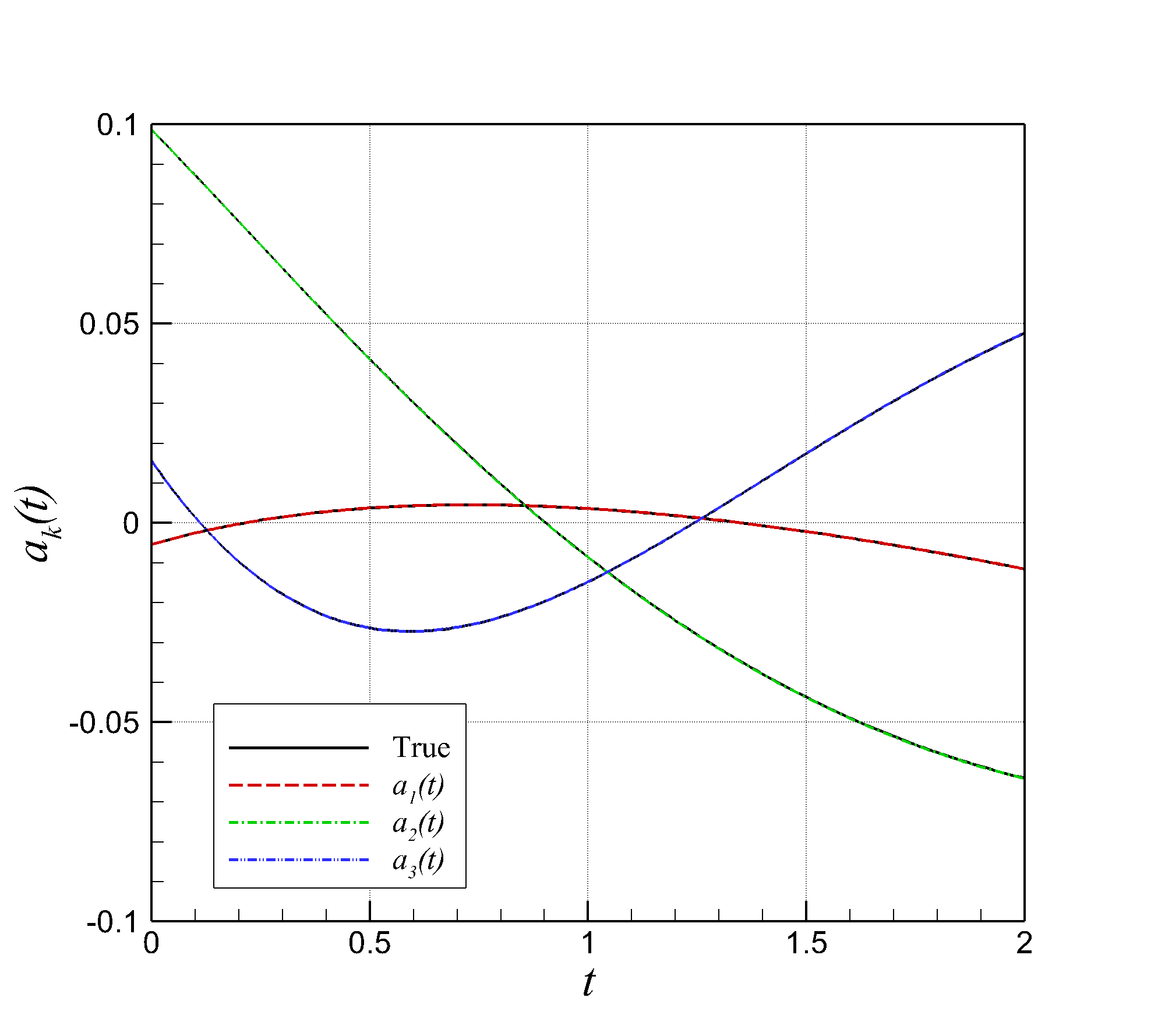}}
}
\caption{Time evolution of first few modes when using the Fourier basis at $Re=750$. Note that this Reynolds number is not included in our training set of $Re \in [200-1200]$. }
\label{fig:11}
\end{figure}

\begin{figure}[!ht]
\centering
\mbox{
\subfigure[BR ($Q=10$)]{\includegraphics[width=0.5\textwidth]{figures/a_ANN_BR10_Fr_re1500.png}}
\subfigure[ELM ($Q=10$)]{\includegraphics[width=0.5\textwidth]{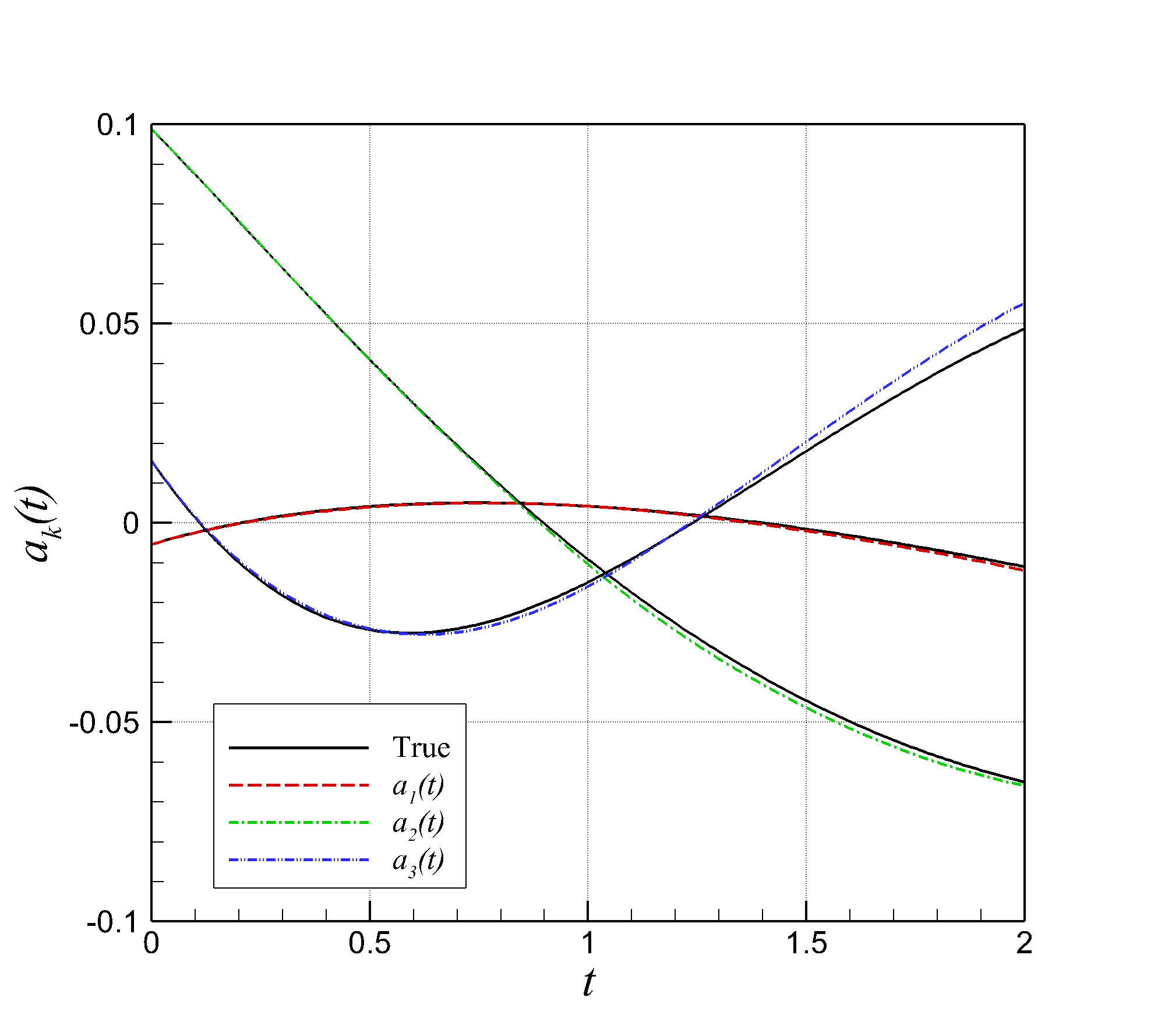}}
}\\
\mbox{
\subfigure[ELM ($Q=20$)]{\includegraphics[width=0.5\textwidth]{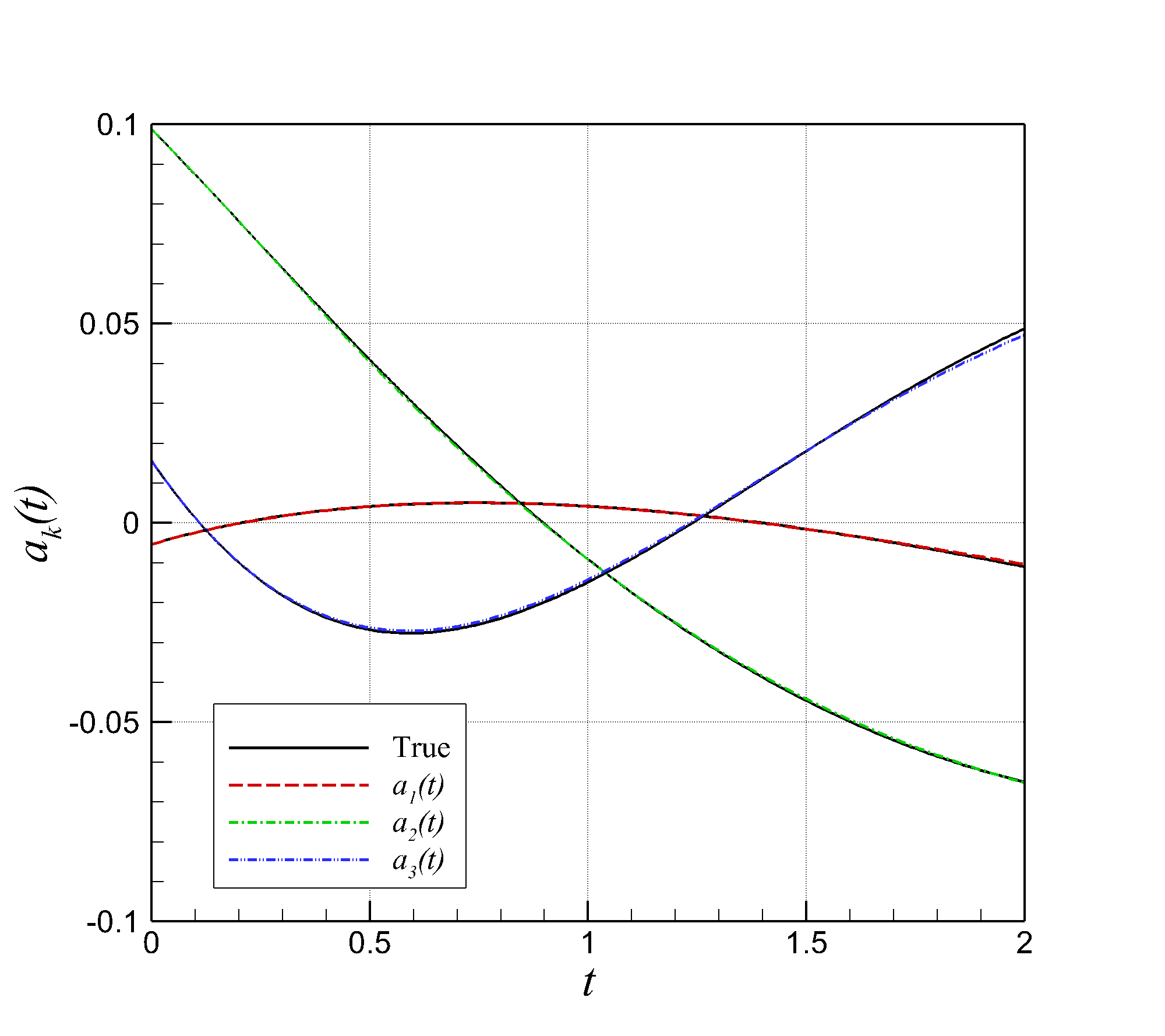}}
\subfigure[ELM ($Q=40$)]{\includegraphics[width=0.5\textwidth]{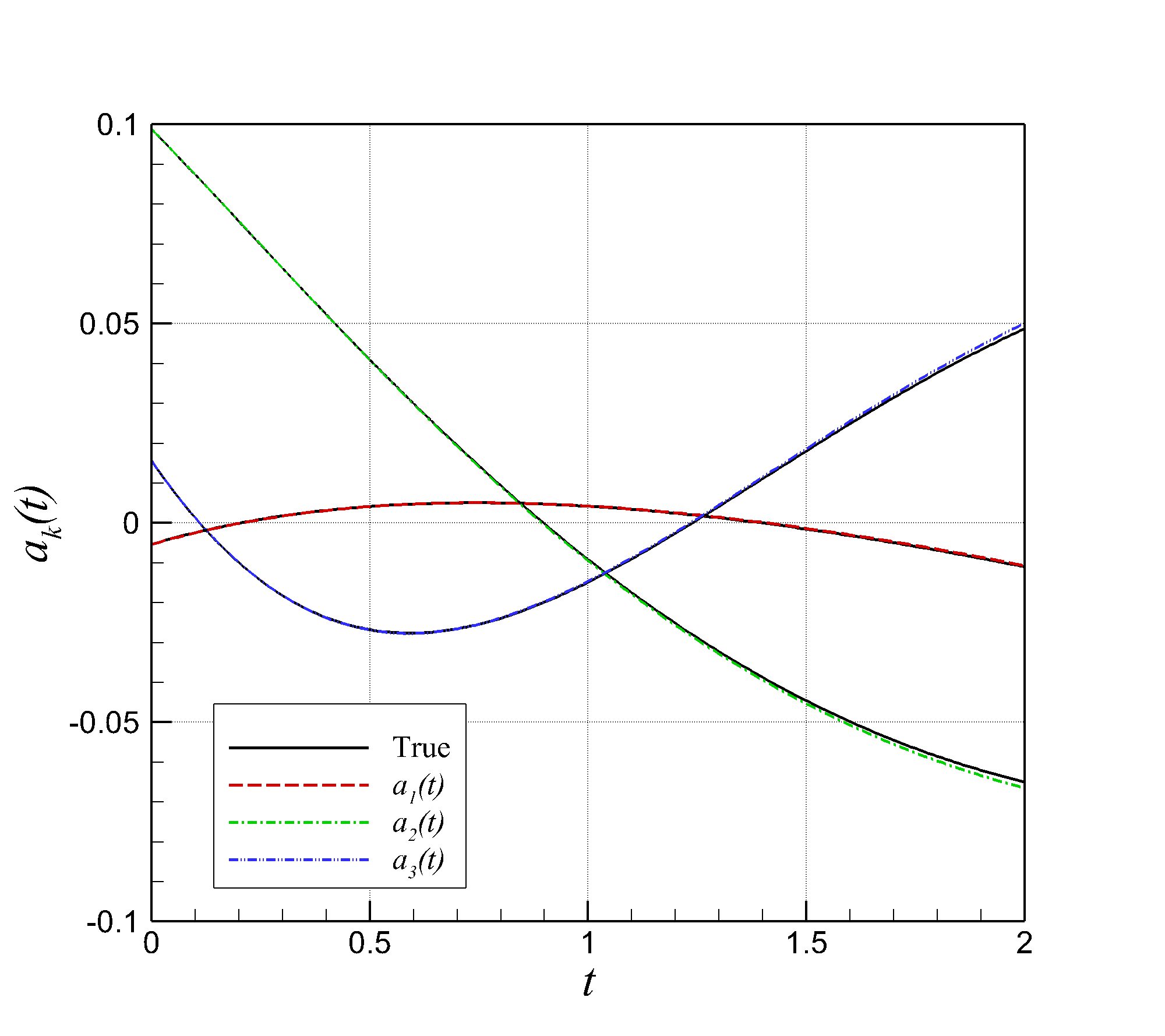}}
}
\caption{Time evolution of first few modes when using the Fourier basis at $Re=1500$. Note that this Reynolds number is not included in our training set of $Re \in [200-1200]$. }
\label{fig:12}
\end{figure}

Finally, we also detail the effect of the training algorithm (i.e., whether BR or ELM) for determining the ANN weights and consequent performance. We have utilized three values of $Re = 100,750,1500$. We remark that none of the values are a part of the training data set. Also, $Re=100$ and $Re=1500$ are outside the \emph{range} of the training data. In particular, we are interested in the performance of ELM since it represents a massive opportunity for fast training requirements in dynamical systems. For the case of $Re=100$, given by Fig.~(\ref{fig:10}), we can see the ELM approach progressively converges to the true evolution with increasing number of neurons $Q$. For $Re=750$, which lies within the range of our training data set, we can see that the ELM and BR approaches are more or less equivalent with excellent agreement with the true evolution (as shown in Fig.~(\ref{fig:11})). Fig.~(\ref{fig:12}) shows the challenging case of $Re=1500$, where a slight deviation from the true solution can be seen for ELM at lower number of hidden neurons. Note that this behavior is expected since the ELM procedure effectively reduces the degrees of freedom of the single layer feedforward ANN which might prove to be a handicap for a lower number of hidden layer neurons. However, the fast training time effectively offsets the slightly higher necessity for neurons.

\section{Conclusions}
\label{sec:cons}

This work demonstrates the value of utilizing a single layer feedforward artificial neural network as a closure for reduced order models of the viscous Burgers equation. A reduced order model for the Burgers equation is created by transforming our partial differential equation using a new set of orthonormal bases (given by either the optimal POD modes or the Fourier modes). While the Fourier modes are predefined, the POD space is constructed by using snapshot data generated from the exact solution of the system we are studying. The Galerkin projection methodology is then used to evolve a severely truncated system with and without our proposed ANN closure. Our proposed closure consists of an ANN architecture with one hidden layer of neurons which is trained using `true' modal evolution values generated from the exact solution of the governing law. The nonlinear relationship approximates the true value of the right hand side of the Galerkin projection system of ordinary differential equations, given the truncated modes at the current time step. It thus acts as a stabilization procedure.

Our numerical assessments reveal that our closure hypothesis proves efficient in the closure of the ROM for both Fourier and POD bases. The ANN closure is also seen to perform robustly for the modal evolution of systems with Reynolds numbers which are different from the training data set. This highlights its potential utility as a closure for ROM-GP based optimal control for fluid systems. In addition, two different training approaches are utilized prior to the deployment of the ANN for the purpose of closure. The BR and ELM training methodologies are selected for this study due to their ability to give \emph{regularized} networks (i.e., networks which are less sensitive to noisy data). While the BR method is exceptional for its accuracy and has become a mainstay for ANN training for noisy data sets, the ELM approach is extremely popular due its high speed in training. It is observed that both approaches are capable of stabilizing modal evolution satisfactorily. In addition, an increasing number of neurons is seen to enhance the accuracy of the ELM training. This is because it has a slightly lower degree of freedom as compared to the BR method. The ELM approach is particularly promising due its application in dynamical systems which may require fast learning. A simple one dimensional training data is also examined with artificial noise using both BR and ELM approaches and the computational benefit of ELM is clearly established.

This investigation leads us to several meaningful insights about the future of hybrid data and physics driven ROM. For the case of the one dimensional Burgers equation, our proposed closure may be considered to be an adequate augmentation to a POD-ROM. However, further investigations are necessary to test the performance of the closure (for example for noisy data). The use of BR and ELM (i.e., regularized training) is motivated by the desire to extract the underlying statistics of the flow (which are ultimately physics driven) and they must be ideally tested against experimental data sets with their inherent background noise. In addition, we also plan to investigate the performance of this closure for higher dimensional systems and ultimately for real world flow control problems.

\bibliographystyle{spmpsci}      
\bibliography{references}   

\begin{thebibliography}{10}
\providecommand{\url}[1]{{#1}}
\providecommand{\urlprefix}{URL }
\expandafter\ifx\csname urlstyle\endcsname\relax
  \providecommand{\doi}[1]{DOI~\discretionary{}{}{}#1}\else
  \providecommand{\doi}{DOI~\discretionary{}{}{}\begingroup
  \urlstyle{rm}\Url}\fi

\bibitem{akhtar2015using}
Akhtar, I., Borggaard, J., Burns, J.A., Imtiaz, H., Zietsman, L.: Using
  functional gains for effective sensor location in flow control: a
  reduced-order modelling approach.
\newblock Journal of Fluid Mechanics \textbf{781}, 622--656 (2015)

\bibitem{akhtar2012new}
Akhtar, I., Wang, Z., Borggaard, J., Iliescu, T.: A new closure strategy for
  proper orthogonal decomposition reduced-order models.
\newblock Journal of Computational and Nonlinear Dynamics \textbf{7}(3),
  034,503 (2012)

\bibitem{amsallem2008interpolation}
Amsallem, D., Farhat, C.: Interpolation method for adapting reduced-order
  models and application to aeroelasticity.
\newblock AIAA Journal \textbf{46}(7), 1803--1813 (2008)

\bibitem{amsallem2012stabilization}
Amsallem, D., Farhat, C.: Stabilization of projection-based reduced-order
  models.
\newblock International Journal for Numerical Methods in Engineering
  \textbf{91}(4), 358--377 (2012)

\bibitem{antoulas2005approximation}
Antoulas, A.C.: Approximation of large-scale dynamical systems.
\newblock SIAM (2005)

\bibitem{aubry1988dynamics}
Aubry, N., Holmes, P., Lumley, J.L., Stone, E.: The dynamics of coherent
  structures in the wall region of a turbulent boundary layer.
\newblock Journal of Fluid Mechanics \textbf{192}(1), 115--173 (1988)

\bibitem{balajewicz2012stabilization}
Balajewicz, M., Dowell, E.H.: Stabilization of projection-based reduced order
  models of the {N}avier--{S}tokes.
\newblock Nonlinear Dynamics \textbf{70}(2), 1619--1632 (2012)

\bibitem{banyay2014proper}
Banyay, G.A., Ahmadpoor, M., Brigham, J.C.: Proper orthogonal decomposition
  based reduced order modeling of the very high temperature reactor lower
  plenum hydrodynamics.
\newblock In: ASME 2014 4th Joint US-European Fluids Engineering Division
  Summer Meeting collocated with the ASME 2014 12th International Conference on
  Nanochannels, Microchannels, and Minichannels, pp. V01DT27A013--V01DT27A013.
  American Society of Mechanical Engineers (2014)

\bibitem{benner2015survey}
Benner, P., Gugercin, S., Willcox, K.: A survey of projection-based model
  reduction methods for parametric dynamical systems.
\newblock SIAM Review \textbf{57}, 483--531 (2015)

\bibitem{benosman2017learning}
Benosman, M., Borggaard, J., San, O., Kramer, B.: Learning-based robust
  stabilization for reduced-order models of 2{D} and 3{D} {B}oussinesq
  equations.
\newblock Applied Mathematical Modelling \textbf{49}, 162--181 (2017)

\bibitem{benosman2016learning}
Benosman, M., Kramer, B., Boufounos, P.T., Grover, P.: Learning-based reduced
  order model stabilization for partial differential equations: Application to
  the coupled {B}urgers' equation.
\newblock In: American Control Conference (ACC), 2016, pp. 1673--1678. IEEE
  (2016)

\bibitem{bergmann2009improvement}
Bergmann, M., Bruneau, C.H., Iollo, A.: Improvement of reduced order modeling
  based on \uppercase{POD}.
\newblock Computational Fluid Dynamics 2008 pp. 779--784 (2009)

\bibitem{borggaard2007interval}
Borggaard, J., Hay, A., Pelletier, D.: Interval-based reduced order models for
  unsteady fluid flow.
\newblock Int. J. Numer. Anal. Model \textbf{4}(3-4), 353--367 (2007)

\bibitem{borggaard2011artificial}
Borggaard, J., Iliescu, T., Wang, Z.: Artificial viscosity proper orthogonal
  decomposition.
\newblock Mathematical and Computer Modelling \textbf{53}(1), 269--279 (2011)

\bibitem{borggaard2016goal}
Borggaard, J., Wang, Z., Zietsman, L.: A goal-oriented reduced-order modeling
  approach for nonlinear systems.
\newblock Computers \& Mathematics with Applications \textbf{71}(11),
  2155--2169 (2016)

\bibitem{buffoni2006low}
Buffoni, M., Camarri, S., Iollo, A., Salvetti, M.V.: Low-dimensional modelling
  of a confined three-dimensional wake flow.
\newblock Journal of Fluid Mechanics \textbf{569}, 141--150 (2006)

\bibitem{bui2007goal}
Bui-Thanh, T., Willcox, K., Ghattas, O., van Bloemen~Waanders, B.:
  Goal-oriented, model-constrained optimization for reduction of large-scale
  systems.
\newblock Journal of Computational Physics \textbf{224}(2), 880--896 (2007)

\bibitem{carlberg2011low}
Carlberg, K., Farhat, C.: A low-cost, goal-oriented `compact proper orthogonal
  decomposition' basis for model reduction of static systems.
\newblock International Journal for Numerical Methods in Engineering
  \textbf{86}(3), 381--402 (2011)

\bibitem{cazemier1997proper}
Cazemier, W.: Proper orthogonal decomposition and low dimensional models for
  turbulent flows.
\newblock Groningen (1997)

\bibitem{cazemier1998proper}
Cazemier, W., Verstappen, R., Veldman, A.: Proper orthogonal decomposition and
  low-dimensional models for driven cavity flows.
\newblock Physics of Fluids (1994-present) \textbf{10}(7), 1685--1699 (1998)

\bibitem{cordier2010calibration}
Cordier, L., Majd, E., Abou, B., Favier, J.: Calibration of {POD} reduced-order
  models using {T}ikhonov regularization.
\newblock International Journal for Numerical Methods in Fluids \textbf{63}(2),
  269--296 (2010)

\bibitem{dawson1998artificial}
Dawson, C.W., Wilby, R.: An artificial neural network approach to
  rainfall-runoff modelling.
\newblock Hydrological Sciences Journal \textbf{43}(1), 47--66 (1998)

\bibitem{demuth2014neural}
Demuth, H.B., Beale, M.H., De~Jess, O., Hagan, M.T.: Neural network design.
\newblock Martin Hagan (2014)

\bibitem{dyke1996modeling}
Dyke, S., Spencer~Jr, B., Sain, M., Carlson, J.: Modeling and control of
  magnetorheological dampers for seismic response reduction.
\newblock Smart Materials and Structures \textbf{5}(5), 565 (1996)

\bibitem{efe2005control}
Efe, M., Debiasi, M., Yan, P., Ozbay, H., Samimy, M.: Control of subsonic
  cavity flows by neural networks-analytical models and experimental
  validation.
\newblock In: 43rd AIAA Aerospace Sciences Meeting and Exhibit, p. 294 (2005)

\bibitem{efe2004modeling}
Efe, M.O., Debiasi, M., Ozbay, H., Samimy, M.: Modeling of subsonic cavity
  flows by neural networks.
\newblock In: Mechatronics, 2004. ICM'04. Proceedings of the IEEE International
  Conference on, pp. 560--565. IEEE (2004)

\bibitem{el2016new}
El~Majd, B.A., Cordier, L.: New regularization method for calibrated {POD}
  reduced-order models.
\newblock Mathematical Modelling and Analysis \textbf{21}(1), 47--62 (2016)

\bibitem{faller1997unsteady}
Faller, W.E., Schreck, S.J.: Unsteady fluid mechanics applications of neural
  networks.
\newblock Journal of Aircraft \textbf{34}(1), 48--55 (1997)

\bibitem{fang2009pod}
Fang, F., Pain, C., Navon, I., Gorman, G., Piggott, M., Allison, P., Farrell,
  P., Goddard, A.: A {POD} reduced order unstructured mesh ocean modelling
  method for moderate {R}eynolds number flows.
\newblock Ocean Modelling \textbf{28}(1), 127--136 (2009)

\bibitem{foresee1997gauss}
Foresee, F.D., Hagan, M.T.: {G}auss-{N}ewton approximation to {B}ayesian
  learning.
\newblock In: Neural Networks, 1997., International Conference on, vol.~3, pp.
  1930--1935. IEEE (1997)

\bibitem{fortuna2012model}
Fortuna, L., Nunnari, G., Gallo, A.: Model order reduction techniques with
  applications in electrical engineering.
\newblock Springer Science \& Business Media (2012)

\bibitem{freund1999reduced}
Freund, R.W.: Reduced-order modeling techniques based on {K}rylov subspaces and
  their use in circuit simulation.
\newblock In: Applied and computational control, signals, and circuits, pp.
  435--498. Springer (1999)

\bibitem{gaspard2005chaos}
Gaspard, P.: Chaos, scattering and statistical mechanics, vol.~9.
\newblock Cambridge University Press (2005)

\bibitem{gillies1995low}
Gillies, E.: Low-dimensional characterization and control of non-linear wake
  flows.
\newblock Ph.D. thesis, PhD. Dissertation, University of Glasgow, Scotland
  (1995)

\bibitem{gillies1998low}
Gillies, E.: Low-dimensional control of the circular cylinder wake.
\newblock Journal of Fluid Mechanics \textbf{371}, 157--178 (1998)

\bibitem{gillies2001multiple}
Gillies, E.: Multiple sensor control of vortex shedding.
\newblock AIAA journal \textbf{39}(4), 748--750 (2001)

\bibitem{hagan1994training}
Hagan, M.T., Menhaj, M.B.: Training feedforward networks with the {M}arquardt
  algorithm.
\newblock IEEE Transactions on Neural Networks \textbf{5}(6), 989--993 (1994)

\bibitem{haykin2009neural}
Haykin, S.S., Haykin, S.S., Haykin, S.S., Haykin, S.S.: Neural networks and
  learning machines, vol.~3.
\newblock Pearson Upper Saddle River, NJ, USA: (2009)

\bibitem{hocevar2004experimental}
Hocevar, M., Širok, B., Grabec, I.: Experimental turbulent field modeling by
  visualization and neural networks.
\newblock Journal of Fluids Engineering \textbf{126}, 316--322 (2004)

\bibitem{holmes1998turbulence}
Holmes, P., Lumley, J.L., Berkooz, G.: Turbulence, coherent structures,
  dynamical systems and symmetry.
\newblock Cambridge University Press (1998)

\bibitem{hotelling1933analysis}
Hotelling, H.: Analysis of a complex of statistical variables into principal
  components.
\newblock Journal of Educational Psychology \textbf{24}(6), 417 (1933)

\bibitem{huang2006extreme}
Huang, G.B., Zhu, Q.Y., Siew, C.K.: Extreme learning machine: theory and
  applications.
\newblock Neurocomputing \textbf{70}(1), 489--501 (2006)

\bibitem{imtiaz2016closure}
Imtiaz, H., Akhtar, I.: Closure modeling in reduced-order model of {B}urgers'
  equation for control applications.
\newblock Journal of Aerospace Engineering pp. 1--15 (2016)

\bibitem{iollo2000stability}
Iollo, A., Lanteri, S., D{\'e}sid{\'e}ri, J.A.: Stability properties of
  {POD}--{G}alerkin approximations for the compressible {N}avier--{S}tokes
  equations.
\newblock Theoretical and Computational Fluid Dynamics \textbf{13}(6), 377--396
  (2000)

\bibitem{kazantzis2010new}
Kazantzis, N., Kravaris, C., Syrou, L.: A new model reduction method for
  nonlinear dynamical systems.
\newblock Nonlinear Dynamics \textbf{59}(1), 183--194 (2010)

\bibitem{khibnik2000analysis}
Khibnik, A., Narayanan, S., Jacobson, C., Lust, K.: Analysis of low dimensional
  dynamics of flow separation.
\newblock In: Continuation Methods in Fluid Dynamics, vol.~74, pp. 167--178.
  Vieweg (2000)

\bibitem{kim2003nonlinear}
Kim, T.W., Vald{\'e}s, J.B.: Nonlinear model for drought forecasting based on a
  conjunction of wavelet transforms and neural networks.
\newblock Journal of Hydrologic Engineering \textbf{8}(6), 319--328 (2003)

\bibitem{kunisch1999control}
Kunisch, K., Volkwein, S.: Control of the {B}urgers equation by a reduced-order
  approach using proper orthogonal decomposition.
\newblock Journal of optimization theory and applications \textbf{102}(2),
  345--371 (1999)

\bibitem{kunisch2001galerkin}
Kunisch, K., Volkwein, S.: Galerkin proper orthogonal decomposition methods for
  parabolic problems.
\newblock Numerische Mathematik \textbf{90}(1), 117--148 (2001)

\bibitem{kunisch2010optimal}
Kunisch, K., Volkwein, S.: Optimal snapshot location for computing {POD} basis
  functions.
\newblock ESAIM: Mathematical Modelling and Numerical Analysis \textbf{44}(3),
  509--529 (2010)

\bibitem{lassila2013model}
Lassila, T., Manzoni, A., Quarteroni, A., Rozza, G.: Model order reduction in
  fluid dynamics: challenges and perspectives.
\newblock In: A.~Quarteroni, G.~Rozza (eds.) Reduced Order Methods for modeling
  and computational reduction. Springer, Milano (2013)

\bibitem{lee1997application}
Lee, C., Kim, J., Babcock, D., Goodman, R.: Application of neural networks to
  turbulence control for drag reduction.
\newblock Physics of Fluids \textbf{9}(6), 1740--1747 (1997)

\bibitem{levenberg1944method}
Levenberg, K.: A method for the solution of certain non-linear problems in
  least squares.
\newblock Quarterly of Applied Mathematics \textbf{2}(2), 164--168 (1944)

\bibitem{loeve1955probability}
Lo{\`e}ve, M.: Probability Theory; Foundations, Random Sequences.
\newblock New York: D. Van Nostrand Company (1955)

\bibitem{lorenz1956empirical}
Lorenz, E.N.: Empirical orthogonal functions and statistical weather prediction
   (1956)

\bibitem{lucia2004reduced}
Lucia, D.J., Beran, P.S., Silva, W.A.: Reduced-order modeling: new approaches
  for computational physics.
\newblock Progress in Aerospace Sciences \textbf{40}(1), 51--117 (2004)

\bibitem{lumley1967structures}
Lumley, J.: The structures of inhomogeneous turbulent flow.
\newblock In: Atmospheric Turbulence and Radio Wave Propagation, A. Yaglom and
  V. Tatarski, eds, pp. 160--178 (1967)

\bibitem{mackay1992bayesian}
MacKay, D.J.: Bayesian interpolation.
\newblock Neural computation \textbf{4}(3), 415--447 (1992)

\bibitem{maleewong2011line}
Maleewong, M., Sirisup, S.: On-line and off-line {POD} assisted projective
  integral for non-linear problems: A case study with {B}urgers' equation.
\newblock International Journal of Mathematical, Computational, Physical,
  Electrical and Computer Engineering \textbf{5}(7), 984--992 (2011)

\bibitem{marquardt1963algorithm}
Marquardt, D.W.: An algorithm for least-squares estimation of nonlinear
  parameters.
\newblock Journal of the society for Industrial and Applied Mathematics
  \textbf{11}(2), 431--441 (1963)

\bibitem{moin2010fundamentals}
Moin, P.: Fundamentals of engineering numerical analysis.
\newblock Cambridge University Press (2010)

\bibitem{moosavi2015efficient}
Moosavi, A., Stefanescu, R., Sandu, A.: Efficient {C}onstruction of {L}ocal
  {P}arametric {R}educed {O}rder {M}odels {U}sing {M}achine {L}earning
  {T}echniques.
\newblock arXiv preprint arXiv:1511.02909  (2015)

\bibitem{moosavi2017multivariate}
Moosavi, A., Stefanescu, R., Sandu, A.: Multivariate predictions of local
  reduced-order-model errors and dimensions.
\newblock arXiv preprint arXiv:1701.03720  (2017)

\bibitem{narayanan1999low}
Narayanan, S., Khibnik, A., Jacobson, C., Kevrekedis, Y., Rico-Martinez, R.,
  Lust, K.: Low-dimensional models for active control of flow separation.
\newblock In: Control Applications, 1999. Proceedings of the 1999 IEEE
  International Conference on, vol.~2, pp. 1151--1156. IEEE (1999)

\bibitem{nguyen1990improving}
Nguyen, D., Widrow, B.: Improving the learning speed of 2-layer neural networks
  by choosing initial values of the adaptive weights.
\newblock In: Neural Networks, 1990., 1990 IJCNN International Joint Conference
  on, pp. 21--26. IEEE (1990)

\bibitem{noack2002low}
Noack, B., Papas, P., Monkewitz, P.: Low-dimensional {G}alerkin model of a
  laminar shear-layer.
\newblock Tech. rep., Tech. Rep. 2002-01. Laboratoire de Mecanique des Fluides,
  Departement de Genie Mecanique, Ecole Polytechnique F{\'e}d{\'e}rale de
  Lausanne, Switzerland (2002)

\bibitem{noack2011reduced}
Noack, B.R., Morzynski, M., Tadmor, G.: Reduced-order modelling for flow
  control, vol. 528.
\newblock Springer Verlag (2011)

\bibitem{ravindran2000reduced}
Ravindran, S.S.: A reduced-order approach for optimal control of fluids using
  proper orthogonal decomposition.
\newblock International Journal for Numerical Methods in Fluids \textbf{34}(5),
  425--448 (2000)

\bibitem{rowley2005model}
Rowley, C.W.: Model reduction for fluids, using balanced proper orthogonal
  decomposition.
\newblock International Journal of Bifurcation and Chaos \textbf{15}(03),
  997--1013 (2005)

\bibitem{rowley2017model}
Rowley, C.W., Dawson, S.T.: Model reduction for flow analysis and control.
\newblock Annual Review of Fluid Mechanics \textbf{49}, 387--417 (2017)

\bibitem{roychowdhury1999reduced}
Roychowdhury, J.: Reduced-order modeling of time-varying systems.
\newblock IEEE Transactions on Circuits and Systems II: Analog and Digital
  Signal Processing \textbf{46}(10), 1273--1288 (1999)

\bibitem{sahan1997artificial}
Sahan, R., Koc-Sahan, N., Albin, D., Liakopoulos, A.: Artificial neural
  network-based modeling and intelligent control of transitional flows.
\newblock In: Control Applications, 1997., Proceedings of the 1997 IEEE
  International Conference on, pp. 359--364. IEEE (1997)

\bibitem{san2013proper}
San, O., Iliescu, T.: Proper orthogonal decomposition closure models for fluid
  flows: {B}urgers equation.
\newblock International Journal of Numerical Analysis and Modeling \textbf{5},
  217--237 (2014)

\bibitem{san2015stabilized}
San, O., Iliescu, T.: A stabilized proper orthogonal decomposition
  reduced-order model for large scale quasigeostrophic ocean circulation.
\newblock Advances in Computational Mathematics \textbf{41}(5), 1289--1319
  (2015)

\bibitem{schmid2010dynamic}
Schmid, P.J.: Dynamic mode decomposition of numerical and experimental data.
\newblock Journal of fluid mechanics \textbf{656}, 5--28 (2010)

\bibitem{serre2002matrices}
Serre, D.: Matrices: Theory and Applications.
\newblock Springer-Verlag, New York (2002)

\bibitem{silveira1996efficient}
Silveira, L.M., Kamon, M., White, J.: Efficient reduced-order modeling of
  frequency-dependent coupling inductances associated with 3-{D} interconnect
  structures.
\newblock IEEE Transactions on Components, Packaging, and Manufacturing
  Technology: Part B \textbf{19}(2), 283--288 (1996)

\bibitem{sirisup2004spectral}
Sirisup, S., Karniadakis, G.E.: A spectral viscosity method for correcting the
  long-term behavior of \uppercase{POD} models.
\newblock Journal of Computational Physics \textbf{194}(1), 92--116 (2004)

\bibitem{sirovich1987turbulence}
Sirovich, L.: Turbulence and the dynamics of coherent structures.
  \uppercase{I-C}oherent structures. \uppercase{II-S}ymmetries and
  transformations. \uppercase{III-D}ynamics and scaling.
\newblock Quarterly of Applied Mathematics \textbf{45}, 561--571 (1987)

\bibitem{taira2017modal}
Taira, K., Brunton, S.L., Dawson, S., Rowley, C.W., Colonius, T., McKeon, B.J.,
  Schmidt, O.T., Gordeyev, S., Theofilis, V., Ukeiley, L.S.: Modal analysis of
  fluid flows: {A}n overview.
\newblock arXiv preprint arXiv:1702.01453  (2017)

\bibitem{ullmann2010pod}
Ullmann, S., Lang, J.: A {POD}-{G}alerkin reduced model with updated
  coefficients for {S}magorinsky {LES}.
\newblock In: V European conference on computational fluid dynamics, ECCOMAS
  CFD, p. 2010 (2010)

\bibitem{wang2012reduced}
Wang, Z.: Reduced-order modeling of complex engineering and geophysical flows:
  analysis and computations.
\newblock Ph.D. thesis, Virginia Tech (2012)

\bibitem{wang2011two}
Wang, Z., Akhtar, I., Borggaard, J., Iliescu, T.: Two-level discretizations of
  nonlinear closure models for proper orthogonal decomposition.
\newblock Journal of Computational Physics \textbf{230}(1), 126--146 (2011)

\bibitem{wang2012proper}
Wang, Z., Akhtar, I., Borggaard, J., Iliescu, T.: Proper orthogonal
  decomposition closure models for turbulent flows: a numerical comparison.
\newblock Computer Methods in Applied Mechanics and Engineering \textbf{237},
  10--26 (2012)

\bibitem{weller2010numerical}
Weller, J., Lombardi, E., Bergmann, M., Iollo, A.: Numerical methods for
  low-order modeling of fluid flows based on {POD}.
\newblock International Journal for Numerical Methods in Fluids \textbf{63}(2),
  249--268 (2010)

\bibitem{wells2017evolve}
Wells, D., Wang, Z., Xie, X., Iliescu, T.: An evolve-then-filter regularized
  reduced order model for convection-dominated flows.
\newblock International Journal for Numerical Methods in Fluids
  \textbf{10.1002/fld.4363}, 1--20 (2017)

\bibitem{widrow1994neural}
Widrow, B., Rumelhart, D.E., Lehr, M.A.: Neural networks: applications in
  industry, business and science.
\newblock Communications of the ACM \textbf{37}(3), 93--106 (1994)

\bibitem{xie2017approximate}
Xie, X., Wells, D., Wang, Z., Iliescu, T.: Approximate deconvolution reduced
  order modeling.
\newblock Computer Methods in Applied Mechanics and Engineering \textbf{313},
  512--534 (2017)

\bibitem{zhang1998forecasting}
Zhang, G., Patuwo, B.E., Hu, M.Y.: Forecasting with artificial neural networks:
  {T}he state of the art.
\newblock International Journal of Forecasting \textbf{14}(1), 35--62 (1998)

\end{thebibliography}

%
%

\end{document}